\documentclass[a4paper,11pt]{article}
\pdfoutput=1
\usepackage{jheppub}

\usepackage{graphicx}
\usepackage[figuresright]{rotating}
\usepackage{bm,amsmath,amssymb}
\usepackage[mathscr]{eucal}
\usepackage{multirow}
\usepackage{xcolor}
\usepackage{mathrsfs}
\usepackage{mathtools}
\usepackage{slashed}
\usepackage{graphics}
\usepackage{graphicx}
\usepackage{subfigure}
\usepackage{dsfont}
\usepackage{longtable}
\usepackage{bbm} 
\usepackage{float}
\usepackage{tcolorbox}
\usepackage{lipsum}
\usepackage{cancel}
\usepackage{enumerate}
\definecolor{lcolor}{rgb}{0.,0.0,0.}
\definecolor{citcolor}{rgb}{0,0.,0.5}

\newcommand{\Hcal}{\mathcal{H}}

\newcommand{\Ccal}{\mathcal{C}}
\newcommand{\Pcal}{\mathcal{P}}
\newcommand{\Rcal}{\mathcal{R}}
\newcommand{\Ical}{\mathcal{I}}
\newcommand{\Jcal}{\mathcal{J}}
\newcommand{\vect}[1]{\boldsymbol{#1}_{\perp}}

\newcommand{\kgt}{\boldsymbol{k_{g\perp}}}

\newcommand{\Pt}{\vect{P}}

\newcommand{\qt}{\vect{q}}
\newcommand{\lt}{\vect{l}}
\newcommand{\vt}{\vect{v}}
\newcommand{\at}{\vect{a}}
\newcommand{\bt}{\vect{b}}
\newcommand{\Kt}{\vect{K}}

\newcommand{\ruupt}{\boldsymbol{r}_{uu'}}
\newcommand{\ktone}{\boldsymbol{k_{1\perp}}}
\newcommand{\kttwo}{\boldsymbol{k_{2\perp}}}

\newcommand{\wt}{\vect{w}}

\newcommand{\xt}{\vect{x}}
\newcommand{\yt}{\vect{y}}
\newcommand{\zt}{\vect{z}}
\newcommand{\ut}{\vect{u}}
\newcommand{\st}{\vect{s}}

\newcommand{\Xt}{\vect{X}}
\newcommand{\Xttilde}{\boldsymbol{\tilde{X}_\perp}}

\newcommand{\rt}{\vect{r}}

\newcommand{\rxyt}{\boldsymbol{r}_{xy}}
\newcommand{\rbbpt}{\boldsymbol{r}_{bb'}}

\newcommand{\rzxt}{\boldsymbol{r}_{zx}}
\newcommand{\rzyt}{\boldsymbol{r}_{zy}}
\newcommand{\rzbt}{\boldsymbol{r}_{zb}}
\newcommand{\rzbpt}{\boldsymbol{r}_{zb'}}
\newcommand{\rzzpt}{\boldsymbol{r}_{zz'}}

\newcommand{\rxytp}{\boldsymbol{r}_{x'y'}}
\newcommand{\rxxtp}{\boldsymbol{r}_{xx'}}
\newcommand{\ryytp}{\boldsymbol{r}_{yy'}}
\newcommand{\rxypt}{\boldsymbol{r}_{xy'}}
\newcommand{\ryxpt}{\boldsymbol{r}_{yx'}}

\newcommand{\rzypt}{\boldsymbol{r}_{zy'}}
\newcommand{\rzxpt}{\boldsymbol{r}_{zx'}}
\newcommand{\rzxtp}{\boldsymbol{r}_{z'x'}}
\newcommand{\rxpyt}{\boldsymbol{r}_{x'y}}

\newcommand{\ptj}{\boldsymbol{p}_J}

\newcommand{\RtS}{\boldsymbol{R}_{\rm SE}}
\newcommand{\RtV}{\boldsymbol{R}_{\rm V}}

\newcommand{\BesselJ}{\textrm{J}}
\newcommand{\der}{\mathrm{d}}

\newcommand{\Tr}{\mathrm{Tr}}
\newcommand{\deltatwo}{\delta^{\rm(2)}_z}
\newcommand{\deltathree}{\delta^{\rm(3)}_z}

\title{Back-to-back inclusive dijets in DIS at small $x$: Sudakov suppression and gluon saturation at NLO}
\author[a]{Paul Caucal,}
\emailAdd{pcaucal@bnl.gov}
\author[b,c,d,e]{Farid Salazar,}
\emailAdd{salazar@physics.ucla.edu}
\author[a]{Bj\"{o}rn Schenke,}
\emailAdd{bschenke@bnl.gov}
\author[a]{Raju Venugopalan}
\emailAdd{raju@bnl.gov}

\affiliation[a]{Physics Department, Brookhaven National Laboratory, Upton, NY 11973, USA}
\affiliation[b]{Department of Physics and Astronomy, University of California, Los Angeles, California 90095, USA}
\affiliation[c]{Mani L. Bhaumik Institute for Theoretical Physics, University of California, Los Angeles, California 90095, USA}
\affiliation[d]{Nuclear Science Division, Lawrence Berkeley National Laboratory, Berkeley, California 94720, USA}
\affiliation[e]{Physics Department, University of California, Berkeley, California 94720, USA}

\abstract{
Back-to-back dijet cross-sections in deeply inelastic scattering (DIS) at small $x_{\rm Bj}$ are suppressed by many-body multiple scattering and screening effects arising from gluon saturation at high parton densities. They are similarly sensitive in these kinematics to large Sudakov logarithms from soft gluon radiation. Uncovering novel physics in this DIS channel therefore requires understanding the interplay of the two phenomena. In this work, we compute the small $x_{\rm Bj}$ inclusive dijet DIS cross-section in back-to-back kinematics at next-to-leading order (NLO) in the Color Glass Condensate effective field theory (CGC EFT). Our result includes, for the first time, all real and virtual NLO contributions to the impact factor. These include all Sudakov double and single logarithm contributions, as well as all other finite  $\mathcal{O}(\alpha_s)$ terms that contribute at this order. We demonstrate explicitly that resummations of small $x$ and Sudakov logarithms can be performed simultaneously in the CGC EFT. This requires that the JIMWLK kernel for small $x$ evolution of the Weizs\"{a}cker-Williams (WW) gluon distribution satisfies a kinematic constraint imposed by lifetime ordering of successive gluon emissions; the corresponding modifications to the kernel, corresponding to resummations of large double transverse logarithms, are precisely of the type required to stabilize JIMWLK evolution beyond leading logarithmic accuracy. We compute the azimuthal harmonics of the NLO back-to-back distributions and show their sensitivity to both the unpolarized and linearly polarized WW gluon distributions. Finally, we discuss how TMD factorization is broken by an emergent saturation scale at small $x$.}
 
\begin{document}
\maketitle
\newpage 
\section{Introduction}

The phenomenon of gluon saturation~\cite{Gribov:1984tu,Mueller:1985wy} refers to the  many-body screening and recombination of gluons that contribute to significantly tame the growth of gluon distributions inside hadron wavefunctions at high energies. It is characterized by an emergent semi-hard saturation scale $Q_s(x)\gg \Lambda_{\rm QCD}$, which grows with increasing energy (or decreasing $x$).
The discovery and characterization of gluon saturation in high energy deeply inelastic electron-proton or electron-nucleus scatterings is one the principal goals of the future Electron-Ion-Collider (EIC)~\cite{Accardi:2012qut,Aschenauer:2017jsk,AbdulKhalek:2021gbh}. 

In this regard, the inclusive production of a dijet (or dihadron) pair in deep-inelastic scattering (DIS) is of great phenomenological interest, especially when the two jets are 
produced with nearly back-to-back transverse momenta. Motivated by experimental signatures \cite{STAR:2006dgg,Braidot:2010zh,PHENIX:2011puq,STAR:2021fgw} and phenomenological studies \cite{Marquet:2007vb,Lappi:2012nh,Albacete:2018ruq,Stasto:2018rci,Benic:2022ixp} of two-particle correlations in hadronic collisions, the observation of the suppression of the back-to-back peak in dijet/dihadron azimuthal correlations at sufficiently low Bjorken $x_{\rm Bj}$ is seen as promising signature of gluon saturation at the EIC \cite{Zheng:2014vka}. A further motivation for this back-to-back dijet/dihadron measurement is its potential sensitivity to the unpolarized and linearly polarized Weizs\"acker-Williams transverse momentum dependent (TMD) gluon distributions~\cite{Dominguez:2010xd,Dominguez:2011wm,Metz:2011wb}.

In addition to the many-body multiple scattering and screening of saturated gluons in the target, there is an additional source of suppression for the back-to-back dijet  cross-section. This is a  purely vacuum-like effect induced by multiple soft gluon radiation that suppresses the formation of a strictly back-to-back pair of jets.  This is the well-known Sudakov effect and its contribution is enhanced to all orders $n\ge 1$ in perturbative QCD by large double and single transverse logarithms,  $\sim \alpha_s^n\ln^{2n}(P_\perp/q_\perp)$ and $\sim \alpha_s^n\ln^{n}(P_\perp/q_\perp)$ respectively, of the ratio between the mean transverse momentum $P_\perp$ of the dijet pair and the ``soft" scale associated with the momentum imbalance $q_\perp$ of the pair \cite{Mueller:2013wwa}. The resummation and exponentiation of such logarithms into a Sudakov form factor is an essential feature of the Collins-Soper-Sterman (CSS) TMD formalism~\cite{Collins:1981uk,Collins:1981uw,Collins:1984kg}.

Leading order studies of the inclusive dijet cross-section in the Regge limit of DIS, including Sudakov effects at double logarithmic accuracy, have been performed in \cite{Zheng:2014vka,Zhao:2021kae,vanHameren:2021sqc}, showing the significant impact of Sudakov suppression on the dijet azimuthal correlations. Therefore any  conclusions regarding the effect of gluon saturation on inclusive dijet production in DIS for back-to-back kinematics depends on the quantitative understanding of the interplay  between both of these phenomena, gluon saturation and the Sudakov effect, in the small $x$ regime of QCD.

We will address here this longstanding problem within the framework of the Color Glass Condensate (CGC) effective field theory~\cite{Iancu:2003xm,Gelis:2010nm,Kovchegov:2012mbw,Albacete:2014fwa,Blaizot:2016qgz,Morreale:2021pnn}. In the CGC, the coherent multiple scatterings  of a high energy colored parton with the dense gluon fields of the hadron target are described by ``shockwave" propagators containing momentum-dependent effective vertices that are proportional to the Fourier transforms of the spatial distribution of lightlike Wilson lines in the strong background fields (of $\mathcal{O}(1/g)$, where $g$ is the QCD coupling) corresponding to saturated gluons in the nuclear target. Using standard perturbative QCD Feynman diagram techniques in the  CGC EFT, we  performed in \cite{Caucal:2021ent} the first complete NLO computation of inclusive dijet production in DIS at small $x$. We demonstrated that the dijet cross-section is infrared and collinear finite, and can be factorized at NLO into a convolution between a perturbatively calculable impact factor and nonperturbative expectation values of correlators of lightlike Wilson lines (which are dubbed ``color correlators" throughout this paper). The rapidity or ``slow gluon" divergence\footnote{Slow refers to the gluon carrying a small longitudinal momentum fraction relative to projectile's (photon) large momentum component while ``soft" in our context will correspond to all of the gluon momenta being small.} in the calculation is absorbed in the renormalization of these color correlators;  the resulting renormalization group (RG) equations  to leading logarithm accuracy (resumming all powers of $\alpha_s\, Y$, where $Y\sim \ln(1/x)$ is the rapidity separation between the projectile and the target), are shown explicitly to be the B-JIMWLK evolution equations~\cite{Balitsky:1995ub,JalilianMarian:1996xn,JalilianMarian:1997dw,Kovner:2000pt,Iancu:2000hn,Iancu:2001ad,Ferreiro:2001qy}. We note that, for large $N_c$  and large mass number $A$, the 2-point dipole correlator in the B-JIMWLK hierarchy is the BK equation~\cite{Balitsky:1995ub,Kovchegov:1999yj}. This nonlinear equation, in turn, reduces to the 
BFKL equation in the dilute limit of low parton densities~\cite{Kuraev:1977fs,Balitsky:1978ic}.

The NLO impact factor is therefore constructed to be free of the logarithms corresponding to the slow rapidity divergences. Further, the expression for the 
impact factor is process-dependent in contrast to the evolution of the color correlators which should be universal. This rapidity factorization~\cite{Balitsky:2010ze} has been shown for a number of processes at the stated leading logarithmic accuracy~\cite{Chirilli:2011km,Altinoluk:2014eka,Boussarie:2016bkq,Boussarie:2016ogo,Beuf:2017bpd,Hanninen:2017ddy,Roy:2019cux,Roy:2019hwr,Beuf:2020dxl,Liu:2020mpy,Shi:2021hwx,Taels:2022tza,Mantysaari:2022kdm,Liu:2022ijp}. Thus far it has been demonstrated explicitly to hold at next-to-leading-logarithmic accuracy in $x$ only for the fully inclusive DIS cross-section \cite{Balitsky:2010ze,Balitsky:2012bs}. 
We note that the NLO BK/JIMWLK Hamiltonian has been derived and discussed at length in~\cite{Balitsky:2007feb,Kovner:2013ona,Caron-Huot:2016tzz,Caron-Huot:2013fea,Dai:2022imf}.

The Sudakov logarithms are part of the NLO impact factor that provide large contributions in back-to-back kinematics. They can therefore be extracted from the NLO impact factor for fully inclusive dijets that we computed in \cite{Caucal:2021ent}. This is however not easy to do ``by inspection" due to the considerable complexity of the 
NLO expressions. However, as we will discuss at length, one can reorganize the NLO impact factor results into three blocks of terms. One of these, in the back-to-back-limit, reduces to  double and single Sudakov logarithms plus finite $\mathcal{O}(\alpha_s)$ pieces; another reduces to just single Sudakov log and finite pieces. The final block only contains finite $\mathcal{O}(\alpha_s)$ pieces. 

The subsequent extraction, and further manipulation, of the double Sudakov logarithms  leads, at first sight, to a surprising result: the coefficient of this contribution has the wrong sign, corresponding to a Sudakov enhancement rather than the suppression required on physical grounds. The recovery of the correct sign follows from a nontrivial interplay between the \textit{soft} gluon radiation contributing to the Sudakov effect  with the \textit{slow} gluon emission contributing to small $x$ evolution. These are separated by a rapidity factorization scale $Y_f$ such that $\alpha_s Y_f \ll 1$ terms ``above the cut" contribute to the impact factor and $\alpha_s Y_f \geq 1$ terms contribute  below this cutoff to the RG evolution. We find that the underlying reason for the wrong sign is the inclusion in the slow gluon RG evolution a piece of the gluon phase-space that properly belongs to the impact factor. This is seen by imposing a physical kinematic constraint on the slow gluon phase-space that enforces both lightcone  momentum and lifetime ordering of the successive gluon emissions described by B-JIMWLK evolution.
Our observations in this regard, and the fact that a kinematically constrained B-JIMWLK evolution of the color correlators is required to recover the correct sign for Sudakov suppression, are in line with those noted recently in the context of inclusive dijet photoproduction at small $x$ \cite{Taels:2022tza}.

This feature of RG evolution is reminiscent of the fully inclusive DIS cross-section at small $x$, and could have therefore been anticipated. For the fully inclusive DIS case, the corresponding large double transverse logarithms are the DGLAP double collinear logarithms in the squared momentum transfer; these are not correctly accounted for by BFKL or BK evolution in the projectile's rapidity. The solution to this problem  \cite{Salam:1998tj,Ciafaloni:1998iv,Ciafaloni:1999yw,Ciafaloni:2003rd,SabioVera:2005tiv,Beuf:2014uia,Iancu:2015vea}  is very similar to the Sudakov problem discussed 
here: one can either modify the BFKL/BK kernel in order to impose lifetime ordering in addition to the projectile's rapidity ordering, or use directly the target rapidity as the evolution variable (``the correct choice of the energy scale" in the terminology of \cite{Salam:1998tj}). However in the latter case, it is difficult to consistently combine the small $x$ RG evolution with the impact factor, since the latter is more conveniently calculated in the dipole frame.

We will compute here the coefficients of the double and single Sudakov logarithms at finite $N_c$ for both longitudinally and transversely polarized virtual photons. 
We recover the results provided in \cite{Mueller:2013wwa,Taels:2022tza} for the double Sudakov logarithms. For the finite $N_c$ single Sudakov logarithm terms, we  recover the single log contributions computed in \cite{Hatta:2020bgy,Hatta:2021jcd} and additionally, a single log term whose coefficient is sensitive to the rapidity factorization scale $Y_f$. The presence of this scale further illustrates the interplay between the Sudakov and small $x$ resummation; this is a novel feature of Sudakov resummation at small $x$ going beyond CSS resummation. A further difference of the CGC EFT with the collinear factorization framework is the absence of a term in the Sudakov form factor proportional to the coefficient of the QCD $\beta$-function~\cite{Catani:1988vd,Catani:1989ne}. The absence of this term has been discussed previously in \cite{Xiao:2017yya,Hentschinski:2021lsh}; while we agree with \cite{Xiao:2017yya} that such a contribution does not exist in general Regge asymptotics, we argue that it can be recovered in a collinear limit of the CGC EFT.

Not least, we  provide complete expressions for the finite (non-logarithmically enhanced)  terms in back-to-back kinematics, which are pure $\alpha_s$ corrections. Expressing our results in terms of the azimuthal harmonics of back-to-back cross-section\footnote{These correspond to the $\cos(n\phi)$ averaged cross-sections, 
where $\phi$ is the relative angle between $\Pt$~and~$\qt$.}, we observe that these are of two kinds. One sort, we will discuss first, are the terms that do not break leading order TMD factorization albeit, as we show, this requires the introduction of additional perturbative hard factors at NLO. For such terms, we show that the azimuthally averaged cross-section is not only sensitive to the unpolarized WW gluon TMD, but to its linearly polarized component as well. This contribution is phenomenologically important since the linearly polarized gluon distribution is large in the small $x$ regime; our computation captures all the $\mathcal{O}(\alpha_s)$ contributions that are sensitive to both the unpolarized and linearly polarized WW and gluon TMDs.

Similarly, the $\langle \cos(2\phi)\rangle$ anisotropy becomes sensitive to the unpolarized WW gluon TMD in addition to the linearly polarized WW gluon TMD. One such contribution was computed in \cite{Hatta:2020bgy,Hatta:2021jcd}; we recover this contribution (up to differences in $1/N_c^2$ relative factors). However we show  that in addition there are several other contributions, not computed previously, that are  proportional to the linearly polarized WW gluon TMD and contribute with equal magnitude to this $\langle \cos(2\phi)\rangle$  anisotropy. 

The other sort of contributions result in the breaking of TMD factorization at NLO. We argue that TMD factorization holds if and only if the saturation scale is the smallest perturbative scale of the problem as compared to $q_\perp$ and $P_\perp$: $Q_s\ll q_\perp\ll P_\perp$. Beyond this very specific kinematic regime, when $Q_s$ becomes comparable to $q_\perp\ll P_\perp$, TMD factorization is violated at NLO due to the increasing importance of higher twist corrections.

The paper is organized as follows. In section~\ref{sec:LO}, we provide a brief outline of the CGC EFT and discuss the leading order inclusive dijet cross-section in back-to-back kinematics. The notations and conventions that we will use in the following sections are also introduced here. In section~\ref{sec:update-if}, we provide the expression for the full NLO inclusive dijet cross-section in DIS previously computed in \cite{Caucal:2021ent}. As noted earlier, we reorganize the various terms in the 
NLO impact factor in a manner that will simplify the extraction of Sudakov logarithms in  back-to-back dijet kinematics. Section~\ref{sec:sudakov} is dedicated to the computation of the back-to-back limit of the NLO impact factor. We identify the double and single Sudakov logarithms and discuss their relations with the small $x$ or rapidity logarithms that are resummed by the B-JIMWLK evolution equation. Finally, the last section is dedicated to the computation of the pure $\alpha_s$ corrections which are not power suppressed in  $q_\perp/P_\perp$ and to the complete $\mathcal{O}(\alpha_s)$ computation of the unpolarized and linearly polarized WW gluon TMDs. Our final results for the zeroth and second harmonics of the back-to-back cross-section are given in Eqs.~\ref{eq:zeroth-moment-final} and \ref{eq:second-moment-final}.

This paper is supplemented by appendices that provide details of the computations for the interested reader. The notations and conventions employed in this work are summarized in Appendix~\ref{app:convention}. Appendix~\ref{app:xs-dec} details the calculation of the various terms in the decomposition of the NLO impact factor presented in section~\ref{sec:update-if} of the main text for the case of a longitudinally polarized virtual photon. The expressions for transversely polarized photons are given in Appendix~\ref{app:transverse}. In Appendix~\ref{app:jet-algo}, we provide the NLO impact factor for inclusive dijet production with jets defined using generalized $k_t$ algorithms. High harmonics ($n\ge 4$) induced by soft gluons in the Fourier decomposition of the azimuthal dependence of the inclusive dijet cross-section are calculated in Appendix~\ref{app:cosnphi}. Finally, Appendix~\ref{app:math-id} gathers useful integral identities relevant to the computation of the back-to-back limit of our NLO impact factor.

\section{Leading order cross-section in back-to-back kinematics}
\label{sec:LO}

In this section, we will derive from the CGC effective field theory the back-to-back limit of the leading order inclusive dijet cross-section in DIS at small Bjorken $x_{\rm Bj}$. In particular, we recover the transverse momentum dependent factorization formula involving the Weizs\"acker-Williams gluon distribution, first derived in \cite{Dominguez:2011wm}.

\subsection{Overview of the CGC effective field theory}

In the CGC effective field theory, the small $x$ gluons with high occupancy number are represented by a classical color field $A^\mu_{\rm cl}$. The classical field is generated by the large-$x$ degrees of freedom of the target nucleus $A$, which are treated as stochastic color sources with color charge density $\rho_A^a$. The sources $\rho_A^a$ and field $A^\mu_{\rm cl}$ are related by the Yang-Mills equations $[D_\mu,F^{\mu\nu}]=J^\nu$, where 
\begin{equation}
    J^\mu(x^-,\xt)=\delta^{\mu+}\rho_A(x^-,\xt)\,,
\end{equation}
the 4-current associated with the large-$x$ sources. Since the target is fast moving along the plus lightcone direction, the current does not depend on $x^+$.
The solution to the Yang-Mills equations is\footnote{The solution in Eq.\,\eqref{eq:CGC_background_field} also satisfies the Lorenz gauge condition $\partial_\mu A^{\mu}_{\rm{cl}} = 0$.}
\begin{equation}
    A^\mu_{\rm cl}(x)=\delta^{\mu+}\alpha(x^-,\xt)\,,\qquad \nabla_\perp^2\alpha(x^-,\xt)=-\rho_A(x^-,\xt)\,,
    \label{eq:CGC_background_field}
\end{equation}
in lightcone gauge $A^-=0$.

In the CGC, the eikonal scattering of a high energy parton moving along the minus lightcone direction in the background field of the small $x$ gluons is described by a lightlike Wilson line. For a fast moving quark propagating in the small $x$ background field, one defines the Wilson line in the fundamental representation as
\begin{equation}
    V_{ij}(\xt)=\mathcal{P}\exp\left(ig\int_{-\infty}^{\infty}\der z^-A^{+,a}_{\rm cl}(z^-,\xt)t^a_{ij}\right)\,,
\end{equation}
which physically corresponds to its color rotation. Here, $t^a$ are the generators of $\rm SU(3)$ in the fundamental representation. The Wilson line resums to all orders multiple scatterings between the quark and
the small $x$ gluons in the target, and ensures that the cross-section satisfies unitarity in the high-energy limit. Analogously, the propagation of a gluon will characterized by a Wilson line in the adjoint representation.

A generic observable $\mathcal{O}$, such as a cross-section, in the CGC effective field theory therefore depends on products of Wilson lines, and consequently, on the background field $A^\mu_{\rm cl}[\rho_a]$ for a given large $x$ color charge configuration $\rho_A$. This color charge configuration is drawn from a stochastic gauge invariant distribution $W_{Y}[\rho_A]$ defined at the rapidity scale $Y=\ln(z)$ for a given typical $z$ fraction of the projectile $q^-$ momentum probed by the observable $\mathcal{O}$. A more precise specification of this scale is discussed in section~\ref{subsub:JIMWLK} when we address the leading logarithmic high energy evolution induced by quantum corrections. Ultimately, any observable needs to be averaged over these color charge configurations:
\begin{equation}
    \langle \mathcal{O}\rangle_{Y}=\int\mathcal{D}[\rho_A]W_{Y}[\rho_A]\mathcal{O}[\rho_A]\,.
\end{equation}
This classical CGC average represents the fact that the large $x$ color sources are frozen on the time scales of the small $x$ gauge field dynamics.

\subsection{Full CGC result}
\label{sub:LO-CGC}

We now provide the formula for the inclusive dijet cross-section within the CGC effective field theory. We work in the dipole frame in which the incoming photon $\gamma_\lambda^\star$ with virtuality squared $Q^2$ has a large $q^-$ component and zero transverse momentum, while the target proton or nucleus has a large $P^+$ component:
\begin{align}
   q^\mu=(-Q^2/(2q^-),q^-,\vect{0}) \,, \quad \quad \quad
   P^\mu=(P^+,M^2/(2P^+),\vect{0}) \,.
\end{align}

The polarization of the photon is denoted by $\lambda$, with $\lambda=0$ for a longitudinally polarized photon and $\lambda=\pm 1$ for a transversely polarized photon. 

At leading order, the virtual photon splits into a quark-antiquark pair that subsequently interacts with the small $x$ gluons of the target before fragmenting into two jets. The 4-momenta of the quark and antiquark are denoted respectively by $k_1^\mu$ and $k_2^\mu$ and the longitudinal momentum fractions with respect to the virtual 
photon are $z_1=k_1^-/q^-$ and $z_2=k_2^-/q^-$. 

The fully differential leading order cross-section for inclusive production of two jets can be written in the compact form\footnote{For a detailed derivation see e.g. Sec.\,2 in \cite{Caucal:2021ent}.}
\begin{align}
    \left.\frac{\der \sigma^{\gamma_{\lambda}^{\star}+A\to q\bar{q}+X}}{ \der^2 \ktone \der^2 \kttwo \der \eta_1 \der \eta_{2}}\right|_{\rm LO}  \!\!\!\!\!\!\! =  \frac{\alpha_{\mathrm{em}} e_f^2 N_c\deltatwo}{(2\pi)^6} \int &\der^8 \Xt e^{-i\ktone\cdot(\xt-\xt')}e^{-i\kttwo\cdot(\yt-\yt')}\nonumber\\
    &\times\Xi_{\mathrm{LO}}(\xt,\yt;\xt', \yt')  \Rcal_{\mathrm{LO}}^{\lambda}(\rxyt,\rxytp)\label{eq:dijet-LO-cross-section} \,.
\end{align}
In this expression, $\alpha_{\rm em}$ is the electromagnetic fine structure constant, $e_f^2$ is the sum of the squares of the light quark fractional charges, and $\deltatwo=\delta(1-z_1-z_2)$ is an overall longitudinal momentum conserving delta function. The cross-section is provided in coordinate space with an 8 dimensional integral, whose differential measure is represented as 
\begin{equation}
    \der^8\Xt=\der^2\xt\der^2\xt'\der^2\yt\der^2\yt' \,,
\end{equation}
with $\xt$ ($\yt$) the transverse coordinate at which the quark (antiquark) crosses the shockwave in the amplitude (and similarly with prime coordinates for the complex conjugate amplitude). We also denote differences of transverse spatial coordinates 
as  
\begin{equation}
    \rxyt\equiv\xt-\yt\,.
\end{equation}

In Eq.\,\eqref{eq:dijet-LO-cross-section}, the integrand is factorized into a perturbative factor $\mathcal{R}_{\rm LO}^\lambda$ corresponding to the QED splitting of the virtual photon into the quark-antiquark pair and the color correlator $\Xi_{\rm LO}$ describing the interaction of the pair with the small $x$ gluons of the target. The perturbative factors are 
\begin{align}
    \Rcal_{\mathrm{LO}}^{\mathrm{L}}(\rxyt,\rxytp) &=  8 z_1^3 z_{2}^3  Q^2 K_0(\bar{Q} r_{xy}) K_0(\bar{Q} r_{x'y'}) \,, \label{eq:dijet-NLO-LLO} \\
    \Rcal_{\mathrm{LO}}^{\mathrm{T}}(\rxyt,\rxytp) &=  2 z_1 z_{2} \left[z_1^2 + z_{2}^2 \right]  \frac{\rxyt \cdot \rxytp}{r_{xy} r_{x'y'}}  \bar{Q}^2K_1(\bar{Q} r_{xy}) K_1(\bar{Q}r_{x'y'})\label{eq:dijet-NLO-TLO} \,,
\end{align}
respectively, for longitudinally and transversely polarized virtual photons. The effective virtuality $\bar Q$ is defined to be $\bar Q^2=z_1z_2 Q^2$. We note $K_n(x)$ is the modified Bessel function of second kind and order $n$.
 
The color correlator is a CGC average of the product of Wilson lines at some projectile rapidity scale $Y$:
\begin{align}
    \Xi_{\mathrm{LO}}(\xt,\yt;\xt', \yt')&=\frac{1}{N_c}\left\langle\Tr\left[\left(V(\xt)V^\dagger(\yt)-\mathbbm{1}\right)\left(V(\yt')V^\dagger(\xt')-\mathbbm{1}\right)\right]\right\rangle_{Y}\label{eq:xiLO-def}\\
    &=\left \langle Q_{xy,y'x'} - D_{xy} -  D_{y'x'} + 1 \right \rangle_{Y} \,,
\end{align}
where the dipole $D$ and quadrupole $Q$ operators are defined as
\begin{align}
 D_{xy}&=\frac{1}{N_c}\Tr\left( V(\xt)V^\dagger(\yt)\right) \,,\\
    Q_{xy,y'x'}&=\frac{1}{N_c}\Tr \left(V(\xt)V^\dagger(\yt)V(\yt')V^\dagger(\xt')\right) \,.
\end{align}
The rapidity $Y$ at which one evaluates the weight functional $W_Y[\rho_A]$ in the CGC average is arbitrary at leading order\footnote{ In phenomenological applications, one typically chooses $Y=\ln(1/x_g)$, where $x_g$ is the typical momentum fraction transferred from the target to the projectile (see for example  \cite{Zheng:2014vka,vanHameren:2021sqc}).}. The appropriate choice of $Y$ at higher orders will be addressed at length in section~\ref{subsub:JIMWLK}.

\subsection{Correlation limit:  TMD factorization}

We turn now to a discussion of the leading order inclusive dijet cross-section in back-to-back kinematics, and its relation with the
so-called ``correlation limit" \cite{Dominguez:2010xd,Dominguez:2011wm} of the ``all-twist" cross-section provided by the CGC formula Eq.\,\eqref{eq:dijet-LO-cross-section}. This correlation limit, defined precisely in the following, admits a TMD-like factorization involving the Weizs\"acker-Williams (WW) gluon distribution.

\subsubsection{Back-to-back kinematics}

To define back-to-back kinematics, we introduce as usual the momentum imbalance $\qt$ and the relative transverse momentum $\Pt$ by
\begin{align}
    \qt&=\ktone+\kttwo \,, \\
    \Pt&=z_2\ktone-z_1\kttwo \,.
\end{align}
The back-to-back limit is defined by $|\qt|\ll |\Pt|$. In order to study this limit at leading order, it is convenient to introduce the transverse coordinates conjugate to $\Pt$ and $\qt$ in the integral \eqref{eq:dijet-LO-cross-section}:
\begin{align}
\ut &= \xt-\yt\,,\\
\bt &= z_1 \xt+z_{2}\yt \,.
\end{align}
The measure $\der^8 \Xt$ is invariant under the change of variable $(\xt,\yt)\to(\ut,\bt)$; the dijet cross-section can therefore be expressed as
\begin{align}
    \left.\frac{\der \sigma^{\gamma_{\lambda}^{\star}+A\to q\bar{q}+X}}{ \der^2 \Pt \der^2 \qt \der \eta_1 \der \eta_{2}}\right|_{\rm LO}  \!\!\!\!\!\!\! = & \frac{\alpha_{\mathrm{em}} e_f^2 N_c\deltatwo}{(2\pi)^6}    \int \der^8 \Xttilde e^{-i\Pt\cdot\ruupt}e^{-i\qt\cdot\rbbpt}\Rcal_{\mathrm{LO}}^{\lambda}(\ut,\ut')\nonumber\\
    &\times\Xi_{\mathrm{LO}}(\bt+z_{2}\ut,\bt-z_1\ut;\bt'+z_{2}\ut',\bt'-z_1\ut') \,, \label{eq:dijet-LO-rtbt}
\end{align}
with
\begin{equation}
    \der^8\Xttilde =\der^2\ut\der^2\ut'\der^2\bt\der^2\bt' \,.
\end{equation}

\subsubsection{Correlation limit in back-to-back kinematics}
\label{subsub:corlimit}

We will now aim to find the leading term in an expansion in powers of $q_\perp/P_\perp$ in  Eq.\,\eqref{eq:dijet-LO-rtbt}. In coordinate space, the standard way to obtain this limit is to work in the ``correlation" limit $|\ut|\ll |\bt|$, $|\ut'|\ll |\bt'|$ inside the integral over these transverse variables, and expand the color correlator up to order $\ut \ut'$. The mathematical justification for this procedure comes from the fact that $\ut$ and $\Pt$ are conjugate variables, as are $\bt$ and $\qt$, through the phases in Eq.\,\eqref{eq:dijet-LO-rtbt}, and therefore $|\ut|\sim 1/|\Pt|\ll|\bt|\sim1/ |\qt|$.

This correlation limit is known to correctly account for the leading (non power suppressed) term in the  $q_\perp/P_\perp$ expansion and all twists in $Q_s/q_\perp$, but fails to capture ``genuine" higher twist corrections to all orders in $Q_s/P_\perp$ as well as sub-leading ``kinematic" twist correction of to all orders in $q_\perp/P_\perp$ \cite{Altinoluk:2019fui,Altinoluk:2019wyu,Boussarie:2020vzf,Mantysaari:2019hkq,Boussarie:2021ybe,Fujii:2020bkl}.
Such kinematic twists can be incorporated using the ITMD framework \cite{Petreska:2018cbf,Kotko:2015ura,vanHameren:2016ftb}, which at LO interpolates between the TMD factorization regime ($Q_s \sim q_\perp \ll P_\perp$) \cite{Belitsky:2002sm,Bomhof:2006dp} and the high-energy factorization regime ($Q_s \ll q_\perp \sim P_\perp$) \cite{Catani:1990eg,Collins:1991ty,Blaizot:2004wu,Blaizot:2004wv}. In this paper, we will focus on the leading term in $q_\perp/P_\perp$ while including all twists in $Q_s/q_\perp$, namely the TMD region. We leave for future work the interesting problems of extending the ITMD framework to NLO, and comparisons with the high-energy factorization framework\footnote{For recent studies within the high-energy factorization framework at NLO see e.g. \cite{Hentschinski:2020tbi,Nefedov:2020ecb,Hentschinski:2021lsh,Celiberto:2022fgx,vanHameren:2022mtk}}.

To expand $\Xi_{\rm LO}$ to lowest order in $\ut$ and $\ut'$, it is convenient to start with its definition in Eq.~\eqref{eq:xiLO-def}; expanding the pair of Wilson lines\footnote{We expand the Wilson lines as $V(\bt) - z_1 \ut^i \partial^i V(\bt) + \mathcal{O}(\ut^2)$, where the minus sign comes from the metric. Furthermore, we also employ $\left(\partial^i V(\bt)\right)V^\dagger(\bt)=-V(\bt)\partial^iV^\dagger(\bt)$, which follows from the unitarity of Wilson lines.}, we first notice that
\begin{align}
    V(\xt)V^\dagger(\yt)-\mathbbm{1}=\ut^iV(\bt)\partial^iV^\dagger(\bt)+\mathcal{O}(\ut^2) \,,\label{eq:VxVy-exp}
\end{align}
so that 
\begin{equation}
    \Xi_{\mathrm{LO}}(\xt,\yt;\xt',\yt')\approx\ut^i\ut'^j\times\frac{1}{N_c}\left\langle\Tr\left[V(\bt)\left(\partial^iV^\dagger(\bt) \right) \left(\partial^jV(\bt')\right)V^\dagger(\bt')\right]\right\rangle_{Y} \,.
    \label{eq:correlation_limit_expansion}
\end{equation}
As shown in \cite{Dominguez:2011wm} (see also Appendix F in \cite{Boussarie:2021ybe}), the operator appearing in this expression is nothing but the operator definition of the Weizs\"acker-Williams gluon TMD distribution~\cite{McLerran:1993ka,McLerran:1993ni}. We define this TMD distribution at small $x$ as
\begin{align}
    G^{ij}_{Y}(\qt)&=\int\frac{\der^2\bt\der^2\bt'}{(2\pi)^4}e^{-i\qt\cdot(\bt-\bt')}\hat G^{ij}_{Y}(\bt,\bt')\,,
\end{align}
where
\begin{align}
    \hat G^{ij}_{Y}(\bt,\bt')&\equiv\frac{-2}{\alpha_s}\left\langle\Tr\left[V(\bt) \left(\partial^iV^\dagger(\bt) \right) V(\bt') \left(\partial^jV^\dagger(\bt') \right)\right]\right\rangle_{Y}\,.
    \label{eq:WWTMD}
\end{align}
At small $x$, the WW gluon TMD implicitly depends on the saturation scale\footnote{
For discussions on the geometrical scaling of the WW distribution with the saturation scale, we refer the reader to  \cite{Dominguez:2011gc,Dominguez:2011br,Dumitru:2015gaa}.} $Q_s$, and resums all powers in $Q_s/q_\perp$.

Employing the expansion in Eq.\,\eqref{eq:correlation_limit_expansion} to the all twist result in Eq.~\eqref{eq:dijet-LO-rtbt}, one finds the following factorized form for the leading term $q_\perp/P_\perp$ (while including all twists in $Q_s/q_\perp$) of the differential cross-section in the back-to-back kinematics
\begin{align}
    \left.\frac{\der \sigma^{\gamma_{\lambda}^{\star}+A\to q\bar{q}+X}}{ \der^2 \Pt \der^2 \qt \der \eta_1 \der \eta_{2}}\right|_{\rm LO}  \!\!\!\!\!\!\! =  \alpha_{\rm em}e_f^2\alpha_s\deltatwo\mathcal{H}_{\rm LO}^{\lambda,ij}(\Pt)\times G^{ij}_{Y}(\qt)+\mathcal{O}\left(\frac{q_\perp}{P_\perp}\right) +\mathcal{O}\left(\frac{Q_s}{P_\perp}\right)\,,
    \label{eq:diff_xsec_TMD_LO}
\end{align}
with the leading order hard factor $\mathcal{H}_{\rm LO}^{\lambda,ij}(\Pt)$ defined 
as
\begin{equation}
    \mathcal{H}_{\rm LO}^{\lambda,ij}(\Pt)=\frac{1}{2}\int\frac{\der^2\ut}{(2\pi)}\frac{\der^2\ut'}{(2\pi)}e^{-i\Pt\cdot\ruupt}\ut^i\ut'^j\Rcal_{\rm LO}^\lambda(\ut,\ut')\label{eq:hard-factor-def} \,.
\end{equation}
Note that the definition of the WW gluon TMD in Eq.\,\eqref{eq:WWTMD} has a $1/\alpha_s$ prefactor corresponding to the high occupancy of WW gluons. As a result, the TMD factorized LO expression in Eq.\,\eqref{eq:diff_xsec_TMD_LO} is of order $\alpha_{\rm em}\alpha_s$.

To proceed further, it is customary to decompose the WW gluon TMD into a trace and a traceless part, defining respectively the unpolarized\footnote{As alluded to earlier, the conventional (unpolarized) WW distribution is the classical non-Abelian gluon distribution in $A^+=0$ lightcone gauge in the CGC EFT~\cite{McLerran:1993ka,McLerran:1993ni,JalilianMarian:1996xn}. For the computation of the linearly polarized WW distribution, see \cite{Metz:2011wb}.}  WW gluon TMD $G^0_Y(\qt)$ and the linearly polarized one $ h^0_Y(\qt)$ as 
\begin{equation}
    G^{ij}_Y(\qt)=\frac{1}{2}\delta^{ij}G^0_Y(\qt)+\frac{1}{2}\left[\frac{2\qt^i\qt^j}{\qt^2}-\delta^{ij}\right]h^0_Y(\qt) \,.
\end{equation}
The computation of the hard factor for the linearly polarized WW gluon TMD involves the azimuthal angle $\phi$ between $\Pt$ and $\qt$. One can then perform the integrals in Eq.\,\eqref{eq:hard-factor-def} analytically for both longitudinally and transversely polarized photons:
\begin{align}
     \left.\frac{\der \sigma^{\gamma_{\rm L}^{\star}+A\to q\bar{q}+X}}{ \der^2 \Pt \der^2 \qt \der \eta_1 \der \eta_{2}}\right|_{\rm LO}  \!\!\!\!\!\!\! &=  \alpha_{\rm em}e_f^2\alpha_s\deltatwo\times\frac{8(z_1z_2)^3Q^2\Pt^2}{(\Pt^2+\bar Q^2)^4}\nonumber\\
     &\times\left[G^0_{Y}(\qt)+\cos(2\phi)h^0_{Y}(\qt)\right]\,,\label{eq:LOb2b-L}\\
     \left.\frac{\der \sigma^{\gamma_{\rm T}^{\star}+A\to q\bar{q}+X}}{ \der^2 \Pt \der^2 \qt \der \eta_1 \der \eta_{2}}\right|_{\rm LO}  \!\!\!\!\!\!\! &=  \alpha_{\rm em}e_f^2\alpha_s\deltatwo \times z_1 z_2 (z_1^2+z_2^2)\frac{\Pt^4+\bar Q^4}{(\Pt^2+\bar Q^2)^4}\nonumber\\     &\times\left[G^0_{Y}(\qt)-\frac{2\bar Q^2\Pt^2}{\Pt^4+\bar Q^4}\cos(2\phi)h^0_{Y}(\qt)\right]\,.\label{eq:LOb2b-T}
\end{align}
For later convenience, we also define the trace component of the hard factor as
\begin{align}
    \Hcal_{\rm LO}^{0,\lambda}(\Pt)\equiv\frac{1}{2}\Hcal_{\rm LO}^{\lambda,ii}(\Pt)=\left\{
    \begin{array}{ll}
        \frac{8(z_1z_2)^3Q^2\Pt^2}{(\Pt^2+\bar Q^2)^4} & \mbox{ for }\lambda=\rm L\\
       z_1 z_2 (z_1^2+z_2^2)\frac{\Pt^4+\bar Q^4}{(\Pt^2+\bar Q^2)^4} & \mbox{ for }\lambda=\rm T \,. 
    \end{array}
\right. 
\end{align}
Averaging over the azimuthal angle of $\Pt$ in Eqs.\,\eqref{eq:LOb2b-L} and \eqref{eq:LOb2b-T}, one sees that the terms proportional to $h^0$ cancel, meaning that the azimuthally averaged dijet cross-section does not depend on the linearly polarized WW gluon TMD. This distribution has however an imprint on the $\langle \cos(2\phi)\rangle$ anisotropy already at leading order. We will discuss the effects of NLO corrections and soft gluon radiation on these expressions in section~\ref{sec:sudakov}.

\section{Update on NLO results for inclusive dijet production in DIS}
\label{sec:update-if}

In this section, we will summarize the principal results of the NLO computation of the inclusive dijet cross-section in DIS presented in \cite{Caucal:2021ent}. We will also take the opportunity to update our results, correcting minor typos and elaborating on our analysis of the real NLO impact factor following the recent study of dijet production in the photoproduction limit \cite{Taels:2022tza}. We will present in the main text the cross-section for longitudinally polarized virtual photons, with the  transversely polarized NLO cross-section provided in the Appendix~\ref{app:transverse}.

\subsection{Jet definition and small-$R$ approximation}
\label{sec:Jet-def}

In order to obtain an infrared and collinear safe cross-section, one needs to define the final state in terms of jets rather than partons. We will employ the jet definition introduced in \cite{Ivanov:2012ms}, as used in previous NLO studies of the jet cross-section at small $x$ \cite{Boussarie:2016ogo,Roy:2019cux,Roy:2019hwr,Caucal:2021ent}. This algorithm is in fact equivalent to the so called cone-jet algorithm \cite{Salam:2007xv,Kang:2017mda}. For a three-parton final state, one first assigns to any pair of particles labeled $i$ and $k$ a four-momentum $p_J^\mu$ using the standard $E$-scheme \cite{Blazey:2000qt,Catani:1993hr}:
\begin{equation}
    p_J^\mu=p_i^\mu+p_k^\mu\,.
\end{equation}
If the distances in the rapidity-azimuth plane between $p_i^\mu$ and $p_J^\mu$ and between $p_k^\mu$ and $p_J^\mu$ are both smaller than the jet radius parameter $R$, then the two partons $i$ and $k$ are combined into a single jet with four-momentum $p_J^\mu$:
\begin{align}
    \Delta \phi_{iJ}^2+\Delta Y_{iJ}^2\le R^2 \textrm{ and }\Delta \phi_{kJ}^2+\Delta Y_{kJ}^2\le R^2 \,,\label{eq:Papadef}
\end{align}
with $\Delta\phi_{iJ}$ and $\Delta Y_{iJ}$ the difference of azimuth and rapidity between the parton $i$ and the jet $J$ in the laboratory frame.

In the small $R$ limit, this condition can be more conveniently written in terms of the collinearity variable $\Ccal_{ik\perp}$ defined as
\begin{equation}
    \Ccal_{ik\perp}= \frac{z_i}{z_J}\boldsymbol{p}_{k\perp}-\frac{z_k}{z_J}\boldsymbol{p}_{i\perp}\,,
    \label{eq:jetcone-smallR}
\end{equation}
where $z_J=p_J^-/q^-$ is the jet longitudinal momentum fraction. One can then show that Eq.\,\eqref{eq:Papadef} is equivalent to 
\begin{equation}
    \Ccal_{ik\perp}^2\le R^2\ptj^2\mathrm{min}\left(\frac{z_i^2}{z_J^2},\frac{z_k^2}{z_J^2}\right)\,,
    \label{eq:jetcone-smallR-2}
\end{equation}
up to powers of $R$ suppressed terms \cite{Boussarie:2016ogo,Roy:2019hwr,Kang:2016mcy}. We will use this jet definition throughout this paper, following our NLO computation in \cite{Caucal:2021ent}. 

However 
one may wish to consider alternative jet algorithms that are more commonly used nowadays
such as the sequential recombination algorithms from the generalized $k_t$ family \cite{Salam:2010nqg,Cacciari:2011ma}. Two well-known examples are the C/A \cite{Dokshitzer:1997in,Wobisch:1998wt} and anti-$k_t$ \cite{Cacciari:2008gp} algorithms. In the small $R$ approximation, the corresponding criterion in terms of the collinearity variable  is \cite{Ellis:2010rwa,Hornig:2016ahz,Kang:2016mcy}
\begin{align}
    \Ccal_{ik\perp }^2&\le R^2\ptj^2\frac{z_i^2z_k^2}{z_J^4}\,,\label{eq:CA-alg}
\end{align}
with the same $E$ recombination scheme. In the narrow jet approximation, this criterion works for all jet algorithms in the generalized $k_t$ family \cite{Ellis:2010rwa,Marzani:2019hun}.
For instance, this condition is the one used in \cite{Taels:2022tza} as a proxy for the C/A algorithm and in \cite{Liu:2022ijp} as the small $R$ limit of the more widely used anti-$k_t$ jet algorithm.

In the small $R$ limit, all of these jet definitions are equivalent up to finite terms in $\alpha_s$ in the NLO impact factor. We provide in Appendix~\ref{app:jet-algo} the corresponding expression for the finite terms for all jet definitions in the generalized $k_t$ family. Note that this calculation also demonstrates the equivalence between the criterion Eq.\,\eqref{eq:Papadef} and the cone-jet algorithm within the narrow jet approximation.

\subsection{NLO cross-section}
\label{sub:NLO-if}

We shall now discuss the NLO impact factor for inclusive dijet production in DIS at small~$x$. The purpose of this subsection is two-fold: (i) we summarize the main results of \cite{Caucal:2021ent},  (ii) we reorganize the different contributions to the NLO impact factor such that it will be simpler to extract the terms that will be enhanced by large Sudakov-like logarithms in the back-to-back limit. For the reader interested in the final result of this subsection, the relevant formula is Eq.\,\eqref{eq:xsec-decomposition}, with individual terms specified by Eqs.\,\eqref{eq:R2R2-soft},\eqref{eq:R2R2'soft},\eqref{eq:sigma_Sud1} and \eqref{eq:sigma-nosud}. To avoid lengthy expressions, Eq.\,\eqref{eq:sigma-nosud} is further decomposed into Eqs.\,\eqref{eq:V-CF},\eqref{eq:V-NLO3},\eqref{eq:V-other} for the virtual component and Eqs.\,\eqref{eq:R-CF},\eqref{eq:R-NLO3},\eqref{eq:dijet-NLO-long-real-other-final} for the real component.

\begin{figure}[tbh]
    \centering
    \includegraphics[width=1\textwidth]{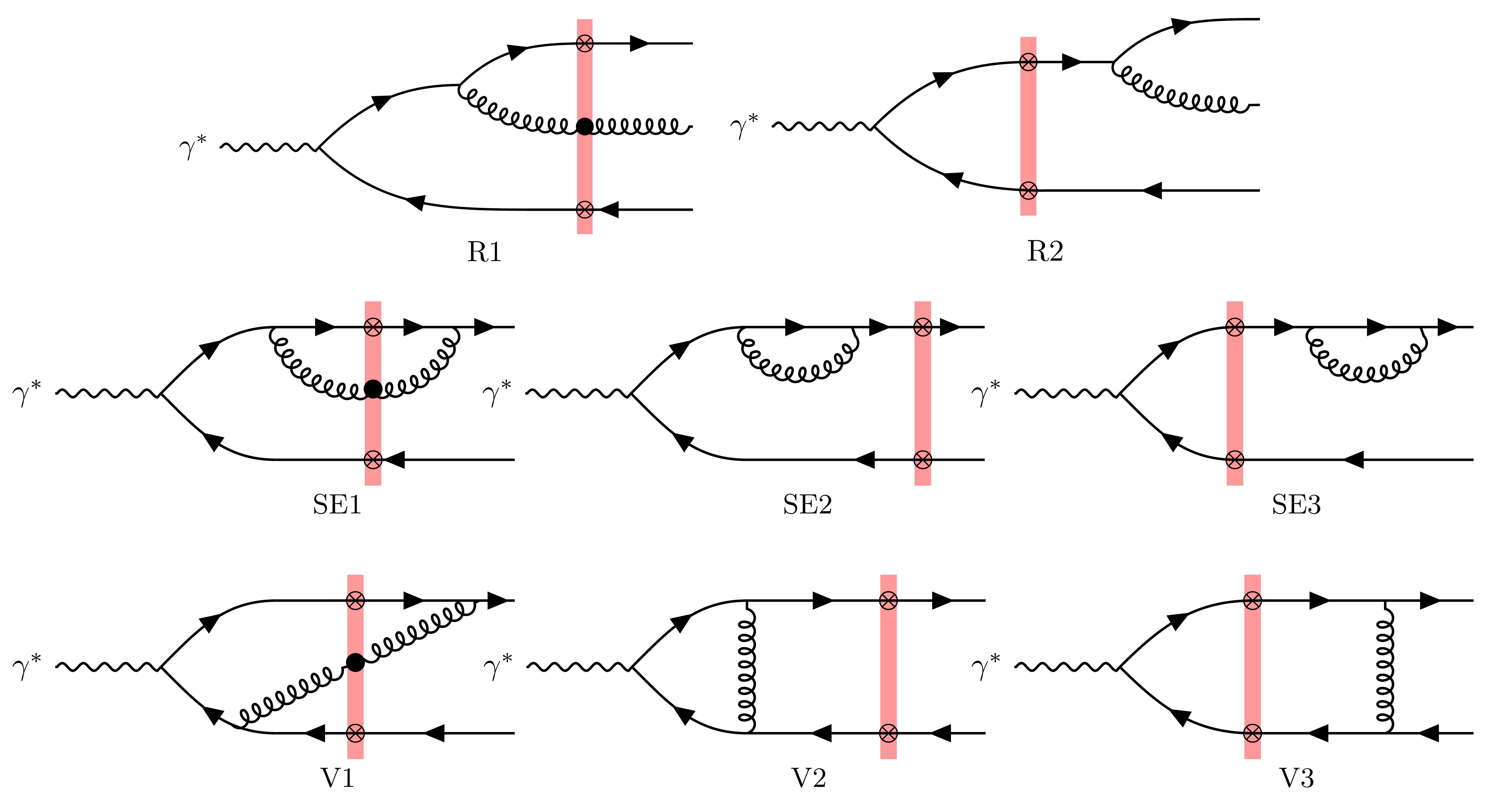}
    \caption{Feynman diagrams that appear in the production of dijets at NLO. Top: real gluon emission diagrams. Middle: self energy  diagrams. Bottom: vertex correction diagrams. Diagrams obtained from $q \leftrightarrow \bar{q}$ interchange are not shown, and are labeled with an additional prime index, for example, $\rm R_2\rightarrow \rm R_2^\prime$. Only the diagrams in which the gluon does not scatter off the shockwave contribute to the Sudakov double and single logarithms.}
    \label{fig:NLO-dijet-all-diagrams}
\end{figure}

The Feynman diagrams corresponding to the real and virtual amplitudes are shown\footnote{\label{footnote:rc-discussion}In addition to these diagrams, there is a diagram corresponding to a one loop correction to the classical shock wave background field. This one loop contribution has a piece that is enhanced by logarithms in $x$ which is absorbed in the rapidity evolution of the cross-section, as we will discuss later in this section. The finite piece, which we shall also discuss further in section~\ref{sec:TMD-NLO}, contains the one loop $\beta$-function; its role is therefore to replace the fixed QCD coupling by the 
running coupling~\cite{Ayala:1995hx}. The scale of the running coupling for this process can however only be set at two loop order, which is part of the NNLO impact factor.} in Fig.\,\ref{fig:NLO-dijet-all-diagrams}. This figure does not display the diagrams that can be obtained from quark-antiquark interchange. We denote these diagrams using the same labels but with an additional prime index. (For instance, diagram $\rm R_2'$ corresponds to a real gluon emission from the antiquark after it scatters off the shockwave.)
These diagrams were computed explicitly in \cite{Caucal:2021ent}. The calculations were performed using dimensional regularization in the transverse plane $d=2\to d=2-\varepsilon$ \footnote{We note that other works follow the convention $d=2\to d=2-2\varepsilon$, which can be obtained from our results by simply $\varepsilon \to 2\varepsilon$.}, and cut-off regularization along the minus lightcone direction with the cut-off scale $\Lambda^-$. The final result of \cite{Caucal:2021ent} (summarized in section~8 of that paper, specifically in Eqs.\,(8.1) to (8.9)) is essentially decomposed into three terms (cf.\,Eq.\,(8.1) in \cite{Caucal:2021ent}). The first term (labeled ``IRC,i.f." in \cite{Caucal:2021ent}) comes from the cancellation between the UV divergent component of $\rm SE_1$, the UV divergent diagrams $\rm SE_2$, $\rm SE_3$ and $\rm V_2$ and the in-cone divergent contributions from $\rm R_2\times R_2$ and $\rm R_2'\times R_2'$. The two other terms come from the other real and virtual finite diagrams (including the finite contribution from the self-energy $\rm SE_1$). This decomposition was sufficient in order to prove the UV and IR finiteness of the NLO cross-section, as well as the factorization of rapidity divergences\footnote{In the Regge limit, rapidity divergences occur since we work with Wilson lines on the lightcone. We regularize such divergence by introducing a longitudinal momentum cut-off $\Lambda^-$, hence the divergence will be traded by a large logarithm $\ln(\Lambda^-)$. Throughout this manuscript, we will use the terms rapidity divergence and large rapidity logarithms interchangeably.}.

That said, in view of isolating the dominant contributions of the NLO impact factor in the back-to-back limit (the large Sudakov logarithms), it will prove more convenient to decompose the NLO cross-section in a different way, as
\begin{align}
    \der\sigma_{\rm NLO}&=\left(\der\sigma_{\rm R_2\times R_2, sud2}+\der \sigma_{\rm R_2\times R_2', sud2}+\mathrm{R}_2\leftrightarrow\mathrm{R}_2'\right)+\der\sigma_{\rm sud1}\nonumber\\
    &+\der\sigma_{\rm R,no-sud}+\der\sigma_{\rm V,no-sud}+\ln\left(\frac{k_f^-}{\Lambda^-}\right)\mathcal{H}_{\rm LL}\otimes \der\sigma_{\rm LO}\,,\label{eq:xsec-decomposition}
\end{align}
where we abbreviate
\begin{equation}
    \der\sigma_{\rm X}\equiv\left.\frac{\der \sigma^{\gamma_{\lambda}^{\star}+A\to \textrm{dijet}+X}}{ \der^2 \ktone \der^2 \kttwo \der \eta_1 \der \eta_{2}}\right|_{\rm X}\,.
\end{equation}
For obvious reasons, we now label the four-momenta of the two jets using the same $k_1^\mu$ and $k_2^\mu$ labels as for the quark-antiquark pair. 

The last term in Eq.\,\eqref{eq:xsec-decomposition} corresponds to the rapidity divergence of the NLO cross-section that will be addressed in the next subsection~\ref{subsub:JIMWLK}. It is proportional to $\ln(k_f^-/\Lambda^-)$ where $\Lambda^-$ is our longitudinal momentum cut-off that regulates the rapidity divergence and $k_f^-$ is an arbitrary rapidity factorization scale (similar in spirit to the transverse momentum scale $\mu_R$ in collinear factorization). Even though it is not manifest in Eq.\,\eqref{eq:xsec-decomposition}, each term in the decomposition Eq.\,\eqref{eq:xsec-decomposition} depends on $k_f^-$, so that the full ($\Lambda^-$ regulator dependent) NLO result given by Eq.\,\eqref{eq:xsec-decomposition} is independent of $k_f^-$. In turn, the NLO impact factor defined by $\der\sigma_{\rm NLO}$ minus the $\ln(k_f^-/\Lambda^-)$ term in Eq.\,\eqref{eq:xsec-decomposition} is factorization scale dependent. We emphasize that when one adds all the terms in Eq.\,\eqref{eq:xsec-decomposition}, except for the rapidity divergence proportional to $\ln(k_f^-/\Lambda^-)$, one gets exactly the same result as Eq.\,(8.1) in \cite{Caucal:2021ent}.

As explained in section~\ref{subsub:JIMWLK}, the remaining logarithmic dependence in the rapidity cut-off $\Lambda^-$ that appears in \eqref{eq:xsec-decomposition} is absorbed by renormalizing the operator $\Xi_{\rm LO}$ in the leading order cross-section. The $k_f^-$ dependence of the latter is then given by the JIMWLK evolution equation.

\begin{table}[tbh]
    \centering
    \begin{tabular}{|c|c|}
    \hline
    $\Xi_{\rm LO}(\xt,\yt;\xt',\yt')$ & $\langle1-D_{xy}-D_{y'x'}+Q_{xy,y'x'}\rangle$\\
    \hline
    $\Xi_{\rm NLO,1}(\xt,\yt,\zt;\xt',\yt')$ & $\frac{N_c}{2}\langle1-D_{y'x'}+Q_{zy,y'x'}D_{xz}-D_{xz}D_{zy}\rangle-\frac{1}{2N_c}\Xi_{\rm LO}$\\
    \hline
    $\Xi_{\rm NLO,2}(\xt,\yt,\zt;\xt',\yt')$ & $\frac{N_c}{2}\langle1-D_{y'x'}+Q_{xz,y'x'}D_{zy}-D_{xz}D_{zy}\rangle-\frac{1}{2N_c}\Xi_{\rm LO}$\\
    \hline 
    $\Xi_{\rm NLO,3}(\xt,\yt;\xt',\yt')$ & $\frac{N_c}{2}\langle1-D_{xy}-D_{y'x'}+D_{xy}D_{y'x'}\rangle-\frac{1}{2N_c}\Xi_{\rm LO}$\\
    \hline  
    $\Xi_{\rm NLO,4}(\xt,\yt,\zt;\xt',\yt',\zt')$ & $\frac{N_c}{2}\langle1-D_{xz}D_{zy}-D_{y'z}D_{zx'}+Q_{xz,z'x'}Q_{y'z',zy}\rangle-\frac{1}{2N_c}\Xi_{\rm LO}$\\
 \hline  
  
  \end{tabular}
   \caption{Color correlators contributing to the next-to-leading order cross-section. }\label{tab:NLO-color}
\end{table}

A detailed derivation of this decomposition from the full NLO computation performed in \cite{Caucal:2021ent} is provided in Appendix~\ref{app:xs-dec}.
The meaning of each term in Eq.~\eqref{eq:xsec-decomposition} can be summarized briefly as follows.  The terms labeled by $\rm sud2$ are the contributions which contain Sudakov \textit{double} logarithms, Sudakov \textit{single} logarithms, as well as terms with no such logs in the back-to-back limit. The double logarithms give the dominant contributions to the NLO impact factor in the back-to-back limit, where $P_\perp \gg q_\perp$. Their expressions are
\begin{align}
    &\der\sigma_{\rm R_2\times R_2,\rm sud2}=\frac{\alpha_{\rm em}e_f^2N_c\deltatwo}{(2\pi)^6}\int\der^8\Xt e^{-i\ktone\cdot\rxxtp-i\kttwo\cdot\ryytp}\Rcal_{\mathrm{LO}}^{\lambda}(\rxyt,\rxytp)\nonumber\\
    \times &C_F\Xi_{\rm LO}(\xt,\yt;\xt',\yt')\times \frac{\alpha_s}{\pi}\int_0^1\frac{\der\xi}{\xi}\left[1-e^{-i\xi\ktone\cdot\rxxtp}\right]\ln\left(\frac{\ktone^2\rxxtp^2R^2\xi^2}{c_0^2}\right)\,,\label{eq:R2R2-soft}\\
    &\der\sigma_{\rm R_2\times R_2',\rm sud2}=\frac{\alpha_{\rm em}e_f^2N_c\deltatwo}{(2\pi)^6}\int\der^8\Xt e^{-i\ktone\cdot\rxxtp-i\kttwo\cdot\ryytp}\Rcal_{\mathrm{LO}}^{\lambda}(\rxyt,\rxytp)\nonumber\\
    \times &\Xi_{\rm NLO,3}(\xt,\yt;\xt',\yt')\times \frac{(-\alpha_s)}{\pi}\int_0^1\frac{\der\xi}{\xi}\left[1-e^{-i\xi\ktone\cdot\rxypt}\right]\ln\left(\frac{\Pt^2\rxypt^2\xi^2}{z_2^2c_0^2}\right)\,,\label{eq:R2R2'soft}
\end{align}
with the $\rm NLO,3$ color correlator defined by
\begin{align}
    \Xi_{\rm NLO,3}(\xt,\yt;\xt',\yt')=\frac{N_c}{2}\left\langle 1-D_{xy}-D_{y'x'}+D_{xy}D_{y'x'}\right\rangle-\frac{1}{2N_c}\Xi_{\rm LO}(\xt,\yt;\xt',\yt')\,,\label{eq:NLO3-correlator}
\end{align}
and the constant 
\begin{equation}
    c_0=2e^{-\gamma_E}\,,
\end{equation}
where $\gamma_E$ is the Euler–Mascheroni constant.

The $\rm R_2'\times R_2'$ and $\rm R_2'\times R_2$ ``sud2" cross-sections can be obtained from Eq.\,\eqref{eq:R2R2-soft}-\eqref{eq:R2R2'soft} using quark-antiquark interchange. As suggested by our notations, these terms mainly come from the product of diagrams $\rm R_2\times \rm R_2$, $\rm R_2'\times \rm R_2'$ or the interference diagrams $\rm R_2\times R_2'$, $\rm R_2'\times R_2$ (cf.\ Fig.\,\ref{fig:NLO-dijet-all-diagrams}). Indeed, among the real amplitudes, only the amplitudes $\rm R_2$ and $\rm R_2'$ have a soft gluon divergence since in the soft gluon limit, the internal quark (or antiquark in the case of the amplitude $\rm R_2'$) propagator goes on-shell. This is a generic consequence of soft gluon factorization in QCD at the amplitude level, which holds in our calculation as can been explicitly checked from the expressions for the amplitude $\rm R_2$ and $\rm R_2'$ provided in appendix~\ref{app:xs-dec}. See also \cite{Roy:2019hwr,Boussarie:2016ogo}. 
 
As shown in appendix~\ref{app:xs-dec}, in our calculation with a rapidity cut-off regulator, the soft divergence appears as a $\ln^2(\Lambda^-)$ divergence which cancels at the cross-section level once virtual corrections are included for IRC safe jet definitions. The finite leftover term depends logarithmically on the ratio between $P_\perp/q_\perp$ in back-to-back kinematics, and therefore blows up when $P_\perp\gg q_\perp$.

The term labeled ``$\rm sud1$" 
contains \textit{single} Sudakov-type logarithms in addition to Sudakov-free terms. It is given by 
\begin{align}
    &\der\sigma_{\rm sud1}=\frac{\alpha_{\rm em}e_f^2N_c\deltatwo}{(2\pi)^6}\int\der^8\Xt e^{-i\ktone\cdot\rxxtp-i\kttwo\cdot\ryytp}\Rcal_{\mathrm{LO}}^{\lambda}(\rxyt,\rxytp)\times \frac{\alpha_s}{\pi} \nonumber\\
     &\times\left\{C_F\Xi_{\rm LO}(\xt,\yt;\xt',\yt')\left[\ln\left(\frac{z_f}{z_1}\right)\ln\left(\frac{\rxxtp^2}{|\rxyt||\rxytp|}\right)+\ln\left(\frac{z_f}{z_2}\right)\ln\left(\frac{\ryytp^2}{|\rxyt||\rxytp|}\right)\right]\right.\nonumber\\
     &\hspace{0.35cm}+\left.\Xi_{\rm NLO,3}(\xt,\yt;\xt',\yt')\left[\ln\left(\frac{z_1}{z_f}\right)\ln\left(\frac{\rxypt^2}{|\rxyt||\rxytp|}\right)+\ln\left(\frac{z_2}{z_f}\right)\ln\left(\frac{\ryxpt^2}{|\rxyt||\rxytp|}\right)\right]\right\}\,.
     \label{eq:sigma_Sud1}
\end{align}
As manifest from this expression, it depends on the leading order and $\rm NLO_3$ color correlators. This term comes from the leftovers from our rapidity divergence subtraction scheme and is due to the nontrivial interplay between slow (aka ``small $x$") gluons and soft gluons.  This is why it depends on $z_f=k_f^-/q^-$, where the scale $k_f^-$ is a longitudinal factorization scale used to isolate the leading logarithmic rapidity divergence. We will comment further on these points in the next section.

Another advantage of the decomposition in  Eq.\,\eqref{eq:xsec-decomposition} is that the terms labeled $\rm sud2$ and $\rm sud1$ depend on the polarization of the virtual photon only through the $\lambda$ dependence of the leading order perturbative factors $\rm \Rcal_{\rm LO}^\lambda$. This factorization implies that the Sudakov logarithms are universal with respect to the polarization of the virtual photon.

In Eq.\,\eqref{eq:xsec-decomposition}, the terms labeled ``$\rm no-sud$"  do not contain any Sudakov logarithm as $P_\perp/q_\perp\to\infty$. For compactness, they are further decomposed according the color correlator upon which they depend:
\begin{align}
    \der\sigma_{\rm R/V, no-sud}\equiv \der\sigma_{\rm R/V,no-sud,LO}+\der\sigma_{\rm R/V,no-sud, NLO_3}+\der\sigma_{\rm R/V, no-sud,other} \,.\label{eq:sigma-nosud}
\end{align}
The term $\der\sigma_{\rm V,no-sud,LO}$ depends on the polarization of the virtual photon through the hard factor $\Rcal_{\rm LO}^\lambda$, and it reads
\begin{align}
       &\der\sigma_{\rm V,no-sud,LO}=\frac{\alpha_{\rm em}e_f^2N_c\deltatwo}{(2\pi)^6}\int\der^8\Xt e^{-i\ktone\cdot\rxxtp-i\kttwo\cdot\ryytp}\Rcal_{\mathrm{LO}}^{\lambda}(\rxyt,\rxytp)\Xi_{\rm LO}(\xt,\yt;\xt',\yt')\nonumber\\
     &\times\frac{\alpha_sC_F}{\pi}\left\{-\frac{3}{4}\ln\left(\frac{\ktone^2\kttwo^2\rxyt^2\rxytp^2}{c_0^4}\right)-3\ln(R)+\frac{1}{2}\ln^2\left(\frac{z_1}{z_2}\right)+\frac{11}{2}+3\ln(2)-\frac{\pi^2}{2}\right\}\,.
     \label{eq:V-CF}
\end{align}
Even though this term does not contain a Sudakov logarithm in the back-to-back limit, it is enhanced by a logarithm of the jet radius from the $-3/2\ln(R^2)$ term inside the curly bracket. This factor of $-3/2$ has a physical origin: it corresponds to the finite part of the $z$-integrated DGLAP quark splitting function associated with hard collinear gluon emissions from the quark and the antiquark. The all order resummation of such logarithms for small jet radii can be performed systematically along the lines of \cite{Dasgupta:2014yra,Kang:2016mcy} using a DGLAP-like evolution equation. The study of the interplay between the $\ln(R)$, $\ln(1/x_{\rm Bj})$ and Sudakov resummations is beyond the scope of the present paper.

All the other terms in Eq.\,\eqref{eq:sigma-nosud} depend on the polarization of the virtual photon in a more complicated way. We provide here the expressions for longitudinally polarized photons, and the formulas for transversely polarized photons are given in Appendix~\ref{app:transverse}. We have for the virtual ``no-sud" term proportional to $\Xi_{\rm NLO,3}$:
\begin{align}
   &\der\sigma^{\lambda=\rm L}_{\rm V,no-sud,NLO_3}=\frac{\alpha_{\rm em}e_f^2N_c\deltatwo}{(2\pi)^6}\int\der^3\Xt e^{-i\ktone\cdot\rxxtp-i\kttwo\ryytp} 8z_1^3z_{2}^3Q^2K_0(\bar Qr_{x'y'})
    \nonumber \\
    & \times \frac{\alpha_s}{\pi} \int_0^{z_1}\frac{\der z_g}{z_g}\left\{  K_0(\bar Q_{\rm V3}r_{xy})\left[\left(1-\frac{z_g}{z_1}\right)^2\left(1+\frac{z_g}{z_2}\right)(1+z_g) e^{i(\Pt+z_g\qt)\cdot \rxyt} K_0(-i\Delta_{\mathrm{V}3}r_{xy}) \right. \right. \nonumber\\
    &\hspace{2cm}\left.\left.-\left(1-\frac{z_g}{2z_1}+\frac{z_g}{2z_2}-\frac{z_g^2}{2z_1z_2}\right) e^{i\frac{z_g}{z_1}\ktone \cdot \rxyt} \Jcal_{\odot}\left(\rxyt,\left(1-\frac{z_g}{z_1}\right)\Pt,\Delta_{\mathrm{V}3}\right)\right]\right.\nonumber\\
    &\hspace{2cm}\left.+  K_0(\bar{Q} r_{xy})\ln\left(\frac{z_g P_\perp r_{xy}}{c_0z_1z_2}\right)+(1\leftrightarrow 2)\right\}\Xi_{\rm NLO,3}(\xt,\yt;\xt',\yt')+c.c. \,, \label{eq:V-NLO3}
\end{align}
with $\bar Q_{\rm V3}^2=z_1z_2(1-z_g/z_1)(1+z_g/z_2)Q^2$ and $\Delta_{\rm V3}^2=(1-z_g/z_1)(1+z_g/z_2)\Pt^2$ and the function $\Jcal_{\odot}$, defined as
\begin{align}
        \Jcal_{\odot}(\rt,\Kt,\Delta)&=\int\frac{\der^2 \lt}{(2\pi)}\frac{2\lt \cdot \Kt \ e^{i\lt \cdot \rt}}{\lt^2\left[(\lt-\Kt)^2-\Delta^2 - i \epsilon\right]}\,.
    \label{eq:Jdot-def}
\end{align}
This integral can be computed using the Schwinger parametrization; we refer the reader to Appendix J in \cite{Caucal:2021ent} for an expression for this integral that is  suitable for numerical evaluation. The term proportional to $\ln(z_gP_\perp r_{xy}/(c_0z_1z_2))$ in Eq.\,\eqref{eq:V-NLO3} ensures that the $z_g$ integral is finite by subtracting the $z_g\to 0$ singularity. We shall also use the notation $(1\leftrightarrow2)$ as a shorthand for quark-antiquark interchange which amounts to switch the four-momenta $k_1 \leftrightarrow k_2$ of the "quark-jet" and the "antiquark jet", switch the transverse coordinates $\xt \leftrightarrow \yt$ ($\xt' \leftrightarrow \yt'$), and take the complex conjugate of the corresponding color correlator.

The term labeled ``no-sud, other" reads\footnote{Note that $C_F\Xi_{\rm LO}$ also appears in $\der\sigma_{\rm V,other}$ as the UV regulator of the self-energy crossing the SW. For this reason, it does not make sense to isolate this contribution from the $\rm SE_1$ diagram which has the color structure $\rm NLO,1$.}
\begin{align}
   &\der\sigma^{\lambda=\rm L}_{\rm V,no-sud, other}= \frac{\alpha_{\rm em}e_f^2N_c\deltatwo}{(2\pi)^6} \int\der^8\Xt e^{-i\ktone\cdot\rxxtp-i\kttwo\ryytp} 8z_1^3z_{2}^3Q^2K_0(\bar Qr_{x'y'}) \int_0^{z_1}\frac{\der z_g}{z_g} \nonumber\\
    & \times \frac{\alpha_s}{\pi}\int\frac{\der^2\zt}{\pi} \left\{\frac{1}{\rzxt^2}\left[\left(1-\frac{z_g}{z_1}+\frac{z_g^2}{2z_1^2}\right) e^{-i\frac{z_g}{z_1}\ktone \cdot \rzxt} K_0(QX_V)-\Theta(z_f-z_g)K_0(\bar Qr_{xy})\right]\right. \Xi_{\rm NLO,1} \nonumber\\
   &-\frac{1}{\rzxt^2}\left[\left(1-\frac{z_g}{z_1}+\frac{z_g^2}{2z_1^2}\right) e^{-\frac{\rzxt^2}{\rxyt^2e^{\gamma_E}}}K_0(\bar Qr_{xy})-\Theta(z_f-z_g) e^{-\frac{\rzxt^2}{\rxyt^2e^{\gamma_E}}} K_0(\bar Qr_{xy})\right] C_F\Xi_{\rm LO}\nonumber\\
   &-\frac{\rzxt\cdot\rzyt}{\rzxt^2\rzyt^2}\Bigg[\left(1-\frac{z_g}{z_1}\right)\left(1+\frac{z_g}{z_2}\right)\left(1-\frac{z_g}{2z_1}-\frac{z_g}{2(z_2+z_g)}\right) e^{-i\frac{z_g}{z_1}\ktone \cdot \rzxt} K_0(QX_V)  \nonumber \\
   & \quad \quad \quad \quad \quad \left. -\Theta(z_f-z_g)K_0(\bar Q r_{xy}) \Bigg] \Xi_{\rm NLO,1} +(1\leftrightarrow 2) \right \}  + c.c. \,. \label{eq:V-other}
\end{align}
This equation involves two additional color correlators labeled $\Xi_{\rm NLO,1}$ and $\Xi_{\rm NLO,2}$ (in the $(1\leftrightarrow2)$ interchange term) given by
\begin{align}
   \Xi_{\rm NLO,1}(\xt,\yt,\zt;\xt',\yt')&=\frac{N_c}{2}\left\langle1-D_{y'x'}+Q_{zy,y'x'}D_{xz}-D_{xz}D_{zy}\right\rangle_{Y_0}\nonumber\\
   &-\frac{1}{2N_c}\Xi_{\rm LO}(\xt,\yt;\xt',\yt')\,,\label{eq:XiNLO1}\\
\Xi_{\rm NLO,2}(\xt,\yt,\zt;\xt',\yt')&=  \frac{N_c}{2}\left\langle1-D_{y'x'}+Q_{xz,y'x'}D_{zy}-D_{xz}D_{zy}\right\rangle_{Y_0}\nonumber\\
   &-\frac{1}{2N_c}\Xi_{\rm LO}(\xt,\yt;\xt',\yt')\,,\label{eq:XiNLO2}
\end{align}
as well as the $q\bar qg$ effective dipole size $X_{\rm V}$ defined as
\begin{equation}
    X_{\rm V}^2=z_{2}(z_1-z_g)\rxyt^2+z_g(z_1-z_g)\rzxt^2+z_{2}z_g\rzyt^2\,.
\end{equation}
It is also infrared finite, and has no rapidity divergence. Indeed, the leading logarithmic $1/z_g$ divergence is systematically subtracted thanks to the $\Theta$-function $\Theta(z_f-z_g)$. This concludes our summary of the virtual cross-section contributing at most finite $\alpha_s$ corrections in the back-to-back limit.

We turn now to the "no-sud" real cross-section. Since this cross-section depends on the selection cuts imposed on the final state to define the dijet cross-section, we will provide the formula for the $q\bar qg$ "no-sud" cross-section and leave the integration over the three-body phase-space (according to the dijet selection) for a future numerical study of the NLO impact factor. We emphasize  nevertheless that the "no-sud" $q\bar q g$ cross-section that we present in this section is free of any divergence, as well as large back-to-back logarithms, for all kinematic regions of the final state.
\begin{align}
    &\der\sigma^{\gamma_{\rm L}^\star+A\to q \bar qg+X}_{\rm R,no-sud,LO}=\frac{\alpha_{\rm em}e_f^2N_c}{(2\pi)^8}\int\der^8\Xt e^{-i\ktone\cdot\rxxtp-i\kttwo\cdot\ryytp}(4\alpha_s C_F)\Xi_{\rm LO}(\xt,\yt;\xt',\yt')\nonumber\\
    &\times\frac{e^{-i\kgt \cdot\rxxtp}}{(\kgt-\frac{z_g}{z_1}\ktone)^2} \left\{ 8z_1z_{2}^3(1-z_{2})^2Q^2\left(1+\frac{z_g}{z_1}+\frac{z_g^2}{2z_1^2}\right)K_0(\bar Q_{\mathrm{R}2}r_{xy})K_0(\bar Q_{\mathrm{R}2}r_{x'y'})\deltathree\right.\nonumber\\
    &\hspace{3.5cm}\left.-\Rcal^{\rm L}_{\rm LO}(\rxyt,\rxytp)\Theta(z_1-z_g)\deltatwo\right\}+(1\leftrightarrow 2)\,,\label{eq:R-CF}
\end{align}
with $\deltathree=\delta(1-z_1-z_2-z_g)$ and $\bar Q_{\rm R2}^2=z_2(1-z_2)Q^2$.
To get the inclusive dijet cross-section, one should ensure that the phase-space integration excludes the domain in which the quark and the gluon lie inside the same jet, since this contribution has already been included in $\der\sigma_{\rm V,no-sud, LO}$. 
The ``no-sud" component of $\rm R_2\times \rm R_2'$ is given by 
\begin{align}
&\der\sigma^{\gamma_{\rm L}^\star+A\to q \bar qg+X}_{\rm R, no-sud,NLO_3}
    =\frac{\alpha_{\rm em}e_f^2N_c}{(2\pi)^8}\int\der^2\Xt e^{-i\ktone\cdot\rxxtp-i\kttwo\cdot\ryytp}(-4\alpha_s)\Xi_{\rm NLO,3}(\xt,\yt;\xt',\yt')\nonumber\\
    &\times \frac{e^{-i\frac{z_g}{z_1}\ktone\cdot\rxypt}}{\lt^2}\left\{8z_1^2z_{2}^2(1-z_{2})(1-z_1)Q^2K_0(\bar{Q}_{\rm R2}r_{xy})K_0(\bar{Q}_{\rm R2'}r_{x'y'})\left[1+\frac{z_g}{2z_1}+\frac{z_g}{2z_{2}}\right]\right.\nonumber\\
    &\left.\times e^{-i\lt\cdot\rxypt}\frac{\lt\cdot(\lt+\Kt)}{(\lt+\Kt)^2}\deltathree-\Rcal^{\rm L}_{\rm LO}(\rxyt,\rxytp)\Theta\left(\frac{c_0^2}{\rxypt^2}\ge \lt^2\ge\Kt^2\right)\Theta(z_1-z_g)\deltatwo\right\}\nonumber\\
    &+(1\leftrightarrow 2)\,,\label{eq:R-NLO3}
\end{align}
with $\lt=\kgt-z_g/z_1\ktone$ and $\Kt=z_g/(z_1z_2)\Pt$.
We will demonstrate in Section~\ref{sec:TMD-NLO} that Eqs.\,\eqref{eq:R-CF} and \eqref{eq:R-NLO3} are power suppressed in the back-to-back limit. The real term without Sudakov logarithms, independent of $\Xi_{\rm LO}$ and $\Xi_{\rm NLO,3}$, coming from real diagrams in which the gluon crosses the shockwave reads\footnote{The prefactors of $1/2$ in the third and fourth terms in the brackets in Eq.\,\eqref{eq:dijet-NLO-long-real-other-final} are introduced to avoid overcounting due to quark-antiquark interchange and complex conjugation.}
\begin{align}
    &\der\sigma_{\rm R,no-sud, other}^{\gamma_{\rm L}^\star+A\to q \bar qg+X}=\frac{\alpha_{\rm em}e_f^2N_c\deltathree}{(2\pi)^8}\,\int\der^8\Xt e^{-i\ktone\cdot\rxxtp-i\kttwo\cdot\ryytp}8z_1^3z_{2}^3Q^2\int\frac{\der^2\zt}{\pi}\frac{\der^2\zt'}{\pi} e^{-i\kgt \cdot\rzzpt}\nonumber\\
    \alpha_s&\left\{
    -\frac{\rzxt\cdot\rzxtp}{\rzxt^2\rzxtp^2}K_0(QX_{\rm R})K_0(\bar Q_{\mathrm{R}2}r_{w'y'})\left(1+\frac{z_g}{z_1}+\frac{z_g^2}{2z_1^2}\right)\Xi_{\rm NLO,1}(\xt,\yt,\zt;\wt',\yt')\right.\nonumber\\
    &+\frac{\rzyt\cdot\rzxtp}{\rzyt^2\rzxtp^2}K_0(QX_{\rm R})K_0(\bar Q_{\rm R2'}r_{w'y'})\left(1+\frac{z_g}{2z_1}+\frac{z_g}{2z_{2}}\right)\Xi_{\rm NLO,1}(\xt,\yt,\zt;\wt',\yt')\nonumber\\
    &+\frac{1}{2}\frac{\rzxt\cdot\rzxtp}{\rzxt^2\rzxtp^2}K_0(QX_{\rm R})K_0(QX'_{\rm R})\left(1+\frac{z_g}{z_1}+\frac{z_g^2}{2z_{ 1}^2}\right)\Xi_{\rm NLO,4}(\xt,\yt,\zt;\xt',\yt',\zt')\nonumber\\
    &-\frac{1}{2}\frac{\rzyt\cdot\rzxtp}{\rzyt^2\rzxtp^2}K_0(QX_{\rm R})K_0(QX'_{\rm R})\left(1+\frac{z_g}{2z_1}+\frac{z_g}{2z_{2}}\right)\Xi_{\rm NLO,4}(\xt,\yt,\zt;\xt',\yt',\zt')\nonumber\\
    &+(1\leftrightarrow2)+c.c.\Bigg\}-\frac{\alpha_{\rm em}e_f^2N_c\deltatwo}{(2\pi)^8}\,\alpha_s\Theta(z_f-z_g)\times \textrm{``slow"}
   \label{eq:dijet-NLO-long-real-other-final}\,.
\end{align}
The subtraction term labeled ``slow" is given by the $z_g\to 0$ limit of the formula; the gluon integrated cross-section is therefore finite when the $\Lambda^-$ regulator goes to 0. The associated counterterm is moved into the $\ln(k_f^-/\Lambda^-)$ component of the full cross-section in Eq.~\eqref{eq:xsec-decomposition}. In Eq.\,\eqref{eq:dijet-NLO-long-real-other-final}, one encounters the color correlator $\Xi_{\rm NLO,4}$ defined by
\begin{align}
     \Xi_{\rm NLO,4}(\xt,\yt,\zt;\xt',\yt',\zt')&=\frac{N_c}{2}\left\langle Q_{z'x';xz}Q_{zy;y'z'}-D_{xz}D_{zy}-D_{y'z'}D_{z'x'}+1\right\rangle_{Y_0}\nonumber\\
    & -\frac{1}{2N_c}\Xi_{\rm LO}(\xt,\yt;\xt',\yt')\,,
\end{align}
as well as the variables
\begin{align}
    X_{\rm R}^2&=z_1z_{2}\rxyt^2+z_1z_g\rzxt^2+z_{2}z_g\rzyt^2\,,\\
   \wt&=(z_1\xt+z_g\zt)/(z_1+z_g)\,.
\end{align}
This concludes our discussion of the terms displayed in the decomposition Eq.\,\eqref{eq:xsec-decomposition} of the inclusive dijet cross-section at NLO.

\subsection{JIMWLK factorization at leading logarithmic accuracy}
\label{subsub:JIMWLK}

In this section, we shall discuss the rapidity divergent term in the full NLO cross-section given in Eq.~\eqref{eq:xsec-decomposition}, 
and the associated rapidity factorization which leads to the JIMWLK evolution equation. This rapidity divergent term is proportional to $\ln(z_f/z_0)=\ln(k_f^-/\Lambda^-)$ and can be extracted diagram-by-diagram or color structure-by-color structure. For instance, the rapidity divergent term associated with $\Xi_{\rm LO}$ and $\Xi_{\rm NLO,3}$ is given by Eq.\,\eqref{eq:sigma_Sud1} with the replacement $z_1,z_2\to z_0$ and an overall minus sign. (See for instance the discussion leading up to Eq.\,\eqref{eq:slowLL-LO} and Eq.~\eqref{eq:slowLL-NLO3} in Appendix~\ref{app:xs-dec}.)

In order to illustrate how the terms proportional to $\ln(z_f/z_0)$ in the NLO cross-section combine to give the leading logarithmic evolution of the leading order color structure, let us focus on the rapidity divergence associated with the LO color structure. Using the identity
\begin{equation}
   \frac{1}{\pi} \int\der^2\zt\left[\frac{\rzxt\cdot\rzxpt}{\rzxt^2\rzxpt^2}-\frac{\rzxt\cdot\rzyt}{\rzxt^2\rzyt^2}\right]=\ln\left(\frac{\rxyt^2}{\rxxtp^2}\right)\,,
   \label{eq:dijet-NLO-jimwlk-kernel-identity}
\end{equation}
in Eq.\,\eqref{eq:sigma_Sud1} (with again $z_1,z_2\to z_0$ and an overall minus sign) to reconstruct JIMWLK kernels, one can write the slow divergence associated with the LO color structure as
\begin{align}
    \der\sigma_{\rm LO, slow}&=\frac{\alpha_{\rm em}e_f^2N_c\deltatwo}{(2\pi)^6}\int\der^8\Xt\Rcal_{\mathrm{LO}}^{\lambda}(\rxyt,\rxytp)e^{-i\ktone\cdot\rxxtp-i\kttwo\cdot\ryytp}\frac{\alpha_sC_F}{\pi^2}\ln\left(\frac{k_f^-}{\Lambda^-}\right)\nonumber\\
    &\times\int\der^2\zt\left[\frac{\rzxt\cdot\rzxpt}{\rzxt^2\rzxpt^2}-\frac{\rzxt\cdot\rzyt}{\rzxt^2\rzyt^2}+\frac{\rzyt\cdot\rzypt}{\rzyt^2\rzypt^2}-\frac{\rzxpt\cdot\rzypt}{\rzxpt^2\rzypt}\right]\Xi_{\rm LO}(\xt,\yt;\xt',\yt')\,,
    \label{eq:dijet-NLO-slow-xsection8}
\end{align}
and similarly for $\der\sigma_{\rm NLO_3, slow}$ for the leading logarithmic rapidity divergence associated with the color structure $\Xi_{\rm NLO,3}$. For the other color correlators which depend explicitly on $\zt$, the kernel is already manifest in the subtracted terms in  Eq.~\eqref{eq:dijet-NLO-long-real-other-final} and Eq.~\eqref{eq:V-other}, which must now be added here with the other rapidity divergent contributions. As shown in \cite{Caucal:2021ent}, combining all the $\ln(z_f/z_0)$ divergent components of the diagrams together, one finds that the rapidity divergence reads as
\begin{align}
    &\der\sigma_{\rm slow}=\frac{\alpha_{\rm em}e_f^2N_c\deltatwo}{(2\pi)^6}\ln\left(\frac{k_f^-}{\Lambda^-}\right)\frac{\alpha_sN_c}{4\pi^2}\int\der^8\Xt e^{-i\ktone\cdot\rxxtp-i\kttwo\cdot\ryytp}\Rcal_{\mathrm{LO}}^{\lambda}(\rxyt,\rxytp)\nonumber\\
    &\times \left\langle\int\der^2\zt\left\{\frac{\rxyt^2}{\rzxt^2\rzyt^2}(2D_{xy}-2D_{xz}D_{zy}+D_{zy}Q_{y'x',xz}+D_{xz}Q_{y'x',zy}-Q_{xy,y'x'}-D_{xy}D_{y'x'})\right.\right.\nonumber\\
    &\hspace{1.6cm}+\frac{\rxytp^2}{\rzxpt^2\rzypt^2}(2D_{y'x'}-2D_{y'z}D_{zx'}+D_{zx'}Q_{xy,y'z}+D_{y'z}Q_{xy,zx'}-Q_{xy,y'x'}-D_{xy}D_{y'x'})\nonumber\\
    &\hspace{1.8cm}+\frac{\rxxtp^2}{\rzxt^2\rzxpt^2}(D_{zx'}Q_{xy,y'z}+D_{xz}Q_{y'x',zy}-Q_{xy,y'x'}-D_{xx'}D_{y'y})\nonumber\\
    &\hspace{1.8cm}+\frac{\ryytp^2}{\rzyt^2\rzypt^2}(D_{y'z}Q_{xy,zx'}+D_{zy}Q_{y'x',xz}-Q_{xy,y'x'}-D_{xx'}D_{y'y})\nonumber\\
    &\hspace{1.8cm}+\frac{\rxypt^2}{\rzxt^2\rzypt^2}(D_{xx'}D_{y'y}+D_{xy}D_{y'x'}-D_{zx'}Q_{xy,y'z}-D_{zy}Q_{y'x',xz})\nonumber\\
    &\hspace{1.8cm}\left.\left.+\frac{\rxpyt^2}{\rzxpt^2\rzyt^2}(D_{xx'}D_{y'y}+D_{xy}D_{y'x'}-D_{y'z}Q_{xy,zx'}-D_{xz}Q_{y'x',zy})\right\}\right\rangle_{Y}\,.
\label{eq:dijet-NLO-full-slow}
\end{align}
Remarkably, as shown in \cite{Dominguez:2011gc}, the full structure of these color correlators can be recovered by the action of the leading log JIMWLK Hamiltonian
\begin{align}
    \mathcal{H}_{\rm LL} \equiv\frac{1}{2} \int \der^2 \ut \der^2 \vt \frac{\delta}{\delta A^{+,a}_{\rm cl}(\ut)} \eta^{ab}(\ut,\vt) \frac{\delta}{\delta A^{+,b}_{\rm cl}(\vt)}\,, \label{eq:JIMWLK-Hamiltonian}
\end{align}
on the leading order dijet cross-section, or more precisely, on the leading order color structure $\Xi_{\rm LO}$, since it is the only object which depends on $A_{\rm cl}$. The $\eta$ kernel\footnote{This kernel corresponds to the $\ln(1/x)$ enhanced piece of the one loop correction to the classical shockwave background field that we discussed previously in section 2.} in the Hamiltonian is defined to be 
\begin{align}
    \eta^{ab}(\ut,\vt)&=\frac{1}{\pi}\int\frac{\der^2\zt}{(2\pi)^2}\frac{(\ut-\zt)(\vt-\zt)}{(\ut-\zt)^2(\vt-\zt)^2}\nonumber\\
    &\times[1+U^\dagger(\ut)U(\vt)-U^\dagger(\ut)U(\zt)-U^\dagger(\zt)U(\vt)]^{ab}\,,
\end{align}
where the $U$ are adjoint lightlike Wilson line counterparts of the fundamental $V$'s we have been working with thus far. 

One can therefore write 
\begin{align}
    \der\sigma_{\rm slow}=\ln\left(\frac{k_f^-}{\Lambda^-}\right)\mathcal{H}_{\rm LL}\otimes\der\sigma_{\rm LO}\,,\label{eq:LLcontribNLO}
\end{align}
completing our discussion of Eq.\,\eqref{eq:xsec-decomposition}.

How one addresses the logarithmic dependence on the cut-off $\Lambda^-$ in the NLO dijet cross-section is similar to the treatment of logarithmic collinear divergences in standard collinear factorization, where they are absorbed into nonperturbative parton distribution functions (PDFs). Likewise, as noted in section~\ref{sub:LO-CGC}, the color correlator $\Xi_{\rm LO}$ at the scale $Y_0$ ($=\ln(z_0)$) is nonperturbative, and model dependent, and the value of $Y_0$ itself is arbitrary at leading order. This nonperturbative color correlator $\Xi_{\rm LO}^{(0)}$ at the scale $Y_0$ absorbs the logarithmic divergence when $\Lambda^-\to 0$, allowing us to define the color correlator at the factorization scale $Y_f=\ln(z_f)$ as
\begin{equation}
    \Xi_{\rm LO}(\Xt|Y_f)=\Xi_{\rm LO}^{(0)}(\Xt|Y_0)+\Hcal_{\rm LL}\otimes \Xi_{\rm LO}^{(0)}(\Xt)\ln\left(\frac{z_f}{z_0}\right)+\mathcal{O}(\alpha_s^2)\,.\label{eq:1step-evol}
\end{equation}
We have used here the short form notation 
$$\Xt=(\xt,\yt;\xt',\yt')\,,$$
to represent all the transverse coordinates. 

The requirement that the NLO cross-section be independent of the factorization scale  $z_f$ at leading logarithmic accuracy in the rapidity evolution generates the JIMWLK renormalization group (RG) equation: 
\begin{equation}
    \frac{\partial \Xi_{\rm LO}(\Xt|Y_f)}{\partial Y_f}=\Hcal_{\rm LL}\otimes \Xi_{\rm LO}(\Xt|Y_f)\,,\label{eq:xiLO-LL}
\end{equation}

More generally, this RG procedure applies to the expectation value of any operator averaged over the CGC weight functional, and the previous equation can be written in the more general form
\begin{equation}
    \frac{\partial W^{\rm LL}_{Y_f}[\rho_A]}{\partial Y_f}=\Hcal_{\rm LL}W^{\rm LL}_{Y_f}[\rho_A]\,,\label{eq:W-evol-LL}
\end{equation}
which can be interpreted as the renormalization, induced by 
small $x$ evolution, of the nonperturbative weight functional specifying the distribution of color sources $\rho_A$ in the gluon saturation regime. 

Thus at leading logarithmic accuracy in $x$, the full LO+NLO inclusive DIS dijet cross-section in the CGC EFT can be expressed as 
\begin{equation}
    \der\sigma_{\rm LO+NLO}=\int\mathcal{D}[\rho_A]W^{\rm LL}_{Y_f}[\rho_A]\left(\der\sigma_{\rm LO}+\alpha_s\der\sigma_{\rm NLO,i.f.}\right)\,,\label{eq:dijet-LL}
\end{equation}
with the NLO impact factor defined as the finite $\mathcal{O}(\alpha_s)$ term after factorization of the rapidity divergence\footnote{Note that the various color correlators entering inside $\der\sigma_{\rm LO}$ and $\der\sigma_{\rm NLO,i.f.}$ should be defined without the $\langle [...]\rangle_{Y}$ CGC average since 
this is then performed explicitly in Eq.\,\eqref{eq:dijet-LL}.}: 
\begin{equation}
    \alpha_s\der\sigma_{\rm NLO,i.f.}\equiv\left(\der\sigma_{\rm R_2\times R_2,sud2}+\der\sigma_{\rm R_2\times R_2',sud2}+\mathrm{R}_2\leftrightarrow\mathrm{R}_2'\right)+\der\sigma_{\rm sud1}+\der\sigma_{\rm R,no-sud}+\der\sigma_{\rm V,no-sud}\,.
\end{equation}
Typical choices \cite{Beuf:2014uia,Beuf:2017bpd,Taels:2022tza} for $Y_f$ in Eq.\,\eqref{eq:dijet-LL} are $Y_f=\ln(\textrm{min}(z_1,z_2))$ or $\ln(z_1z_2)$, with the only requirement that  $\alpha_s\ln(z_1/z_f), \alpha_s\ln(z_2/z_f)\ll 1$, ensuring the remaining logarithms in the impact factor are pure $\mathcal{O}(\alpha_s)$ corrections. With this imposed, the sensitivity of the cross-section to the choice $Y_f$ is parametrically of higher order in $\alpha_s$ and can be used to gauge the theoretical uncertainties of our result.

Assuming that rapidity factorization holds at next-to-leading logarithmic accuracy, Eq.\,\eqref{eq:dijet-LL} can be promoted to NLL accuracy using the available NLL JIMWLK evolution equation and the resulting resummed $W^{\rm NLL}_{Y_f}$ weight function~\cite{Balitsky:2013fea,Kovner:2014lca}. 

Finally, we comment on the initial conditions to the evolution equation Eq.\,\eqref{eq:W-evol-LL}.
A physically motivated model of the nonperturbative distribution of color sources at small $x$ is the McLerran-Venugopalan (MV) model \cite{McLerran:1993ni,McLerran:1993ka}, which has a robust justification for very large nuclei\footnote{An alternative approach to determine the initial conditions for smaller nuclei and at larger values of $x$ has been followed in \cite{Dumitru:2018vpr,Dumitru:2021tvw}, where the two-point, three-point and four-point function of color charge correlators is determined perturbatively from the lightcone wave-functions of its valence quarks.}. 

In practice, one solves Eq.\,\eqref{eq:xiLO-LL} up to $Y_f$ using the MV model or a different initial nonperturbative distribution at the rapidity scale $Y_0=\ln(z_0)$. The minus  momentum fraction $z_0$ can be related to a fractional plus momentum of the target $x_0 P^+$ (with $x_0\ll 1$), by noting that for  real gluon emission from the projectile dipole, the on-shell condition gives $2 k_g^+ k_g^-=\kgt^2$. The condition $k_g^+\le x_0 P^+$ then gives 
\begin{equation}
    k_g^-=\frac{\kgt^2}{2k_g^+}\ge \frac{Q_0^2}{2x_0P^+}\Longrightarrow z_g\ge z_0=\frac{Q_0^2}{Q^2}\frac{x_{\rm Bj}}{x_0}\label{eq:z0model}
\end{equation}
where we have used $2q^-P^+=Q^2/x_{\rm Bj}$ and assumed that the transverse momentum of the gluon is larger than a fixed transverse scale $Q_0$ of order of $\Lambda_{\rm QCD}$ for a proton or the initial saturation momentum of a large nucleus provided that $x_0$ is small enough\footnote{A typical  choice in small $x$ studies is $x_0=0.01$, corresponding to a value where logarithms in $x_0$ are large enough to ensure $\alpha_S \ln(1/x_0)\sim \mathcal{O}(1)$.}. Eq.\,\eqref{eq:z0model} provides the typical value for the initial scale $Y_0=\ln(z_0)$ of the rapidity evolution, which accounts for the $x_{\rm Bj}$ dependence of the cross-section.

We now turn to discussing the further improvement in computational accuracy of the back-to-back dijet cross-section at this order that results from the identification and resummation of large Sudakov logarithms.

\section{Sudakov logarithms in back-to-back kinematics at NLO}
\label{sec:sudakov}

In the previous section, we provided a detailed update on our results in \cite{Caucal:2021ent} for the NLO dijet impact factor. Our purpose was two-fold: firstly, we corrected typos. Secondly, we wrote the impact factor in a manner that makes it easier to extract the large Sudakov logarithms that appear in the back-to-back limit of the full NLO inclusive dijet cross-section. 

In this section, we turn to the principal focus of this paper, the NLO inclusive dijet cross-section in DIS at small $x$ in the back-to-back limit. We will  compute, for finite $N_c$, the large double and single Sudakov logarithms
\begin{equation}
    \alpha_s\ln^2\left(\frac{P_\perp}{q_\perp}\right)\,,\qquad \alpha_s\ln\left(\frac{P_\perp}{q_\perp}\right)\,\,,
\end{equation}
that arise when $P_\perp\gg q_\perp$ in the NLO impact factor. These contributions are the dominant ones in the back-to-back limit of the impact factor and are particularly important in order to address the relative importance of saturation effects and soft gluon radiation in the suppression of the back-to-back peak in dijet azimuthal correlations.

This section is divided into four subsections. We will first discuss the state-of-the-art literature on the back-to-back limit of inclusive dijet production at NLO at small $x$. We will then present our computation of the Sudakov logarithms. Somewhat surprisingly, we first find that the coefficient of the double logarithm is positive, which is at odds with the physical expectation of Sudakov suppression which would give the opposite sign. We trace the root of the problem to the proper treatment of the BFKL kernel generating the leading logarithmic evolution in $x$ of the WW gluon TMD, when one imposes $k_g^-$ and $k_g^+$ ordering \cite{Beuf:2014uia,Iancu:2015vea,Iancu:2015joa,Ducloue:2019ezk}, as required in the back-to-back kinematics. Without proper scale choice, the BFKL equation describing the leading twist (in inverse powers of $Q^2$) energy evolution, to leading logarithmic accuracy in $x$ in perturbative QCD, generates 
large double transverse logarithms; this problem becomes manifest in the unphysical behavior of next-to-leading-log (NLL) BFKL evolution~\cite{Fadin:1998py,Ciafaloni:1998gs}. However the BFKL kernel can be modified so as to resum such large double logs, leading to a NLL kernel that is significantly more stable~\cite{Salam:1998tj,Ciafaloni:1999yw,Ciafaloni:2003rd}. Likewise, the positive Sudakov contribution originates from  improper scale choice in small $x$ evolution; this becomes manifest already at leading log on account of the presence of a hard scale in back-to-back kinematics. We will discuss in section~\ref{sub:kcJIMWLK} the interplay of the collinear improvement of the rapidity evolution of the WW TMD  with Sudakov effects. Lastly, we discuss our results for Sudakov logarithms at NLO and comment on their resummation in the TMD formalism at small $x$.

\subsection{Sudakov state-of-the-art at small $x$}
\label{sub:history}

We begin by reviewing the state of the art on the back-to-back limit of inclusive dijet production in DIS at small Bjorken $x$. The Sudakov double logarithms  in these kinematics were first obtained in \cite{Mueller:2013wwa}; their derivation was based on the similarities with the problem of Higgs boson production in proton-nucleus collisions in the limit of large ratio $M^2/k_\perp^2$ where $M$ and $k_\perp$ are the mass and the transverse momentum of the Higgs boson, respectively. For the dijet case, the authors found that the NLO cross section up to the leading double Sudakov logarithm takes the form
\begin{align}
    \int \frac{\der \phi}{2\pi} \left.\frac{\der \sigma^{\gamma_{\rm \lambda}^{\star}+A\to \rm dijet+X}}{ \der^2 \Pt \der^2 \qt \der \eta_1 \der \eta_{2}}\right|_{\rm NLO}&= \alpha_{\rm em}e_f^2\alpha_s\deltatwo\mathcal{H}_{\rm LO}^{0,\lambda}(\Pt)\times\int\frac{\der^2\bt\der^2\bt'}{(2\pi)^4}e^{-i\qt\cdot\rbbpt}\nonumber\\
     &\times\left[1-\frac{\alpha_sN_c}{4\pi}\ln^2\left(\frac{\Pt^2\rbbpt^2}{c_0^2}\right)+\mathcal{O}\left(\alpha_s\ln\right)\right] \hat G^0_Y(\bt,\bt') \,,\label{eq:sudakov-history}
\end{align}
where the $\mathcal{O}(\alpha_s\ln)$ term contains the sub-leading single logarithmic corrections $\propto \alpha_s\ln(\Pt^2\rbbpt^2)$ as well as the finite terms $\propto \alpha_s$. With reference to our previous discussion in 
section \ref{subsub:corlimit}, note that this is the cross-section averaged over the azimuthal angle of $\Pt$ w.r.t.\ $\qt$, which explains why only the unpolarized WW gluon TMD appears. The derivation of this result is performed starting from the TMD factorization framework in the back-to-back limit. 

In our approach, we begin with general kinematics and do not a priori assume TMD factorization at NLO. We find that the manner in which infrared divergences cancel between real and virtual diagrams in our case differs from the approach in~\cite{Mueller:2013wwa}. In particular, we do not find that the virtual diagrams  $\rm V_1$ and $\rm SE_1$ (in which the virtual gluon crosses the shockwave) contribute to the Sudakov logarithms via the cancellation of infrared divergences. In our calculation, the soft infrared divergences cancel separately among the virtual (between diagrams $\rm SE_2, V_2$ and the UV singular component of $\rm SE_1$) and real corrections (between in-cone and out-cone contributions) for the diagrams proportional to the leading order color correlator (the leading $N_c$ contribution in the terminology of \cite{Mueller:2013wwa}). For the diagrams proportional to $\Xi_{\rm NLO,3}$ (the $1/N_c$ suppressed terms in the terminology of \cite{Mueller:2013wwa}), the cancellation occurs between the virtual diagram $\rm V_3\times LO$ and the real diagram $\rm R_2\times R_2'$. We refer the reader to appendix~\ref{app:xs-dec} for more details on the mechanism of cancellations in our computation.

In this respect, our approach is closer to the recent study in \cite{Taels:2022tza}, which addresses the back-to-back limit of inclusive dijet photoproduction at small $x$. We reach similar conclusions as the authors of this paper, namely the importance of the interplay between the need to go beyond leading logarithmic JIMWLK factorization with the imposition of kinematic constraints on the rapidity evolution that are necessary to recover the correct structure of the Sudakov double logarithm. We however go beyond \cite{Taels:2022tza} by computing a more 
general process (their results in the photoproduction limit are obtained from the $Q^2\to 0$ limit of our $\gamma_{\rm T}^\star\to\rm dijet +X$ cross-section) which allows us to compute the back-to-back results for both transversely and longitudinally polarized virtual photons.  We also go beyond their large $N_c$ results by obtaining not only the Sudakov double logarithm at finite $N_c$ but the single logarithms as well, in the small $R$ approximation. In the more general kinematics that we consider, we are further able to compute\footnote{These contributions vanish in the photoproduction limit (at leading order) because of the $\bar Q$ factor in front of the $\cos(2\phi)$ asymmetry in Eq.\,\eqref{eq:LOb2b-T}.} the Sudakov logarithms associated with the linearly polarized WW gluon TMD $h^0_Y(\qt)$.

\subsection{Correlation limit of the NLO impact factor}
\label{sub:NLOb2b}

In this section, we will compute the back-to-back limit $|\Pt|\gg|\qt|$ of the NLO impact factor $\alpha_s\der\sigma_{\rm NLO,i.f.}$ given by the sum of the terms 
\begin{equation}
    \left(\der\sigma_{\rm R_2\times R_2,sud2}+\der\sigma_{\rm R_2\times R_2',sud2}+\mathrm{R}_2\leftrightarrow\mathrm{R}_2'\right)+\der\sigma_{\rm sud1}+\der\sigma_{\rm R,no-sud}+\der\sigma_{\rm V,no-sud}\,.\label{eq:xsec-dec-recap}
\end{equation}
Since we anticipate the Sudakov logarithms to be proportional to the LO operator $G^{ij}_Y$, we will focus on the terms which are proportional to the color correlators $\Xi_{\rm LO}$ and $\Xi_{\rm NLO,3}$ since these are the only correlators which reduce to the WW gluon distribution in the correlation limit (without further assumptions). We will now proceed by first computing the contributions in the impact factor coming from the ``sud2" term in Eq.\,\eqref{eq:xsec-dec-recap} and subsequently the ``sud1" term.

\subsubsection{Calculation of the Sudakov double logarithms}
\label{subsub:XiLO}

\paragraph*{Contributions depending on $\Xi_{\rm LO}$.} We first take the back-to-back limit of the term labeled $\der\sigma_{\rm R_2\times R_2,sud2}$ in our NLO impact factor. The color structure of this contribution is proportional to $\Xi_{\rm LO}$.

In the aforementioned term, we perform the change of variables $$(\ktone,\kttwo)\to(\Pt,\qt) \,\,;\,\, (\xt,\yt)\to(\ut,\bt)$$ in the integrands, and then take the correlation limit, to extract the leading term in powers of then  $q_\perp/P_\perp$ expansion of the cross-section. We  neglect corrections of order $Q_s/P_\perp$ as well.

In particular, we have already seen that to leading power in $q_\perp/P_\perp$, one can approximate
\begin{align}
    \Xi_{\rm LO}(\xt,\yt;\xt',\yt')&\approx\frac{\alpha_s}{2N_c}\ut^i\ut'^j \times \hat G^{ij}_{Y_f}(\bt,\bt')\,,\\
    \hat G^{ij}_{Y_f}(\bt,\bt') &= \frac{-2}{\alpha_s}\left\langle\Tr\left(V(\bt)\partial^iV^\dagger(\bt)V(\bt')\partial^jV^\dagger(\bt')\right)\right\rangle_{Y_f}\,,
\end{align}
inside the NLO impact factor. The color correlator is evaluated at the rapidity factorization scale $Y_f=\ln(z_f)$ following our discussion in section~\ref{subsub:JIMWLK}.
Finally, one can replace $\ktone$ by $\Pt$, $\kttwo$ by $-\Pt$, and $\rxxtp$, $\ryytp$ by $\rbbpt$ in the back-to-back limit.

After these manipulations, we obtain for the so-called ``sud2" contribution:
\begin{align}
    \der\sigma_{\rm R_2\times R_2,sud2}&=\alpha_{\rm em}\alpha_se_f^2\deltatwo\Hcal^{\lambda,ij}_{\rm LO}(\Pt)\int\frac{\der^2\bt\der^2\bt'}{(2\pi)^4} e^{-i\qt\cdot\rbbpt}\hat G^{ij}_{Y_f}(\bt,\bt')\nonumber\\
    &\times \frac{\alpha_sC_F}{\pi}\int_0^1\frac{\der\xi}{\xi}\left[1-e^{-i\xi\Pt\cdot\rbbpt}\right]\ln\left(\frac{\Pt^2\rbbpt^2R^2\xi^2}{c_0^2}\right)+\mathcal{O}\left(R^2,\frac{q_\perp}{P_\perp},\frac{Q_s}{P_\perp}\right)\,.\label{eq:R2R2-soft-b2b} 
\end{align}
For simplicity, we shall drop from now on the order of magnitude of the neglected terms but one should keep in mind that these expressions neglect powers of $R$, powers of $q_\perp/P_\perp$ and powers of $Q_s/P_\perp$.

To proceed further, it is convenient to simplify the tensor structure of this expression. Firstly due to translational invariance, $\hat G^{ij}$ 
only depends on $\rbbpt=\bt-\bt'$ and does not depend on the impact parameter $\st =\frac{1}{2}(\bt+\bt')$. One can then write, with $S_\perp$ denoting the transverse area of the target,
\begin{equation}
    \int\frac{\der^2\bt\der^2\bt'}{(2\pi)^4} e^{-i\qt\cdot\rbbpt}\hat G^{ij}_{Y_f}(\bt,\bt')=S_\perp\int\frac{\der^2\rbbpt}{(2\pi)^4}e^{-i\qt\cdot\rbbpt}\hat G^{ij}_{Y_f}(\rbbpt)\,,
\end{equation}
and decompose the gluon distribution $\hat G^{ij}_{Y_f}(\rbbpt)$ into trace and traceless components,
\begin{equation}
    \hat G^{ij}_{Y_f}(\rbbpt)=\frac{1}{2}\delta^{ij}\hat G^0_{Y_f}(\rbbpt)+\frac{1}{2}\left(\frac{2\rbbpt^i\rbbpt^j}{\rbbpt^2}-\delta^{ij}\right)\hat h^0_{Y_f}(\rbbpt)\,.\label{eq:Gij-decomp-bt}
\end{equation}
One should keep in mind that the unpolarized and linearly polarized WW gluon distributions in $\bt$-space depend only on the modulus of $\rbbpt$. Note also that $\hat G^0(\rbbpt)$ and $\hat h^0(\rbbpt)$ are the Fourier transforms of their momentum space counterparts $G^0(\qt)$ and $h^0(\qt)$ (discussed previously in section~\ref{subsub:corlimit}), respectively, 
\begin{align}
    G_Y^0(\qt)&=\frac{S_\perp}{(2\pi)^2}\int\frac{\der^2\rbbpt}{(2\pi)^2} e^{-i\qt\cdot\rbbpt}\hat G_Y^0(\rbbpt)\,,\\
    h_Y^0(\qt)&=\frac{S_\perp}{(2\pi)^2}\int\frac{\der^2\rbbpt}{(2\pi)^2} e^{-i\qt\cdot\rbbpt}\cos\left(2\theta\right)\hat h_Y^0(\rbbpt)\,,
\end{align}
with $\theta$ the angle between $\qt$ and $\rbbpt$.

It is then straightforward to obtain the NLO corrections from the ``sud2" term, for both the unpolarized and linearly polarized WW gluon TMDs:
\begin{align}
    \der\sigma^{\lambda=\rm L}_{\rm R_2\times R_2,sud2}&= \alpha_{\rm em}\alpha_s e_f^2\deltatwo\Hcal_{\rm LO}^{0,\lambda=\rm L}(\Pt)\int\frac{\der^2\rbbpt}{(2\pi)^4}e^{-i\qt\cdot\rbbpt}\left[\hat G^0_{Y_f}(\rbbpt)+\cos(2\hat\phi)\hat h^0_{Y_f}(\rbbpt)\right]\nonumber\\
    &\times \frac{\alpha_sC_F}{\pi}\int_0^1\frac{\der\xi}{\xi}\left[1-e^{-i\xi|\Pt||\rbbpt|\cos(\hat\phi)}\right]\ln\left(\frac{\Pt^2\rbbpt^2R^2\xi^2}{c_0^2}\right)\,,
    \label{eq:R2R2b2b-phi}
\end{align}
where $\hat\phi$ is the azimuthal angle between $\Pt$ and $\rbbpt$. (We absorbed here the target area $S_\perp$ in the TMD distributions.)
In order to simplify the $\hat\phi$ dependence of this expression, we shall expand the azimuthal dependence of the cross-section into its  Fourier harmonics:
\begin{equation}
    \frac{\der \sigma^{\gamma_{\lambda}^{\star}+A\to \textrm{dijet}+X}}{ \der^2 \Pt \der^2 \qt \der \eta_1 \der \eta_{2}}= \der \sigma^{(0),\lambda}(P_\perp,q_\perp,\eta_1,\eta_2)+2\sum_{n=1}^{\infty} \der \sigma^{(n),\lambda}(P_\perp,q_\perp,\eta_1,\eta_2)\cos(n\phi)\,,
\end{equation}
with $\phi$ the azimuthal angle between $\Pt$ and $\qt$.
Because of the $\phi\leftrightarrow-\phi$ symmetry, only cosine terms appear in the Fourier decomposition. Furthermore, due to quark anti-quark symmetry, the odd harmonics in the Fourier expansion vanish identically\footnote{Recall that under quark anti-quark interchange, $\ktone \leftrightarrow \kttwo$ and $z_1 \leftrightarrow z_2$; thus $\Pt \to -\Pt$ and $\qt \to \qt$ (or equivalently $\phi \to \phi + \pi$).}.

To illustrate the emergence of Sudakov type logarithms, and the interplay between the contributions to the azimuthal anisotropy due to the linearized WW TMD, and that due to soft gluon radiation, we compute the first nonvanishing $\der \sigma^{(0),\rm L}$ and $\der \sigma^{(2),\rm L}$ coefficients for a longitudinally polarized virtual photon. The former is simply the inclusive dijet cross-section averaged over the azimuthal angle of $\Pt$ w.r.t.\ $\qt$:
\begin{equation}
    \der \sigma^{(0),\lambda}=\frac{1}{2\pi}\int_0^{2\pi}\der\phi \ \frac{\der \sigma^{\gamma_{\lambda}^{\star}+A\to \textrm{dijet}+X}}{ \der^2 \Pt \der^2 \qt \der \eta_1 \der \eta_{2}}\,,
\end{equation}
while the latter is the $\langle\cos(2\phi)\rangle$ anisotropy (related to the familiar $v_2$ coefficient in heavy-ion physics through $v_2=\der \sigma^{(2)}/\der \sigma^{(0)}$):
\begin{equation}
     \der \sigma^{(2),\lambda}=\frac{1}{2\pi}\int_0^{2\pi} \der\phi \  \frac{\der \sigma^{\gamma_{\lambda}^{\star}+A\to \textrm{dijet}+X}}{ \der^2 \Pt \der^2 \qt \der \eta_1 \der \eta_{2}} \cos(2\phi) \,.
\end{equation}
Higher $\langle \cos(n\phi)\rangle$ harmonics are defined similarly to $\der \sigma^{(2)}$ with the replacement $\cos(2\phi)\to\cos(n\phi)$.
Using Eq.\,\eqref{eq:R2R2b2b-phi} and the Jacobi-Anger identity
\begin{equation}
    e^{iz\cos(x)}=\BesselJ_0(z)+2\sum_{n=1}^\infty \ i^n\BesselJ_n(z)\cos(nx)\,,\label{eq:Jacobi-Anger}
\end{equation}
to decompose the phase $e^{-i\xi|\Pt||\rbbpt|\cos(\hat\phi)}$ into cosine harmonics, one obtains the following expression for the cross-section averaged over $\phi$:
\begin{align}
    &\der \sigma^{(0),\lambda=\rm L}_{\rm R_2\times R_2,sud2}= \alpha_{\rm em}\alpha_s e_f^2\deltatwo\Hcal_{\rm LO}^{0,\lambda=\rm L}(\Pt)\nonumber\\
    &\times \frac{\alpha_sC_F}{\pi}\int\frac{\der^2\rbbpt}{(2\pi)^4}e^{-i\qt\cdot\rbbpt}\left\{\hat G^0_{Y_f}(\rbbpt)\int_0^1\frac{\der\xi}{\xi}\left[1-\BesselJ_0\left(\xi|\Pt||\rbbpt|\right)\right]\ln\left(\frac{\Pt^2\rbbpt^2R^2\xi^2}{c_0^2}\right)\right.\nonumber\\
    &\left.+\hat h^0_{Y_f}(\rbbpt)\int_0^1\frac{\der\xi}{\xi}\BesselJ_2\left(\xi|\Pt||\rbbpt|\right)\ln\left(\frac{\Pt^2\rbbpt^2R^2\xi^2}{c_0^2}\right)\right\}\label{eq:R2R2b2b-c0}\,,
\end{align}
while the $\langle \cos(2\phi)\rangle$ anisotropy reads
\begin{align}
    &\der \sigma^{(2),\lambda=\rm L}_{\rm R_2\times R_2,sud2}= \alpha_{\rm em}\alpha_s e_f^2\deltatwo\Hcal_{\rm LO}^{0,\lambda=\rm L}(\Pt)\nonumber\\
    &\times \frac{\alpha_sC_F}{\pi}\int\frac{\der^2\rbbpt}{(2\pi)^4}e^{-i\qt\cdot\rbbpt}\cos(2\theta)\left\{\hat G ^0_{Y_f}(\rbbpt)\int_0^1\frac{\der\xi}{\xi}\,\BesselJ_2\left(\xi|\Pt||\rbbpt|\right)\ln\left(\frac{\Pt^2\rbbpt^2R^2\xi^2}{c_0^2}\right)\right\}\nonumber\\
    &\left.+\frac{1}{2}\hat h ^0_{Y_f}(\rbbpt)\int_0^1\frac{\der\xi}{\xi}\left[1-\BesselJ_0(\xi|\Pt||\rbbpt|)-\BesselJ_4(\xi|\Pt||\rbbpt|)\right]\ln\left(\frac{\Pt^2\rbbpt^2R^2\xi^2}{c_0^2}\right)\right\}\,,\label{eq:R2R2b2b-c2}
\end{align}
where we recall that $\theta$ is the angle between $\qt$ and $\rbbpt$. 

Before computing the $\xi$ integral, we note that the $\langle\cos(2\phi)\rangle$ anisotropy is sensitive to the unpolarized gluon distribution as a consequence of the azimuthal anisotropy generated by soft gluon radiation. This feature was  discussed previously in the collinear and TMD factorization framework \cite{Hatta:2020bgy,Hatta:2021jcd}. Higher order even harmonics can be computed in a similar fashion; this is worked out in Appendix~\ref{app:cosnphi}.

It is easy to understand how the convergent $\xi$ integrals in Eqs.\,\eqref{eq:R2R2b2b-c0}-\eqref{eq:R2R2b2b-c2} give rise to large Sudakov logarithms $\propto \ln(|\Pt||\rbbpt|)$ when $|\Pt|\gg1/|\rbbpt|$;  since $\rbbpt$ and $\qt$ are conjugate to each other, such logarithms are of the type $\ln(P_\perp/q_\perp)$. The Bessel functions decay typically for $\xi>1/(|\Pt||\rbbpt|)$. To extract the logarithmic terms, it suffices to approximate 
\begin{equation}
    1-\BesselJ_0(\xi |\Pt||\rbbpt|)\approx \Theta\left(\xi-\frac{c_0}{ |\Pt||\rbbpt|}\right)\,.
\end{equation}
Therefore when $|\Pt||\rbbpt|$ becomes large, the integral over $\xi$ becomes strongly sensitive to the logarithmic singularity in $\xi=0$, giving contributions of order $\ln^2(|\Pt||\rbbpt|)$ and $\ln(|\Pt||\rbbpt|)$. Using the identities in Appendix~\ref{app:math-id}, one can perform explicitly all these $\xi$ integrals up to corrections of order $q_\perp/P_\perp$. Our final results are then
\begin{align}
    &\der \sigma^{(0),\lambda=\rm L}_{\rm R_2\times R_2, sud2}=\alpha_{\rm em}\alpha_s e_f^2\deltatwo\Hcal_{\rm LO}^{0,\lambda=\rm L}(\Pt)\nonumber\\
    &\times \frac{\alpha_sC_F}{\pi}\int\frac{\der^2\rbbpt}{(2\pi)^4}e^{-i\qt\cdot\rbbpt}\left\{\hat G^0_{Y_f}(\rbbpt)\left[\frac{1}{4}\ln^2\left(\frac{\Pt^2\rbbpt^2}{c_0^2}\right)+\ln(R)\ln\left(\frac{\Pt^2\rbbpt^2}{c_0^2}\right)\right]\right.\nonumber\\
    &\hspace{4.5cm} \left.+\hat h^0_{Y_f}(\rbbpt)\left[\frac{1}{2}+\ln(R)\right]\right\}\,,\label{eq:c0b2b-final}\\
    &\der \sigma^{(2),\lambda=\rm L}_{\rm R_2\times R_2, sud2}=\alpha_{\rm em}\alpha_s e_f^2\deltatwo\Hcal_{\rm LO}^{0,\lambda=\rm L}(\Pt)\nonumber\\
    &\times \frac{\alpha_sC_F}{\pi}\int\frac{\der^2\rbbpt}{(2\pi)^4}e^{-i\qt\cdot\rbbpt}\frac{\cos(2\theta)}{2}\left\{\hat h^0_{Y_f}(\rbbpt)\left[\frac{1}{4}\ln^2\left(\frac{\Pt^2\rbbpt^2}{c_0^2}\right)+\ln(R)\ln\left(\frac{\Pt^2\rbbpt^2}{c_0^2}\right)\right.\right.\nonumber\\
    &\hspace{4.5cm} \left.\left.-\frac{5}{8}-\frac{1}{2}\ln(R)\right]+\hat G^0_{Y_f}(\rbbpt)\left[1+2\ln(R)\right]\right\}\,.
    \label{eq:c2b2b-final}
\end{align}
It is useful to compare our results with those given in \cite{Hatta:2021jcd} for the azimuthally symmetric term $\der \sigma^{(0)}$ and the $\langle \cos(2\phi)\rangle$ anisotropy. Firstly, unlike the LO result, we observe\footnote{An analogous observation was made in \cite{delCastillo:2021znl} within the TMD factorization formalism when supplemented by angular dependent soft functions.} that at NLO the azimuthally averaged cross-section is sensitive to the linearly polarized WW gluon TMD $\hat h^0$.  Physically, this term comes from the azimuthal anisotropy induced by soft gluon radiation, combined with the anisotropy induced by the gluon distribution itself. Secondly, for the same reason, we have a pure $\alpha_s$ contribution proportional to $-5/8-1/2\ln(R)$ that multiplies the linearly polarized TMD $\hat h^0$ in the $\langle \cos(2\phi)\rangle$ anisotropy. This term is not given in \cite{Hatta:2021jcd}; it is parametrically of the same order as the one proportional to the unpolarized TMD $\hat G^0$.

This concludes our calculation of $\der\sigma_{\rm R_2\times R_2, sud2}$ in the back-to-back limit.  We  observed the emergence of large Sudakov double and single logarithms from the ``sud2" terms in the fully inclusive NLO cross-section computed in the previous section.  The contribution $\der\sigma_{\rm R_2'\times R_2'}$ can be obtained using  quark-antiquark interchange, and therefore gives exactly the same results as Eqs.\,\eqref{eq:c0b2b-final}-\eqref{eq:c2b2b-final}.

\paragraph*{Contributions depending on $\Xi_{\rm NLO,3}$.}

In the previous paragraph, we computed the back-to-back limit of the ``sud2" term in the NLO impact factor which is proportional to the LO color correlator $\Xi_{\rm LO}$. We found that some of these terms develop large logarithms of the form $\ln(|\Pt||\rbbpt|)$ when $P_\perp/q_\perp\to\infty$. We will show now that similar features occur for the terms in the cross-section proportional to the color correlator $\Xi_{\rm NLO,3}$, that arise from the interference diagram $\rm R_2\times R_2'$ or the virtual graph $\rm V_3\times \rm LO$. 

It is important to note that this color correlator also naturally reduces to the WW gluon TMD in the correlation limit. To see this, we first express the color correlator as 
\begin{equation}
    \Xi_{\rm NLO,3}(\xt,\yt;\xt',\yt')=\frac{1}{N_c}\left\langle\Tr \left[t^aV(\xt)V^\dagger(\yt)t_a-C_F\mathbbm{1}\right]\left[V(\yt')V^\dagger(\xt') -\mathbbm{1} \right]\right\rangle_{Y_f}\,.
\end{equation}
Using the expansion Eq.\,\eqref{eq:VxVy-exp}, and the Fierz identity, one gets to leading power in $\ut$ and $\ut'$:
\begin{align}
   \Xi_{\rm NLO,3}(\xt,\yt;\xt',\yt')&=\frac{1}{2}\ut^i\ut'^j\frac{1}{N_c}\left\langle \Tr\left[V(\bt)\partial_iV^\dagger(\bt)\right]\Tr\left[\left(\partial_j V(\bt') \right) V^\dagger(\bt')\right]\right\rangle_{Y_f}\nonumber\\
   &-\frac{1}{2}\ut^i\ut'^j\frac{1}{N_c^2}\left\langle\Tr\left[\left(\partial_i V^\dagger(\bt) \right) V(\bt') \left(\partial_j V^\dagger(\bt') \right) V(\bt)\right]\right\rangle_{Y_f}\,.
\end{align}
Due to unitarity of the Wilson lines one has $\Tr[V(\bt)\partial^iV^\dagger(\bt)]=0$,  we see that this color correlator reduces to the WW gluon TMD in the correlation limit with a $-1/(2N_c^2)$ prefactor,
\begin{equation}
    \Xi_{\rm NLO,3}(\xt,\yt;\xt',\yt')\approx-\frac{\alpha_s}{4N_c^2}\ut^i\ut'^j \times \hat G^{ij}_{Y_f}(\bt,\bt')\,.\label{eq:XiNLO3b2b}
\end{equation}
The calculation of the $P_\perp/q_\perp\to\infty$ limit of $\der\sigma_{\rm R_2\times R_2',sud2}$ is then similar to the one of $\der\sigma_{\rm R_2\times R_2,sud2}$. Using Eq.\,\eqref{eq:R2R2'soft} and Eq.\,\eqref{eq:XiNLO3b2b}, as well as the standard change of variable from $(\ktone,\kttwo)$ to $(\Pt,\qt)$, we get
\begin{align}
    \der\sigma_{\rm R_2\times R_2',sud2}&=\alpha_{\rm em}\alpha_se_f^2\deltatwo\Hcal^{\lambda,ij}_{\rm LO}(\Pt)\int\frac{\der^2\bt\der^2\bt'}{(2\pi)^4} e^{-i\qt\cdot\rbbpt}\hat G^{ij}_{Y_f}(\bt,\bt')\nonumber\\
    &\times \frac{\alpha_s}{2\pi N_c}\int_0^1\frac{\der\xi}{\xi}\left[1-e^{-i\xi\Pt\cdot\rbbpt}\right]\ln\left(\frac{\Pt^2\rbbpt^2\xi^2}{z_2^2c_0^2}\right)\,,\label{eq:R2R2'-soft-b2b}
\end{align}
where we have also used $\rxypt\simeq \rbbpt$ in the correlation limit. Decomposing the WW gluon TMD $\hat G_{Y_f}^{ij}$ and the azimuthal dependence of the cross-section in Fourier modes, as done in the previous paragraph on $\Xi_{\rm LO}$ contributions, we find that the azimuthally averaged cross-section and the $\langle \cos(2\phi)\rangle$ anisotropy read
\begin{align}
    &\der \sigma^{(0),\lambda=\rm L}_{\rm R_2\times R_2',sud2}=\alpha_{\rm em}\alpha_s e_f^2\deltatwo\Hcal_{\rm LO}^{0,\lambda=\rm L}(\Pt)\nonumber\\
    &\times \frac{\alpha_s}{2\pi N_c}\int\frac{\der^2\rbbpt}{(2\pi)^4}e^{-i\qt\cdot\rbbpt}\left\{\hat G^0_{Y_f}(\rbbpt)\left[\frac{1}{4}\ln^2\left(\frac{\Pt^2\rbbpt^2}{c_0^2}\right)-\ln(z_2)\ln\left(\frac{\Pt^2\rbbpt^2}{c_0^2}\right)\right]\right.\nonumber\\
    &\hspace{4.5cm} \left.+\hat h^0_{Y_f}(\rbbpt)\left[\frac{1}{2}-\ln(z_2)\right]\right\}\,,\label{eq:c0_R2R2'_b2b-final}\\
    &\der \sigma^{(2),\lambda=\rm L}_{\rm R_2\times R_2',sud2}=\alpha_{\rm em}\alpha_s e_f^2\deltatwo\Hcal_{\rm LO}^{0,\lambda=\rm L}(\Pt)\nonumber\\
    &\times \frac{\alpha_s}{2\pi N_c}\int\frac{\der^2\rbbpt}{(2\pi)^4}e^{-i\qt\cdot\rbbpt}\frac{\cos(2\theta)}{2}\left\{\hat h^0_{Y_f}(\rbbpt)\left[\frac{1}{4}\ln^2\left(\frac{\Pt^2\rbbpt^2}{c_0^2}\right)-\ln(z_2)\ln\left(\frac{\Pt^2\rbbpt^2}{c_0^2}\right)\right.\right.\nonumber\\
    &\hspace{4.5cm} \left.\left.-\frac{5}{8}+\frac{1}{2}\ln(z_2)\right]+\hat G^0_{Y_f}(\rbbpt)\left[1-2\ln(z_2)\right]\right\}\,.\label{eq:c2_R2R2'_b2b-final}
\end{align}
These results are very similar to Eqs.\eqref{eq:c0b2b-final} and \eqref{eq:c2b2b-final} modulo the replacement $C_F\to 1/(2N_c)$ coming from the color correlator $\Xi_{\rm NLO,3}$ and $\ln(R)\to-\ln(z_2)$ coming from the $\xi$ integral itself. The contribution from the diagram $\rm R_2'\times \rm R_2$ can be obtained by replacing $z_2\to z_1$ in these expressions.

\subsubsection{Calculation of the Sudakov single logarithms}

The computation of the back-to-back limit of $\der\sigma_{\rm R_2\times R_2,sud2}$ and $\der\sigma_{\rm R_2\times R_2',sud2}$ in the previous subsection is sufficient to get the double Sudakov logarithms, with the full $N_c$ dependence of the coefficient of the double logarithm. The expressions Eqs.\,\eqref{eq:c0b2b-final},\eqref{eq:c0_R2R2'_b2b-final} and Eqs.\,\eqref{eq:c2b2b-final},\eqref{eq:c2_R2R2'_b2b-final} also have single Sudakov logarithms. In order to collect all such single logarithmic terms, 
we need to examine the other term in the NLO impact factor labeled $\der\sigma_{\rm sud1}$. We shall demonstrate that this term gives rise to the required \textit{single} Sudakov logarithms (plus finite pieces) in the back-to-back limit, as claimed in section~\ref{sec:update-if}.

Making the replacements discussed in the introduction to subsection \ref{subsub:XiLO} in Eq.~\eqref{eq:sigma_Sud1} and using the correlation limit of $\Xi_{\rm LO}$ and $\Xi_{\rm NLO,3}$, one gets
\begin{align}
\der\sigma_{\rm sud1}&=\frac{\alpha_{\rm em}\alpha_se_f^2\deltatwo}{2(2\pi)^6}\int\der^8\Xttilde e^{-i\Pt\cdot\ruupt-i\qt\cdot\rbbpt}\ut^i\ut'^j\Rcal_{\mathrm{LO}}^{\lambda}(\ut,\ut')\hat G^{ij}_{Y_f}(\qt,\qt')\nonumber\\
    &\times\frac{\alpha_sN_c}{2\pi}\left\{-\ln\left(\frac{z_1}{z_f}\right)\ln\left(\frac{\rbbpt^2}{|\ut||\ut'|}\right)-\ln\left(\frac{z_2}{z_f}\right)\ln\left(\frac{\rbbpt^2}{|\ut||\ut'|}\right)\right\}\,.
    \label{eq:IRC-final-b2b}
\end{align}
From the phase factor, we have $|\Pt|\sim |\ut|\sim |\ut'|$, indicating that the single logarithms inside the curly bracket are Sudakov-like. We then decompose 
\begin{equation}
   \ln\left(\frac{\rbbpt^2}{|\ut||\ut'|}\right)=\ln\left(\frac{\Pt^2\rbbpt^2}{c_0^2}\right)+\ln(c_0^2)-\frac{1}{2}\ln\left(\Pt^4\ut^2\ut'^2\right)\,.
\end{equation}
By examining carefully the resulting expression, we notice that it cannot be cast into the same factorized form as the LO cross-section or the back-to-back contributions computed in the previous paragraph. This is due to the $\ln(\Pt^4\ut^2\ut'^2)$ term coming from the equation above. This is remedied by introducing another hard factor $\Hcal_{\rm NLO,1}^{\lambda,ij}$,
\begin{equation}
    \Hcal_{\rm NLO,1}^{\lambda,ij}(\Pt)\equiv\frac{1}{2}\int\frac{\der^2\ut}{(2\pi)}\int\frac{\der^2\ut'}{(2\pi)}e^{-i\Pt\cdot\ruupt}\ut^i\ut'^j\Rcal_{\rm LO}^\lambda(\ut,\ut')\ln(\Pt^4\ut^2\ut'^2)\,.\label{eq:hard-NLO1}
\end{equation}
Since we are interested here in Sudakov-like logarithms in various contributions to the cross-section, we will leave the evaluation of this hard factor to section~\ref{sec:TMD-NLO}, where we discuss TMD factorization at NLO. For the moment, using the formal expression for the  new hard factor, we can rewrite Eq.\,\eqref{eq:IRC-final-b2b} as 
\begin{align}
    \der\sigma_{\rm sud1}&=\alpha_{\rm em}\alpha_se_f^2\deltatwo\Hcal_{\rm LO}^{\lambda,ij}(\Pt)\times\frac{\alpha_sN_c}{2\pi}\int\frac{\der^2 \rbbpt}{(2\pi)^4}e^{-i\qt\cdot\rbbpt}\hat G^{ij}_{Y_f}(\rbbpt)\nonumber\\
    &\times\left\{\ln\left(\frac{z_f^2}{z_1z_2}\right)\ln\left(\frac{\Pt^2\rbbpt^2}{c_0^2}\right)+\ln\left(\frac{z_f^2}{z_1z_2}\right)\ln(c_0^2)\right\}\nonumber\\
    &+\alpha_{\rm em}\alpha_se_f^2\deltatwo\Hcal_{\rm NLO,1}^{\lambda,ij}(\Pt)\times\frac{\alpha_sN_c}{4\pi}\ln\left(\frac{z_1z_2}{z_f^2}\right)\int\frac{\der^2 \rbbpt}{(2\pi)^4}e^{-i\qt\cdot\rbbpt}\hat G^{ij}_{Y_f}(\rbbpt)\,,\label{eq:sud1-final}
\end{align}
which is our final result for the back-to-back limit of $\der\sigma_{\rm sud1}$. This expression contains a single Sudakov logarithm of the form 
\begin{align}
    \frac{\alpha_sN_c}{2\pi}\ln\left(\frac{z_f^2}{z_1z_2}\right)\ln\left(\frac{\Pt^2\rbbpt^2}{c_0^2}\right)\,.
\end{align}
The dependence of the coefficient of the Sudakov logarithm upon the rapidity factorization scale $z_f$ is intriguing and will be further discussed in section~\ref{sub:kcJIMWLK}. 
The single Sudakov logarithm Eq.\,\eqref{eq:sud1-final} and those computed in the previous section and included in $\der\sigma_{\rm sud2}$ are the main results of this section.

\subsubsection{Summary and discussion}

We are now ready to combine together Eqs.\,\eqref{eq:c0b2b-final},\eqref{eq:c0_R2R2'_b2b-final} and \eqref{eq:sud1-final}, for the azimuthally averaged cross-section and Eqs.\,\eqref{eq:c2b2b-final},\eqref{eq:c2_R2R2'_b2b-final} and \eqref{eq:sud1-final} for the $\langle \cos(2\phi)\rangle $ anisotropy.  We shall here focus on the Sudakov logarithms and systematically discard the finite terms of order $\mathcal{O}(\alpha_s)$. The final result that includes these finite terms will be presented in section~\ref{sub:final-resum}. 

For the azimuthally averaged cross-section ($\der \sigma^{(0)}$ in the Fourier decomposition) we find
\begin{align}
    \der \sigma^{(0),\lambda=\rm L}&=\alpha_{\rm em}\alpha_s e_f^2\deltatwo\Hcal_{\rm LO}^{0,\lambda=\rm L}(\Pt)\int\frac{\der^2\rbbpt}{(2\pi)^4}e^{-i\qt\cdot\rbbpt}\left\{1+\frac{\alpha_s N_c}{4\pi}\ln^2\left(\frac{\Pt^2\rbbpt^2}{c_0^2}\right)\right.\nonumber\\
    &\left.-\frac{\alpha_s }{\pi}\left[C_F\ln\left(\frac{1}{z_1 z_2R^2}\right)-N_c \ln\left(\frac{z_f}{z_1z_2}\right)\right]\ln\left(\frac{\Pt^2\rbbpt^2}{c_0^2}\right)\right\}\hat G^0_{Y_f}(\rbbpt)+\mathcal{O}(\alpha_s)\,,\label{eq:c0-final}
\end{align}
and for the $\langle \cos(2\phi)\rangle$ anisotropy, 
\begin{align}
    \der \sigma^{(2),\lambda=\rm L}&=\alpha_{\rm em}\alpha_s e_f^2\deltatwo\Hcal_{\rm LO}^{0,\lambda=\rm L}(\Pt)\int\frac{\der^2\rbbpt}{(2\pi)^4}e^{-i\qt\cdot\rbbpt}\frac{\cos(2\theta)}{2}\left\{1+\frac{\alpha_s N_c}{4\pi}\ln^2\left(\frac{\Pt^2\rbbpt^2}{c_0^2}\right)\right.\nonumber\\
    &\left.-\frac{\alpha_s }{\pi}\left[C_F\ln\left(\frac{1}{z_1 z_2R^2}\right)-N_c\ln\left(\frac{z_f}{z_1z_2}\right)\right]\ln\left(\frac{\Pt^2\rbbpt^2}{c_0^2}\right)\right\}\hat h^0_{Y_f}(\rbbpt)+\mathcal{O}(\alpha_s)\label{eq:c2-final} \,.
\end{align}
Note that in these expressions, we have added the small $x$ evolved LO result; this gives the $``1"$ term in the curly brackets.
One further notices that the coefficients of the double and single Sudakov logarithms are the same for the unpolarized and linearly polarized WW gluon distributions. In the resummation (\`{a} la Collins-Soper) we will discuss in section~\ref{sub:final-resum}, this means that the all order resummation of soft gluon logarithms in the NLO impact factor is identical for both TMD distributions ($\hat G^0$ and $\hat h^0$) in coordinate space.

Comparing the result in Eq.~\eqref{eq:c0-final} with Eq.\,\eqref{eq:sudakov-history}  derived previously in \cite{Mueller:2013wwa}, the reader will observe that
the coefficient of our \textit{double} logarithmic term has the opposite sign relative to the Sudakov double log in \cite{Mueller:2013wwa}. This is obviously unphysical: one expects soft gluon radiation to reduce the cross-section in the back-to-back limit since emitted soft gluons contribute to a transverse momentum imbalance $\kgt\sim \qt$. The wrong sign is due to a missing soft contribution in Eq.\,\eqref{eq:xsec-decomposition} 
for $\der\sigma_{\rm NLO}$ that was absorbed (due to a particular choice of the rapidity factorization scale) in the term corresponding to JIMWLK leading logarithmic rapidity evolution. We shall see in the next subsection that, by putting a constraint on this leading logarithmic evolution to exclude the soft gluon phase-space, one recovers the correct Sudakov double logarithm. With this new evolution, the LO term in Eqs.\,\eqref{eq:c0-final} and \eqref{eq:c2-final} will be shifted by a correction of order $\alpha_s$, namely $\hat G^0_{Y_f}\to \hat G^0_{Y_f}-\#\alpha_s\ln^2(\Pt^2\rbbpt)G^0_{Y_f}$ that changes the sign of the Sudakov double logarithm.
The interplay between the ``slow" gluon and soft gluon regimes is also manifest in the dependence of the term proportional to $N_c$ where one observes that the coefficient of the single Sudakov logarithm is sensitive to the rapidity factorization scale $z_f$.

Before turning to a detailed discussion of the kinematic improvement of JIMWLK leading logarithmic evolution, we should note that the coefficient of the single Sudakov logarithm which is proportional to $C_F$,
\begin{equation}
    -\frac{\alpha_s}{\pi}\left[C_F\ln\left(\frac{1}{z_1z_2R^2}\right)+\mathcal{O}(R^2)\right]\,,
    \label{eq:coeff-CF-single-log}
\end{equation}
can be reexpressed in terms of the rapidity difference between the two jets $\Delta \eta_{12}$, 
given by 
\begin{equation}
    \Delta \eta_{12}=\frac{1}{2}\ln\left(\frac{z_1^2\kttwo^2}{z_2^2\ktone^2}\right)\approx\frac{1}{2}\ln\left(\frac{z_1^2}{z_2^2}\right)\,,\label{eq:dY12-id}
\end{equation}
where the second equality holds in the back-to-back limit. Employing the identity
\begin{equation}
    -\ln(z_1z_2)= \ln(2(1+\cosh(\Delta \eta_{12})))\,,
\end{equation}
Eq.~\eqref{eq:coeff-CF-single-log} can be written as 
\begin{equation}
    -\frac{\alpha_sC_F}{\pi}\ln\left(\frac{2(1+\cosh(\Delta \eta_{12}))}{R^2}\right)\,,
\end{equation}
in agreement with \cite{Hatta:2021jcd}.

\subsection{Sudakov suppression from slow gluons and kinematically improved rapidity evolution}
\label{sub:kcJIMWLK}

It is well-known that ``naive" BFKL factorization and evolution in $Y_g=\ln(k_g^-/q^-)$ in fully inclusive DIS does not reproduce the standard collinear (or DGLAP~\cite{Gribov:1972ri,Altarelli:1977zs,Dokshitzer:1977sg}) regime; recall that in the latter the transverse momenta along the evolution are strongly ordered from the large virtuality of the photon down to the typical transverse scale $Q_0\gtrsim \Lambda_{\rm QCD}$ of the target $Q^2\gg \boldsymbol{k}_{g\perp, 1}^2\gg ... \gg Q_0^2 $ with the $k_{g,i}^+$ of the same order \cite{Kwiecinski:1997ee,Salam:1998tj,Ciafaloni:1998iv,Ciafaloni:1999yw,Ciafaloni:2003rd,SabioVera:2005tiv,Beuf:2014uia,Iancu:2015vea}. As noted previously, this failure, and its likely resolution, is well documented in the small $x$ literature. When using the NLL BK or BFKL equation with $Y$ evolution, large double collinear logarithms in the NLL kernel spoil the convergence of the resummation and lead to instabilities impacting the predictive power of the small $x$ resummation program beyond leading logarithmic accuracy~\cite{Salam:1998tj,Ciafaloni:1999yw,Ciafaloni:2003rd}. These problems are cured by employing an improved LL evolution equation in $Y$, resumming to all orders the large double collinear logarithms.

We will demonstrate in this section that a similar improvement in the LL evolution kernel of the WW gluon TMD --- namely, an additional constraint that enforces lifetime ordering of successive emissions --- solves the positive Sudakov sign problem we noted in the previous subsection. This observation was also arrived at independently recently in \cite{Taels:2022tza}. Thus strikingly, a generic problem with BFKL evolution becomes manifest already at leading log for back-to-back jet final states, with its resolution via the resummed kernel proving essential for recovering the physical result of Sudakov suppression. 

The nontrivial interplay between rapidity evolution and Sudakov logarithms can be appreciated by realizing that in the decomposition of the cross-section in Eq.\,\eqref{eq:xsec-decomposition}, we  separated the contribution from the eikonal factor due to soft gluon emission into several pieces, namely, in the ``sud2", ``sud1", and leading log $\Hcal_{\rm LL}$ terms contributing to JIMWLK small $x$ evolution. This suggests that there is a soft gluon contribution hidden in the leading rapidity log term that needs to be extracted in order to obtain the Sudakov contribution with the right sign; we will now detail how a robust separation between soft and slow gluons is achieved in practice.

\subsubsection{Lifetime ordering from the NLO impact factor}

When isolating the logarithmic rapidity divergence in the NLO cross-section, given by
\begin{equation}
    \ln\left(\frac{k_f^-}{\Lambda^-}\right)\Hcal_{\rm LL}\otimes \der\sigma_{\rm LO}\,,
\end{equation}
we set $z_g=0$ everywhere except in the $1/z_g$ divergence piece. While this procedure gives the correct leading logarithmic divergence, we will now demonstrate that it overestimates the phase-space associated with slow gluons. The discussion below is similar to that for the fully inclusive DIS cross-section \cite{Beuf:2014uia}, but the ensuing constraint turns out to be different. To see this, let us consider the ``no-sud, other" terms in the impact factors $\der\sigma_{\rm R, no-sud, other}$ and $\der\sigma_{\rm V, no-sud, other}$ given by Eqs\,\eqref{eq:dijet-NLO-long-real-other-final} and \eqref{eq:V-other}. These terms have an explicit $\zt$ integration --- where $\zt$ is the transverse coordinate of the gluon crossing the shockwave -- and the NLO wavefunctions involve modified Bessel functions of the form
\begin{equation}
    K_0(QX_{\rm R})\,,\quad K_0(QX_{\rm V})\,,
\end{equation}
with 
\begin{align}
    X_{\rm R}^2&=z_1z_{2}\rxyt^2+z_1z_g\rzxt^2+z_{2}z_g\rzyt^2 \,, \\
    X_{\rm V}^2&=z_{2}(z_1-z_g)\rxyt^2+z_g(z_1-z_g)\rzxt^2+z_{2}z_g\rzyt^2\,.
\end{align}
To obtain the slow gluon limit, we set $z_g=0$ in $X_{\rm R/V}$ to recover the LO wavefunction $K_0(\bar Q r_{xy})$. However since $\zt$ is integrated over, then, regardless of how small $z_g$ is, there is always a domain in which the large transverse sizes $|\rzyt|$ or $|\rzxt|$ compensate for the smallness of $z_g$ \cite{Beuf:2014uia}.

To derive the criterion for which the slow gluon limit is valid, let us simplify the expression for $X_{\rm R}$ and $X_{\rm V}$ in the back-to-back limit:
\begin{equation}
    X_{\rm R}^2\approx X_{\rm V}^2\approx z_1z_2\ut^2+z_g\rzbt^2\,.
\end{equation}
When $z_g\rzbt^2\gg z_1z_2\ut^2$, one cannot approximate $K_0(QX_{\rm R/V})$ by $K_0(\bar Q u_\perp)$ anymore. Thus 
\begin{equation}
    z_g\rzbt^2\le z_1z_2\ut^2\,,
\end{equation}
in the slow gluon divergent term. This excludes then phase-space in the $\zt$ integrals which, if also subtracted, would lead to the oversubtraction of the rapidity divergent phase-space we alluded to.

It is enlightening to understand physically the impact of this constraint $z_g\rzbt^2\le z_1z_2\ut^2$ on  rapidity evolution. Parametrically, $\ut^2\sim 1/Q^2$ and $\rzbt^2\sim 1/\kgt^2$, with $\kgt$ the transverse momentum of slow gluons. Hence the constraint is equivalent to
\begin{equation}
    \frac{z_g}{\kgt^2}\le\frac{z_1z_2}{Q^2}\Longleftrightarrow \frac{1}{k_g^+}\le z_1z_2\frac{2q^-}{Q^2}\,.\label{eq:lifetime-ordering}
\end{equation}
Since $1/k_g^+$ is the lifetime of the gluon fluctuation, one sees that the constraint amounts to imposing ($z_1$ and $z_2$ being $\mathcal{O}(1)$ numbers in this context) an ordering of lifetimes in the evolution of the projectile: $1/q^+\gg 1/k_g^+$ .

Note that our argument applied specifically only to the diagrams in which the gluon scatters off the shockwave. However in order to obtain a consistent evolution equation with a probabilistic interpretation, the lifetime ordering constraint needs to be imposed for all diagrams, or in other words, at the level of the LL kernel itself. As we shall see now, this constraint is  crucial in order to recover the correct Sudakov logarithm, since the diagrams in which the additional gluon does not scatter off the shockwave are those which contribute to the Sudakov double logarithm.

\subsubsection{Improved rapidity evolution of the Weizs\"acker-Williams TMD gluon distribution}

The basic idea of collinearly improved BK or BFKL evolution in the rapidity of the projectile $Y$ is to modify the kernel in order to impose $1/k^+$ ordering for successive gluon emissions in the ladder\footnote{For recent discussions of the matching between DGLAP and small $x$ evolution at the operator level, and the connection with lifetime ordering in this context, see \cite{Boussarie:2020fpb,Boussarie:2021wkn}.}. Unfortunately, the collinearly improved JIMWLK evolution equation is available\footnote{Since the evolution becomes nonlocal in rapidity, there is no Hamiltonian formulation of this equation. Further,its Langevin formulation is not suitable when the various transverse sizes in the correlators are very different from each other, as is the case in the back-to-back limit since $\ut^2\ll \rbbpt^2$.} only in Langevin form  \cite{Hatta:2016ujq}.  Nevertheless, in the back-to-back limit, the problem simplifies considerably since one can focus on the evolution of the WW gluon TMD alone, as we shall discuss. In this sense, our approach differs from the computation in \cite{Taels:2022tza} where the authors implement the kinematic constraint at the level of the JIMWLK kernels written in momentum space. While their method works for the color correlators which do not depend on $\zt$ ($\Xi_{\rm LO}$ and $\Xi_{\rm NLO,3}$ in our notations), it is not obvious how to generalize this implementation to  the other correlators.

We will address here the evolution equation for the WW gluon distribution in coordinate space, where it is natural to implement the kinematic constraint \cite{Beuf:2014uia}. Taking the back-to-back limit of Eq.\,\eqref{eq:dijet-NLO-full-slow}, the unconstrained (or ``naive") rapidity evolution equation for the WW gluon TMD in integral form reads
\begin{align}
    \hat G^{ij}_{Y_f}(\bt,\bt') =\hat G^{ij}_{Y_0}+\frac{\alpha_sN_c}{2\pi^2}\int_{Y_0}^{Y_f}\der Y \int\der^2\zt\left\{-\frac{\rbbpt^2}{\rzbt^2\rzbpt^2}\hat G^{ij}_{Y}(\bt,\bt')+\textrm{ other correlators}\right\}\,.\label{eq:WW-TMD-evolution}
\end{align}
The full expression specifying the other correlators (the evolution of the WW gluon TMD is not closed) can be found in \cite{Dominguez:2011gc}. Only the terms specified in Eq.\,\eqref{eq:WW-TMD-evolution} will be important for this discussion since we already know\footnote{The $\hat G^{ij}$ dependent term in Eq.\,\eqref{eq:WW-TMD-evolution} can be obtained in a straightforward way from Eq.\,\eqref{eq:dijet-NLO-slow-xsection8} by taking the correlation limit of the slow gluon divergence proportional to $\Xi_{\rm LO}$ and $\Xi_{\rm NLO,3}$ labeled  $\der\sigma_{\rm LO, LL}$ and $\der\sigma_{\rm NLO_3,\rm LL}$ (cf.\ Eqs.\,\eqref{eq:slowLL-LO}-\eqref{eq:slowLL-NLO3}).} that the Sudakov logarithms accompany the WW gluon TMD. 

Let us now implement the kinematic constraint from lifetime ordering. In addition to the $k_g^-$ ordering $k_g^-\le k_f^-$, one also imposes $k_g^+\ge k_f^+$ with the plus factorization scale related to the minus one through the relation 
\begin{equation}
    2k_f^+k_f^- \equiv Q_f^2\,.
\end{equation}
From Eq.\,\eqref{eq:lifetime-ordering}, and the condition $k_f^-\sim z_1z_2q^-$ (which ensures that $\alpha_s\ln(z_{1,2}/z_f)\ll 1$ in the NLO impact factor), we have $Q_f^2\sim Q^2\sim \Pt^2$. In coordinate space, the condition $1/k_g^+\le 1/k_f^+$ can be rewritten as 
\begin{equation}
    k_g^-\,\textrm{min}(\rzbt^2,\rzbpt^2)\le \frac{k_f^-}{Q_f^2}\,,
\end{equation}
since as discussed above, $\kgt^2\sim 1/\rzbt^2\sim 1/\rzbpt^2$; the function $\textrm{min}(\rzbt^2,\rzbpt^2)$ is a convenient choice which symmetrizes the role of $\bt$ and $\bt'$. The kinematically improved rapidity evolution of $\hat G^{ij}_{Y_f}$ then reads
\begin{align}
   \hat G^{ij}_{Y_f}=\hat G^{ij}_{Y_0}-\frac{\alpha_sN_c}{2\pi^2}\int_{Y_0}^{Y_f}\der Y\int\der^2\zt\Theta\left(Y_f-Y-\ln(\textrm{min}(\rzbt^2,\rzbpt^2)Q_f^2)\right)\frac{\rbbpt^2}{\rzbt^2\rzbpt^2}\hat G^{ij}_{Y}+...\,,
   \label{eq:kc-WW-TMD-evolution}
\end{align}
where, for simplicity, we omitted the argument of $\hat G^{ij}_Y(\bt,\bt')$ as well as the other correlators which depend on $\zt$. It is important to realize that, contrary to the collinearly improved BK/BFKL equation which has the additional constraint
\begin{equation}
   \sim \Theta\left(Y_f-Y-\ln(\textrm{min}(\rzbt^2,\rzbpt^2)/\rbbpt^2)\right)\,,
\end{equation}
for the evolution of the dipole operator $\langle \Tr[ U(\bt)U^\dagger(\bt')]\rangle_Y$, the kinematic constraint in the case of the evolution of the WW gluon distribution involves the ``external" kinematic variable $Q_f^2\sim \Pt^2\sim 1/\rxyt^2$, and \textit{not} $1/\rbbpt$. This is a priori not obvious since it is well known that in the dilute limit, the unconstrained rapidity evolution of the WW TMD follows the BFKL equation \cite{Dominguez:2011gc,Dominguez:2011br}. It implies therefore that, after kinematic improvement, the dilute limit of the WW rapidity evolution is \textit{not} given by the kinematically improved BFKL equation. 

Using this kinematically improved evolution amounts to subtracting off the term 
\begin{equation}
    -\frac{\alpha_s N_c}{2\pi^2}G^{ij}_{Y_f}(\bt,\bt')\int_{z_0}^{z_f}\frac{\der z_g}{z_g}\int\der^2\zt\Theta\left(\textrm{min}(\rzbt^2,\rzbpt^2)Q_f^2-\frac{z_f}{z_g}\right)\frac{\rbbpt^2}{\rzbt^2\rzbpt^2}\,,\label{eq:counterterm-kLL}
\end{equation}
from the naive LL evolution and adding it to the NLO impact factor. 
This integral (in the limit $z_0\to 0$,  as the integral is convergent) is computed in detail in Appendix~\ref{app:math-id}, where we demonstrate that Eq.\,\eqref{eq:counterterm-kLL} is equal to
\begin{equation}
    -\frac{\alpha_sN_c}{2\pi}\ln^2\left(Q_f^2\rbbpt^2\right)G^{ij}_{Y_f}(\bt,\bt')+\mathcal{O}\left(\frac{1}{\rbbpt^2 Q_f^2}\right)\,.
    \label{eq:Sudakov-slow}
\end{equation}
Since $Q_f^2\sim \Pt^2$, this is a Sudakov-like logarithm. It is convenient to break the expression above into three pieces
\begin{align}
    \underbrace{-\frac{\alpha_sN_c}{2\pi}\ln^2\left(\frac{\Pt^2\rbbpt^2}{c_0^2}\right)G^{ij}_{Y_f}(\bt,\bt')}_{\textrm{Sudakov double log.}} &\underbrace{-\frac{\alpha_sN_c}{\pi}\ln\left(\frac{Q_f^2 c_0^2}{\Pt^2}\right)\ln\left(\frac{\Pt^2\rbbpt^2}{c_0^2}\right)G^{ij}_{Y_f}(\bt,\bt')}_{\textrm{Sudakov single log.}}\nonumber\\
    &\underbrace{-\frac{\alpha_sN_c}{2\pi}\ln^2\left(\frac{Q_f^2 c_0^2}{\Pt^2}\right)}_{\textrm{finite}}+\mathcal{O}\left(\frac{1}{\rbbpt^2 Q_f^2}\right)\,.
    \label{eq:Sudakov-slow2}
\end{align}
 
We have thus achieved our goal of extracting the Sudakov double logarithm hidden in the naive leading logarithmic evolution. In particular, we agree with the result of \cite{Taels:2022tza}, provided the scale $Q_f$ is chosen equal to $P_\perp/c_0$. The presence of $c_0$ in their calculation comes from the particular implementation of the kinematic constraint in the momentum representation of the kernel $\rbbpt^2/(\rzbt^2\rzbpt^2)$ that cannot be easily generalized to the other terms in Eq.~\eqref{eq:kc-WW-TMD-evolution}. In any case, since the factorization scale $k_f^+$ is arbitrary, we have the freedom to make such a choice or any other as long as $Q_f$ is parametrically of order $P_\perp$. In general, two different prescriptions for $Q_f$ change the coefficient of the {\it single} Sudakov logarithm. Since we compute this coefficient as well, we shall use $Q_f$ instead of making a particular choice for the factorization scale that would then correspond to a specific value for the coefficient.

\subsection{Summary of results for Sudakov resummation}
\label{sub:final-resum}

We can now combine the Sudakov logarithms from the impact factor given by Eqs.\,\eqref{eq:c0-final}-\eqref{eq:c2-final} with the Sudakov logarithm extracted from the constrained LL rapidity evolution given by Eq.\,\eqref{eq:Sudakov-slow2}. The latter being proportional to $\hat G^{ij}$, the decomposition into Fourier harmonics is trivial (and no higher harmonics than $n=0,2$ are generated by the kinematic improvement). Our final result for the Sudakov double and single logarithms is 
\begin{align}
    \der \sigma^{(0),\lambda=\rm L}&=\alpha_{\rm em}\alpha_s e_f^2\deltatwo\frac{8(z_1z_2)^3Q^2\Pt^2}{(\Pt^2+\bar Q^2)^4}\int\frac{\der^2\rbbpt}{(2\pi)^4}e^{-i\qt\cdot\rbbpt}\hat G^0_{Y_f}(\rbbpt)\left\{1-\frac{\alpha_s N_c}{4\pi}\ln^2\left(\frac{\Pt^2\rbbpt^2}{c_0^2}\right)\right.\nonumber\\
    &\left.-\frac{\alpha_s}{\pi}\left[C_F\ln\left(\frac{1}{z_1 z_2R^2}\right)-N_c\ln\left(\frac{z_f\Pt^2}{z_1z_2c_0^2Q_f^2}\right)\right]\ln\left(\frac{\Pt^2\rbbpt^2}{c_0^2}\right)\right\}+\mathcal{O}(\alpha_s)\,,\label{eq:c0-final-2}
\end{align}
and similarly for the $\langle \cos(2\phi)\rangle$ anisotropy with the replacement $\hat G^0\to \frac{1}{2}\cos(2\theta) \hat h^0$. It is crucial to note that in  Eq.\,\eqref{eq:c0-final-2}, in contrast to  Eqs\,\eqref{eq:c0-final},\eqref{eq:c2-final}, the $Y_f$ dependence of the TMDs $\hat G^{ij}$ is given by the kinematically constrained rapidity evolution in Eq.~\eqref{eq:kc-WW-TMD-evolution}. After this modification, the coefficient of the double logarithm is negative as it should be.

In particular, in comparison to the discussion in \cite{Mueller:2013wwa}, we emphasize that the separation between the Sudakov ``soft" logarithms and the rapidity (or ``slow") logarithms is manifest in the projectile rapidity variable provided that one improves the leading logarithmic rapidity evolution. This separation is not clear at single logarithmic accuracy for the Sudakov contribution since we observe a factorization scheme ($z_f, Q_f$) dependence of the coefficient of the single logarithm in Eq.\,\eqref{eq:c0-final-2}. There is a particular choice of the factorization scale $k_f^+$ for which the (factorization scheme dependent) coefficient vanishes. In terms of $x_f=k_f^+/P^+$, it is given by
\begin{equation}
    x_f = \frac{Q^2}{c_0^2\Pt^2}\frac{x_1x_2}{x_{\rm Bj}}\,,\label{eq:xf-choice}
\end{equation}
where $x_1=k_1^+/P^+\sim x_{\rm Bj}$ and $x_2=k_2^+/P^+\sim x_{\rm Bj}$.
The fact that the condition can be written in terms of $k_f^+$ suggests that an evolution in terms of the target rapidity $\eta_g=\ln(k_g^+/P^+)$ would be more natural. However as alluded to above, such an evolution complicates the separation between evolution and the impact factor, which is naturally computed from the projectile side in the dipole picture of DIS. For such an evolution in $Y$, the choices $k_f^-=(k_1^-k_2^-)/q^-$ and $Q_f=P_\perp/c_0$ in the kinematically constrained rapidity evolution enable one to cancel this single logarithmic coefficient.

The resummation of the Sudakov logarithms in TMD factorization relies on the Collins-Soper formalism \cite{Collins:1981uk,Collins:1981uw}. It amounts to an exponentiation of these dominant soft contributions convoluted with the WW gluon TMD. This exponentiation property has also been derived at double logarithmic accuracy from the study of the WW gluon TMD from low to moderate $x$ in \cite{Balitsky:2015qba}. Since we have shown that the Sudakov logarithms are the same for the unpolarized and linearly polarized WW TMDs, this exponentiation can be written as $ \hat G^{ij}_{Y_f}(\rbbpt)\to\hat G^{ij}_{Y_f}(\rbbpt)\mathcal{S}(\Pt^2,\rbbpt^2)$ with
\begin{equation}
   \mathcal{S}(\Pt^2,\rbbpt^2)=\exp\left(-\int_{c_0^2/\rbbpt^2}^{\Pt^2}\frac{\der\mu^2}{\mu^2}\frac{\alpha_s(\mu^2)N_c}{\pi}\left[\frac{1}{2}\ln\left(\frac{\Pt^2}{\mu^2}\right)+\frac{C_F}{N_c}s_0-s_f\right]\right) \,,
   \label{eq:CSS}
\end{equation}
where the factorization scheme independent single log coefficient $s_0$ reads
\begin{equation}
    s_0 = \ln\left(\frac{2(1+\cosh(\Delta \eta_{12}))}{R^2}\right)+\mathcal{O}(R^2)\,,\label{eq:s0}
\end{equation}
and the factorization scheme dependent one reads
\begin{equation}
    s_f=\ln\left(\frac{\Pt^2x_{\rm Bj}}{z_1z_2 Q^2 c_0^2x_f}\right)\,.
\end{equation}
Finally, we included in Eq.\,\eqref{eq:CSS} the running of the strong coupling constant at the scale $\mu^2$. As explained in footnote \ref{footnote:rc-discussion}, our calculation accounts for the one-loop running of the strong coupling, but the choice of the scale for the running can only be addressed at two-loop order. For back-to-back kinematics, one expects that a two-loop computation would favor the natural choice $\mu^2=\Pt^2$ since $P_\perp$ is the hardest scale in this problem. Expanding $\alpha_s(\mu^2)$ in power of $\alpha_s(\Pt^2)$, one gets
\begin{equation}
    \alpha_s(\mu^2)\simeq \alpha_s(P_\perp^2)\left(1-\alpha_s(P_\perp^2)\beta_0\ln\left(\frac{\mu^2}{\Pt^2}\right)+\mathcal{O}(\alpha_s^2)\right)\,,\label{eq:1loop-rc-sudakov}
\end{equation}
with $\beta_0=(11C_A-4n_fT_R)/(12\pi)$.
This shows that the scale of the running coupling would be fixed by computing the two-loop impact factor, as the logarithmic scale dependence $\ln(\mu^2/\Pt^2)$ enters at order $\alpha_s^2(P_\perp^2)$.
 
The running of the coupling resums in the argument of the exponential both leading Sudakov logarithms of the form $\alpha_s^n\ln^{n+1}(\Pt^2\rbbpt^2)$ and next-to-leading logarithms of the form $\alpha_s^n\ln^{n}(\Pt^2\rbbpt^2)$, to all orders $n$. This can be checked by doing the integral over $\mu^2$ in Eq.\,\eqref{eq:CSS} for the first few terms in the $\alpha_s(\mu^2)$ expansion given by Eq.\,\eqref{eq:1loop-rc-sudakov}. At order $\alpha_s^2$, this gives a contribution of the form 
\begin{equation}
    \frac{\alpha_s^2(\Pt^2)N_c}{\pi}\times\beta_0\left[ \frac{1}{6}\ln^3\left(\frac{\Pt^2\rbbpt^2}{c_0^2}\right)+\left(\frac{C_F}{N_c}s_0-s_f\right)\frac{1}{2}\ln^2\left(\frac{\Pt^2\rbbpt^2}{c_0^2}\right)\right]\,,
\end{equation}
and so forth for the higher orders terms, demonstrating the resummation structure alluded above. Hence, to control all the next-to-leading logarithms, a 2-loop running coupling should be employed.  Note that the integration of $\mu$ in Eq.\,\eqref{eq:CSS} with a one-loop or two-loop running coupling can be done analytically (see e.g.\ \cite{Catani:1992ua,Marzani:2019hun}).

Comparing our Sudakov derivation with those in the non-Abelian exponentiation framework 
of \cite{Catani:1988vd,Catani:1989ne} and those \cite{Ji:2004wu,Ji:2005nu,Sun:2014gfa,Hatta:2021jcd} derived within the equivalent  Collins-Soper-Sterman (CSS) formalism \cite{Collins:1981uk,Collins:1981uw,Collins:1984kg}, we notice two significant  differences.  Firstly, as already outlined, the coefficient of the single logarithm is dependent on the factorization scheme for the rapidity factorization. This result is new. Secondly, there is no contribution proportional to the finite part of the gluon DGLAP splitting function (which is equal to $\pi\beta_0/C_A$) in the coefficient of the Sudakov single logarithm. 
 
The reason for this absence has been discussed in the literature. In \cite{Xiao:2017yya} it is argued that this term is not there because it is presumed that the running of the coupling is treated differently in the CGC EFT than in the collinear factorization formalism and that the running of the coupling only appears at NLO in the BK evolution. In \cite{Hentschinski:2021lsh}, it is argued that such a  contribution is not present in the CGC framework but present in Lipatov's reggeon field theory framework~\cite{Lipatov:1996ts}. However as observed in footnote~\ref{footnote:rc-discussion}, and shown previously in \cite{Ayala:1995hx}, the running coupling appears in an one-loop ``polarization diagram" correction to the classical shockwave field. The $x$-dependent piece of the corresponding diagram is absorbed in JIMWLK evolution;  its $x$-independent piece  running coupling is part of the NLO impact factor. In this systematic power counting, one sees indeed that the  running of the coupling in JIMWLK evolution appears at $\mathcal{O}(\alpha_s^2 \ln(x))$. This 
follows from the combination of the $x$-independent piece with $\mathcal{O}(\alpha_s\ln(x))$ contributions in the two-loop diagrams contributing to the NLO JIMWLK kernel.
 
As we will discuss further in the Section~\ref{sub:other-corr}, the $\beta_0$-dependent term should be  there in the CGC EFT at one loop and can be recovered by respectively taking ``dilute" and collinear limits in this framework--leading to a smooth matching to the collinear results. It is not robust beyond these particular limits which explains why it is not seen in the general expression for Sudakov resummed contributions to the 
NLO impact factor at small $x$.
 
\section{TMD factorization at NLO from the CGC formalism}
\label{sec:TMD-NLO}

In this section, we shall compute some of the pure (not enhanced by large Sudakov logs) $\alpha_s$ corrections in the back-to-back limit. We shall also explore the existence of a TMD-like factorization at NLO in the small $x$ CGC formalism. This factorization is obvious for all the terms in the impact factor that are proportional to $\Xi_{\rm LO}$ and $\Xi_{\rm NLO,3}$ ; these  naturally reduce to the WW gluon TMD in the correlation limit. For the other terms, it is not a priori clear if they reduce to the WW gluon TMD.

In the first subsection, we will calculate all the finite $\alpha_s$ corrections in the back-to-back limit that are proportional to the WW gluon TMD. We will obtain a simple expression for the factorized cross-section with all finite terms in $\alpha_s$. This expression depends on two new ``hard factors" that differ from $\Hcal_{\rm LO}^{\lambda,ij}(\Pt)$.

In the second subsection, we will discuss the other terms in the impact factor which involve more complicated color correlators that arise from the emitted gluon crossing the shockwave. We will argue that these terms break TMD factorization at NLO in the back-to-back limit. In the dilute limit, where one assumes that $Q_s \ll q_\perp \ll P_\perp$, one recovers TMD factorization as expected.

\subsection{Contributions proportional to the Weizs\"{a}cker-Williams gluon TMD }

We shall first address the back-to-back limit of the terms in the impact factor that we did not discuss in the previous section, namely $\der\sigma_{\rm R, no-sud}$ and $\der\sigma_{\rm V, no-sud}$; as the labels imply, they do not contain Sudakov logs. In these terms, we will focus in particular on the contributions which are proportional to $\Xi_{\rm LO}$ and $\Xi_{\rm NLO,3}$ since, once again, those correlators simply give the WW gluon TMD for back-to-back kinematics.

\paragraph{Real contributions without Sudakov enhancement.} In the unintegrated real cross-section $\der\sigma_{\rm R,no-sud}$ defined by Eq.\,\eqref{eq:sigma-nosud}, the term proportional to $\Xi_{\rm LO}$ is $\der\sigma_{\rm R,no-sud,LO}$, given by the expression in Eq.\,\eqref{eq:R-CF}. It can be written in the back-to-back limit as
\begin{align}
    &\der\sigma^{\gamma_{\rm L}^\star+A\to q \bar qg+X}_{\rm R,no-sud,LO}=\frac{\alpha_{\rm em}\alpha_se_f^2}{2(2\pi)^8}\int\der^8\Xttilde e^{-i\Pt\cdot\ruupt-i\qt\cdot\rbbpt}\ut^i\ut'^j\hat G^{ij}_{Y_f}(\rbbpt)\frac{e^{-i\kgt \cdot\rbbpt}}{(\kgt-\frac{z_g}{z_1}\Pt)^2}\nonumber\\
    &\times4\alpha_s C_F \left\{ 8z_1z_{2}^3(1-z_{2})^2Q^2\left(1+\frac{z_g}{z_1}+\frac{z_g^2}{2z_1^2}\right)K_0(\bar Q_{\mathrm{R}2} u_\perp )K_0(\bar Q_{\mathrm{R}2} u_\perp')\deltathree\right.\nonumber\\
    &\hspace{7cm}\left.-\Rcal^{\rm L}_{\rm LO}(\ut,\ut')\Theta(z_1-z_g)\deltatwo\right\}\,.\label{eq:reg-final-b2b}
\end{align}
The primary purpose of our discussion here is to explain why, despite being proportional to $\Xi_{\rm LO}$, this contribution contains neither Sudakov logarithms nor any finite $\alpha_s$ terms. It contributes a power suppressed correction in the limit $P_\perp\gg q_\perp$. 

As mentioned in subsection \ref{sub:NLO-if}, the gluon phase-space integration depends on the experimental definition of the dijet cross-section. Since our aim is to show that this regular term is finite and even power suppressed, we shall follow a drastic path and integrate over the full gluon phase-space without taking into account configurations for which one of the tagged jets is the gluon, in which case one should integrate over the quark or antiquark phase-space. Even though the  resulting cross-section is not IRC safe and not accessible experimentally (since experiments would have to tag the quark and the antiquark in the final state), the result of this exercise should suffice to convince ourselves that the "no-sud" $\rm R_2\times R_2$ term does not contain any large Sudakov logarithms, nor finite $\mathcal{O}(\alpha_s)$ terms in the limit $q_\perp/P_\perp\to0$.

The only $\kgt$ dependence is inside the prefactor of the curly bracket. Integrating this factor over $\kgt$, for a gluon outside the quark-jet cone, we get
\begin{align}
    \int\frac{\der^2\kgt}{(2\pi)^2}\frac{e^{-i\kgt \cdot\rbbpt}}{(\kgt-\xi\Pt)^2}\Theta\left(\Ccal_{qg\perp}^2-R^2\Pt^2\mathrm{min}(1,\xi^2)\right)=&-\frac{1}{4\pi}e^{-i\xi\Pt\cdot\rbbpt}\ln\left(\frac{\Pt^2\rbbpt^2R^2\xi^2}{c_0^2}\right) \nonumber \\
    &+ \mathcal{O}(R^2)\,,
\end{align}
with $\xi=z_g/z_1$. Integrating over $z_2$ to get rid of the delta-functions in 
Eq.~\eqref{eq:reg-final-b2b}, we can express the gluon-integrated ``no-sud" $\rm R_2\times R_2$ cross-section as
\begin{align}
    &\der\sigma^{\gamma_{\rm L}^\star+A\to \textrm{dijet}+X}_{\rm R,no-sud,LO}=-\frac{\alpha_{\rm em}\alpha_se_f^2}{2(2\pi)^6}\int\der^8\Xttilde e^{-i\Pt\cdot\ruupt-i\qt\cdot\rbbpt}\ut^i\ut'^j\hat G^{ij}_{Y_f}(\rbbpt)\nonumber\\
    &\times\frac{\alpha_s C_F}{\pi} \int_0^1\frac{\der\xi}{\xi}e^{-i\xi\Pt\cdot\rbbpt}\left\{\mathcal{I}_{\rm NLO}(\xi,z_1,\ut,\ut')-\mathcal{I}_{\rm LO}(z_1,\ut\ut')\right\}\ln\left(\frac{\Pt^2\rbbpt^2R^2\xi^2}{c_0^2}\right)\,,
\end{align}
where we introduced the notation $\Ical_{\rm LO/NLO}$ for the terms in the integrand inside the curly bracket in Eq.\,\eqref{eq:reg-final-b2b}, 
with the important property (for all $z_1$, $\ut$ and $\ut'$),
\begin{equation}
    \lim\limits_{\xi\to 0}\mathcal{I}_{\rm NLO}(\xi,z_1,\ut,\ut')=\mathcal{I}_{\rm LO}(z_1,\ut,\ut')\,.\label{eq:+prescriptin}
\end{equation}
After the change of variable $\tilde\xi=\xi|\Pt||\rbbpt|$, the above integral transforms into
\begin{equation}
    \int_0^\infty\frac{\der\tilde\xi}{\tilde\xi}e^{-i\tilde \xi\frac{\Pt\cdot\rbbpt}{|\Pt||\rbbpt|}}\left\{\mathcal{I}_{\rm NLO}\left(\frac{\tilde\xi}{|\Pt||\rbbpt|},z_1,\ut,\ut'\right)-\mathcal{I}_{\rm LO}(z_1,\ut\ut')\right\}\ln\left(\frac{\tilde\xi^2R^2}{c_0^2}\right)\,.
\end{equation}
Both $\mathcal{I}_{\rm LO}$ and $\mathcal{I}_{\rm NLO}$ contain step functions that constrain the longitudinal momentum fraction $z_g$ (or $\xi$) of the gluon. Because the phase suppresses this expression for  $\tilde\xi\gtrsim 1$, in the limit $P_\perp/q_\perp\to\infty$ (or equivalently $P_\perp|\rbbpt|\to\infty$) the integral vanishes thanks to the property Eq.~\eqref{eq:+prescriptin}. This proves, as stated, that the real ``no-Sud-LO" cross-section associated with the regular component of diagram $\rm R_2\times R_2$ contributes as a power correction in $q_\perp/P_\perp$ and can therefore be neglected in the back-to-back limit.

Exactly the same reasoning applies for the ``no-sud" real contribution $\der\sigma_{\rm R,no-sud,NLO_3}$ proportional to $\Xi_{\rm NLO,3}$. In conclusion, the real cross-section $\der\sigma_{\rm R,no-sud,LO}+\der\sigma_{\rm R,no-sud,NLO_3}$ vanishes in the back-to-back limit and does not receive any finite $\alpha_s$ contribution. Physically, it means that non-soft real gluon emissions are necessarily power suppressed.

\paragraph*{Virtual contributions without Sudakov enhancement.} We now turn to the contribution to virtual cross-section without Sudakov logs, and in particular the terms proportional to $\Xi_{\rm LO}$ and $\Xi_{\rm NLO,3}$ labeled $\der\sigma_{\rm V,no-sud,LO}$ and $\der\sigma_{\rm V,no-sud,NLO_3}$. Taking the correlation limit of Eq.\,\eqref{eq:V-CF}, it is straightforward to obtain
\begin{align}
    \der\sigma_{\rm V,no-sud,LO}&=\alpha_{\rm em}\alpha_se_f^2\deltatwo\Hcal_{\rm LO}^{\lambda,ij}(\Pt)\times\frac{\alpha_sC_F}{\pi}\int\frac{\der^2 \rbbpt}{(2\pi)^4}e^{-i\qt\cdot\rbbpt}\hat G^{ij}_{Y_f}(\rbbpt)\nonumber\\
    &\times\left\{\frac{3}{2}\ln(c_0^2)-3\ln(R)+\frac{1}{2}\ln^2\left(\frac{z_1}{z_2}\right)+\frac{11}{2}+3\ln(2)-\frac{\pi^2}{2}+\mathcal{O}(R^2)\right\}\nonumber\\
    &-\alpha_{\rm em}\alpha_se_f^2\deltatwo\Hcal_{\rm NLO,1}^{\lambda,ij}(\Pt)\times\frac{\alpha_sC_F}{2\pi}\times\frac{3}{2}\int\frac{\der^2 \rbbpt}{(2\pi)^4}e^{-i\qt\cdot\rbbpt}\hat G^{ij}_{Y_f}(\rbbpt)\,,\label{eq:V-noS-LO-finite}
\end{align}
where the hard factor $\Hcal^{\lambda,ij}_{\rm NLO,1}$ was defined in Eq.\,\eqref{eq:hard-NLO1}.

The virtual correction without Sudakov enhancement and proportional to $\Xi_{\rm NLO,3}$ is also finite and nonvanishing in the back-to-back limit. Taking the correlation limit of $\der\sigma_{\rm V,no-sud,NLO_3}$, we find
\begin{align}
    \der\sigma_{\rm V,no-sud,NLO_3}=\alpha_{\rm em}\alpha_se_f^2\deltatwo\Hcal_{\rm NLO,2}^{\lambda,ij}(\Pt)\times\left(\frac{-\alpha_s}{2\pi N_c}\right)\int\frac{\der^2 \rbbpt}{(2\pi)^4}e^{-i\qt\cdot\rbbpt}\hat G^{ij}_{Y_f}(\rbbpt)+c.c.\label{eq:V-noS-NLO3-finite}
\end{align}
with the hard factor $\Hcal_{\rm NLO,2}^{\lambda,ij}$ defined as
\begin{align}
    &\Hcal_{\rm NLO,2}^{\lambda=\textrm{L},ij}(\Pt)\equiv   \frac{1}{2}\int\frac{\der^2\ut}{(2\pi)}\int\frac{\der^2\ut'}{(2\pi)}e^{-i\Pt\cdot\ruupt}\ut^i \ut'^j \Rcal^{\rm \lambda}_{\rm LO}(\ut,\ut')\nonumber\\
    &\times \int_0^{z_1}\frac{\der z_g}{z_g}\left\{\frac{K_0(\bar Q_{\rm V3}u_\perp)}{K_0(\bar Qu_\perp)}\left[\left(1-\frac{z_g}{z_1}\right)^2\left(1+\frac{z_g}{z_2}\right)(1+z_g)e^{i\Pt\cdot\ut}K_0(-i\Delta_{\rm V3}u_\perp)\right.\right.\nonumber\\
    &\left.-\left(1-\frac{z_g}{z_1}\right)\left(1+\frac{z_g}{z_2}\right)\left(1-\frac{z_g}{2z_1}+\frac{z_g}{2z_2}-\frac{z_g}{2z_1z_2}\right)e^{i\frac{z_g}{z_1}\Pt\cdot\ut}\Jcal_{\odot}\left(\ut,\left(1-\frac{z_g}{z_1}\right)\Pt,\Delta_{\rm V3}\right)\right]\nonumber\\
    &\left.+\ln\left(\frac{z_g P_\perp u_\perp}{c_0z_1z_2}\right)\right\}+(1\leftrightarrow2) \,.
    \label{eq:Hard_NLO2}
\end{align}
Here $\bar Q_{\rm V3}^2=z_1z_2(1-z_g/z_1)(1+z_g/z_2)Q^2$ and $\Delta_{\rm V3}^2=(1-z_g/z_1)(1+z_g/z_2)\Pt^2$.
The expression for transversely polarized photons is given in Appendix~\ref{app:transverse} (see Eq.\,\eqref{eq:HNLO2-transverse}).

The hard factor in Eq.\,\eqref{eq:Hard_NLO2} cannot be computed analytically and should be evaluated numerically to obtain the finite term in Eq.\,\eqref{eq:V-noS-NLO3-finite}.

\subsection{TMD factorization breaking contributions}
\label{sub:other-corr}

We now turn to the finite terms coming from more complex color correlators that do not naturally collapse to the WW gluon distribution in the correlation limit. Let us first consider the term proportional to $\Xi_{\rm NLO,1}$ in the virtual cross-section
$\der\sigma_{\rm V,no-sud,other}$ (without Sudakov logs)  given by Eq.\,\eqref{eq:V-other}. In contrast to the terms proportional to $\Xi_{\rm LO}$ and $\Xi_{\rm NLO,3}$ (that do not contain an integral over the transverse coordinate $\zt$ of the gluon crossing the shockwave) this contribution has an explicit $\zt$ integral and the phases that control the coordinates conjugate to $\Pt$ and $\qt$ are now
\begin{equation}
   \approx e^{-i\Pt\cdot\left(\ut+\frac{z_g}{z_1}\rzbt\right)-i\qt\cdot\bt}\,.
\end{equation}
Therefore it is not clear whether the correlation limit 
corresponding to the expansion of the color correlators in powers of $\ut$ and $\ut'$
best captures the contribution from this term in back-to-back kinematics. In principle, instead of the expansion in powers of $|\ut|\ll |\bt|$, one should expand for $|\ut+z_g/z_1\rzbt|\ll |\bt|$ \cite{Taels:2022tza}. 

On the other hand, since the rapidity evolution of the WW gluon TMD does not have a  closed form (even in the large $N_c$ limit) and mixes under evolution with other operators, there is no reason for our impact factor at small $x$ in the back-to-back limit to depend on the WW gluon TMD alone. From this argument, one can conclude that leading order TMD factorization at small $x$ is violated at NLO by terms in the impact factor proportional to color correlators that do not collapse to the WW TMD. 

A typical example\footnote{This discussion extends straightforwardly to the other $\zt$ dependent color correlators.} of such a color correlator is $\Xi_{\rm NLO,1}$ defined by Eq.\,\eqref{eq:XiNLO1}. Employing the brute force $|\ut|\ll |\bt|$ (and $|\ut'|\ll |\bt'|$) correlation limit at leading order, this operator simplifies to 
\begin{align}
    &\Xi_{\rm NLO,1}(\xt,\yt,\zt;\xt',\yt')\nonumber\\
    &=\ut'^j \times\frac{1}{2N_c}\left\langle\Tr\left[V(\bt)V^\dagger(\zt)\right]\Tr\left[V(\zt)V^\dagger(\bt)\partial^jV(\bt')V^\dagger(\bt')\right]\right\rangle+\mathcal{O}( \ut^2)\,,
\end{align}
 neglecting higher orders in $\mathcal{O}(|\ut||\ut'|)$ but without restrictions on 
 the range of the integrated variable $\zt$. This new operator appears in the rapidity evolution of the WW gluon TMD --- it is one the terms not spelled out in Eq.\,\eqref{eq:WW-TMD-evolution}).  To understand how the kernel of this operator in the rapidity evolution of the WW gluon TMD emerges from our full NLO result,  consider  the leading logarithmic divergence in $z_g$ of the real and virtual terms proportional to $\Xi_{\rm NLO,1}$.
 One finds\footnote{Table 3 in \cite{Caucal:2021ent} spells out the kernels for the different color correlators},
\begin{align}
    \der\sigma_{\rm NLO_1,slow}&=\frac{\alpha_{\rm em}e_f^2N_c\deltatwo}{(2\pi)^6} \int\der^8\Xttilde e^{-i\Pt\cdot\ruupt-i\qt\cdot\rbbpt} \ut'^j \Rcal^{\rm L}_{\rm LO}(\ut,\ut')  \nonumber\\
    & \times \frac{\alpha_s}{2\pi N_c}\int_{z_0}^{z_f}\frac{\der z_g}{z_g}\int\frac{\der^2\zt}{\pi} \left[\frac{\rxyt^2}{\rzxt^2\rzyt^2}-\frac{\rxpyt^2}{\rzxpt^2\rzyt^2}+\frac{\rxxtp^2}{\rzxt^2\rzxpt^2}\right]\nonumber\\
    &\times
    \left\langle\Tr(V(\bt)V^\dagger(\zt))\Tr(V(\zt)V^\dagger(\bt)\partial^jV(\bt')V^\dagger(\bt'))\right\rangle\,.
\end{align}
Writing the transverse coordinates $\xt$, $\yt$ and $\xt'$ in terms of $\ut$, $\ut'$, $\bt$ and $\bt'$, and expanding to leading nontrivial order in $\ut$ and $\ut'$, the kernel in the square brackets becomes
\begin{align}
    \frac{\rxyt^2}{\rzxt^2\rzyt^2}-\frac{\rxpyt^2}{\rzxpt^2\rzyt^2}+\frac{\rxxtp^2}{\rzxt^2\rzxpt^2}=2\frac{\rbbpt^2}{\rzbt^2\rzbpt^2}\left[\frac{\rbbpt^i}{\rbbpt^2}+\frac{\rzbt^i}{\rzbt^2}\right]\ut^i+\mathcal{O}(\ut^2)\,.
\end{align}
Thus the new NLO TMD distribution defined as
\begin{align}
    &\hat G^j_{Y,\rm NLO_1}(\bt,\bt',\zt)\equiv\frac{-2}{\alpha_s}\left\langle\Tr[V(\bt)V^\dagger(\zt)]\Tr[V(\zt)V^\dagger(\bt)\partial^jV(\bt')V^\dagger(\bt')]\right\rangle_{Y}\,,
    \label{eq:NLO-TMD-distrib}
\end{align}
contributes to the rapidity evolution of the WW gluon TMD $G_Y^{ij}(\qt)$ with the kernel
\begin{equation}
    -\frac{1}{N_c}\frac{\rbbpt^2}{\rzbt^2\rzbpt^2}\left[\frac{\rbbpt^i}{\rbbpt^2}+\frac{\rzbt^i}{\rzbt^2}\right]\,,
\end{equation}
inside equation Eq.\,\eqref{eq:WW-TMD-evolution} in agreement with the results of \cite{Dominguez:2011gc}. This result only relies on the limit $|\ut|\ll |\bt|$, since no assumption is made on the magnitude of the integrated transverse coordinate $|\zt|$ relative to either $|\ut|$ or $|\bt|$.

If one further expands $\zt$ around $\bt$ in  Eq.\,\eqref{eq:NLO-TMD-distrib} using
\begin{equation}
    V(\zt)=V(\bt)-\rzbt^i\partial^iV(\bt)+\mathcal{O}(\rzbt^2)\,,\label{eq:Vz-expansion}
\end{equation}
with $\rzbt=\zt-\bt$, we find that the TMD defined by Eq.\,\eqref{eq:NLO-TMD-distrib} collapses to the WW gluon TMD:
\begin{align}
   \hat G^j_{Y,\rm NLO_1}(\bt,\bt',\zt)&\approx -N_c\times\rzbt^i\hat G^{ij}_{Y}(\bt,\bt') \,.
\end{align}
This is not surprising since the rapidity evolution of the WW gluon TMD becomes closed in the dilute limit and reduces to the BFKL equation. The meaning of ``dilute" can be made precise from the regime of validity of Eq.~\eqref{eq:Vz-expansion}.
Since parametrically  $\kgt^2\sim 1/\rzbt^2$, and  since the derivative of $V(\bt)$ is typically of order $Q_s$, the truncation of the Taylor series is justified in the limit
\begin{equation}
    Q_s^2\ll \kgt^2\lesssim \qt^2\,,
\end{equation}
thereby quantifying the dilute limit.

This discussion sheds light on the kinematics for which TMD factorization is valid. At leading order, it is known that ``kinematic"  and ``genuine" power corrections break TMD factorization beyond the kinematics  $P_\perp \gg \max(q_\perp, Q_s)$. (Recall the discussion in section \ref{subsub:corlimit}.). At NLO, it is not sufficient to assume $P_\perp\gg Q_s$ in order to recover factorization: one also needs to impose $q_\perp\gg Q_s$. In back-to-back kinematics $P_\perp \gg q_\perp$, one should then distinguish three physical regimes depending on the magnitude of $Q_s$:
\begin{align}
    P_\perp\gg q_\perp \gg Q_s \label{eq:dilute-regime}\,,\\
    P_\perp \gg Q_s \gg q_\perp\label{eq:correlation-regime}\,,\\
    Q_s \gg P_\perp \gg q_\perp\,.\label{eq:Qs-regime}
\end{align}
In the dilute regime corresponding to Eqs.\,\eqref{eq:dilute-regime}, we expect TMD factorization with the WW gluon TMD to hold\footnote{Indeed, a TMD factorization for inclusive dijet production in DIS was recently proposed in the Soft-Collinear Effective Theory (SCET)~\cite{delCastillo:2020omr,Kang:2020xez} framework albeit not at small $x$.}; we leave for future work the evaluation of the terms proportional to the nontrivial color correlators in the impact factor in this limit. In the correlation regime given by Eqs.\,\eqref{eq:correlation-regime}, one cannot neglect the saturation corrections associated with new gluon TMDs at NLO, and  leading order factorization in terms of the  WW gluon TMD is violated. The evaluation of the cross-section requires one to solve the full nonlinear (and non-closed form) evolution equation for the WW gluon TMD and in the computation of the factorization breaking terms in the impact factor. Finally, in the saturation dominated regime given by Eqs.\,\eqref{eq:Qs-regime}, as noted, factorization is violated already at leading order and the expansion which enabled us to extract the WW gluon TMD from the leading order correlator $\Xi_{\rm LO}$ is not allowed \cite{Boussarie:2021ybe}.

\begin{figure}
    \centering
    \includegraphics[width=0.43\textwidth]{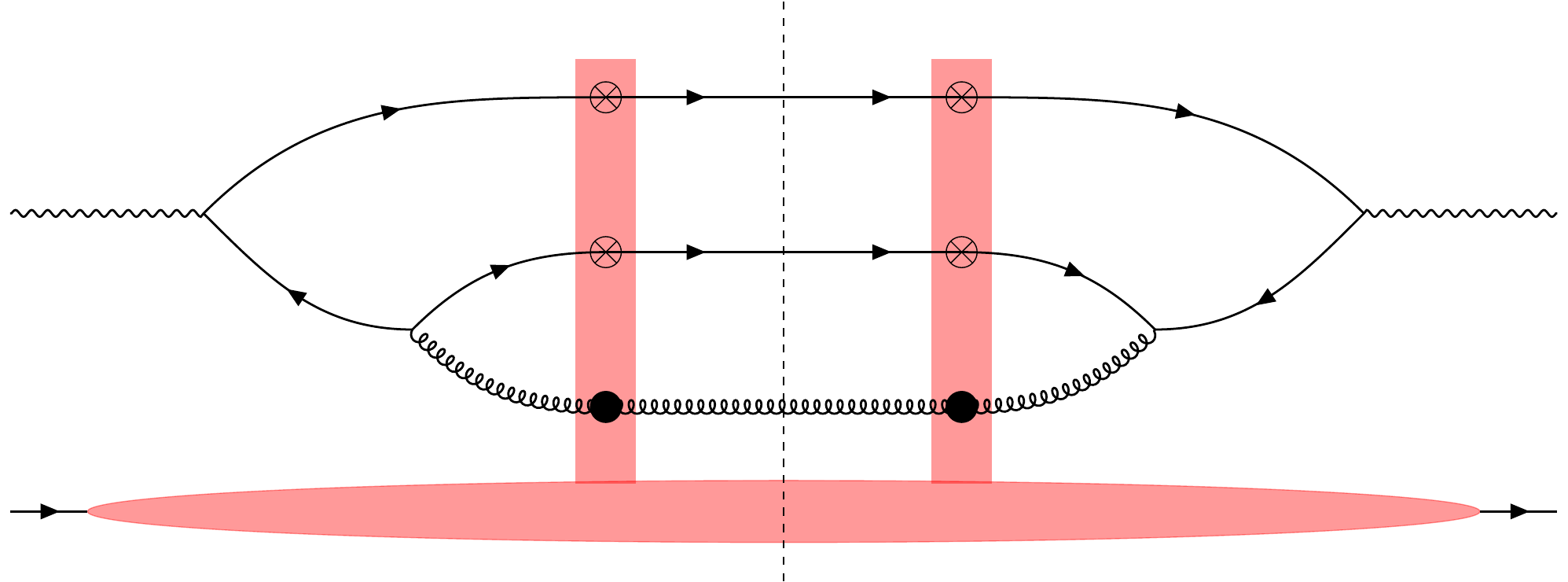}\hfill
    \includegraphics[width=0.43\textwidth]{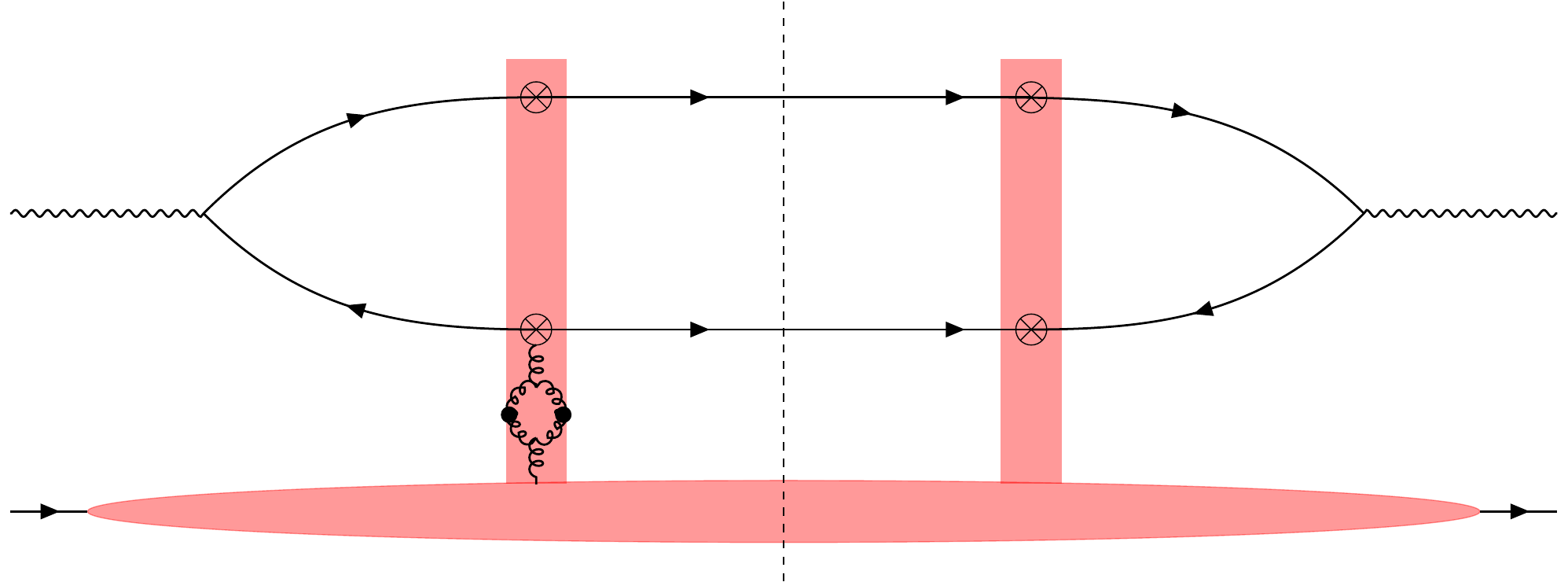}
    \caption{(Left) Diagram ($\rm R_1'\times R_1'$) for a real gluon emission crossing the shockwave both in the amplitude and in the complex conjugate amplitude. (Right) Diagram from the one-loop correction to the shock-wave ``above the cut" (contributing to the NLO impact factor) times the leading order amplitude. See discussion in the text.}
    \label{fig:R1xR1-1loopxLO}
\end{figure}

\begin{figure}
    \centering
    \includegraphics[width=0.43\textwidth]{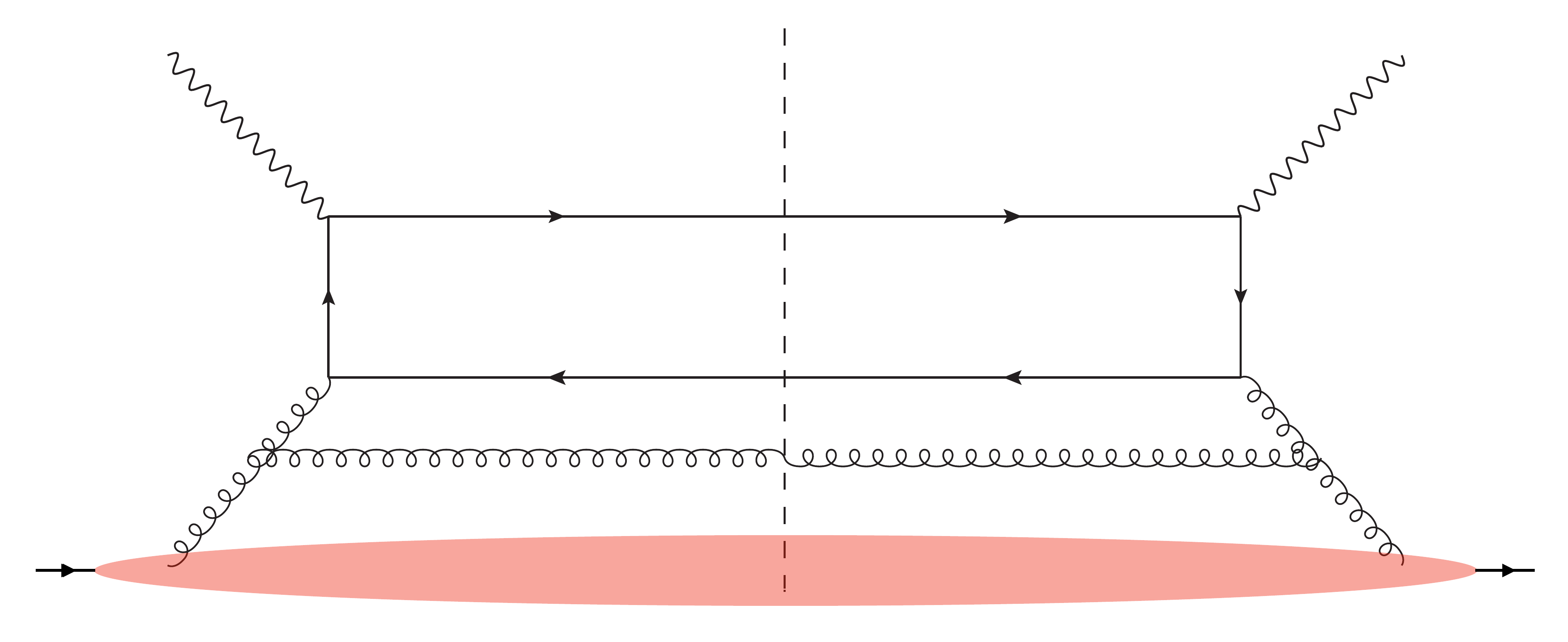}\hfill
    \includegraphics[width=0.43\textwidth]{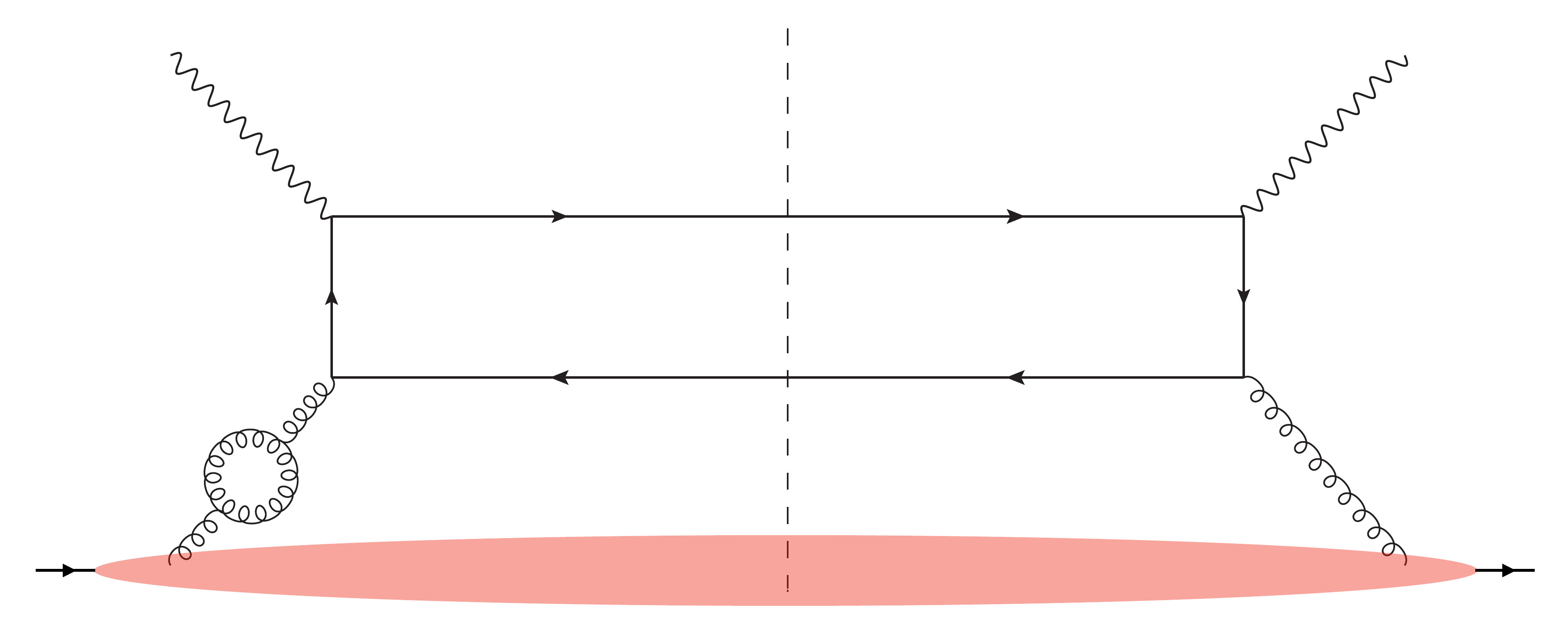}
    \caption{Diagrams appearing in the dilute limit of the respective contributions in Figs.\,\ref{fig:R1xR1-1loopxLO}.}
    \label{fig:R1xR1-1loopxLO-dilute}
\end{figure}

Our discussion also sheds light on the presence of an additional $\beta_0$-dependent single logarithmic contribution in Eq.~\ref{eq:CSS}. In Fig.\,\ref{fig:R1xR1-1loopxLO} (left), we show a cut real diagram corresponding to a gluon crossing the shockwave in the amplitude and the complex conjugate amplitude. Fig.\,\ref{fig:R1xR1-1loopxLO} (right) is the interference diagram contribution of the leading order dijet in the amplitude and the one-loop correction to the shockwave in the complex conjugate amplitude. This particular diagram is of order 
$\mathcal{O}(\alpha_s)$ and contributes to the running coupling, as does a diagram involving a fermion loop that contributes at the same order.
Both Figs.\,\ref{fig:R1xR1-1loopxLO} (left and right) contribute to the NLO impact factor. As we discussed above, the left diagram does not contain the WW gluon TMD. However in the dilute limit, these diagrams contain the contributions shown in Fig.\,\ref{fig:R1xR1-1loopxLO-dilute} (left) and Fig.\,\ref{fig:R1xR1-1loopxLO-dilute} (right). If one further takes the collinear limit, these diagrams should contain the contributions that provide the $\beta_0$-dependent piece of the gluon splitting function that will 
then appear in Eq.~\ref{eq:CSS}. Away from this particular limit, there is no $\beta_0$-dependent contribution.  We note that computations at tree level, have shown how one recovers the collinear limit in the CGC EFT~\cite{Gelis:2003vh,Benic:2016uku}. We will leave a detailed discussion of the matching to collinear limits at one-loop accuracy to future work.

\subsection{Final results for back-to-back inclusive dijets}
\label{sec:final_results}

In this section we collect our final results including the Sudakov resummation, and gather all the finite terms from the expressions in  Eqs.\,\eqref{eq:c0b2b-final},\eqref{eq:c2b2b-final}\,,\eqref{eq:c0_R2R2'_b2b-final}\,,\eqref{eq:c2_R2R2'_b2b-final}\,,\eqref{eq:sud1-final}\,,\eqref{eq:V-noS-LO-finite}\,,\eqref{eq:V-noS-NLO3-finite}\, and \eqref{eq:Sudakov-slow2}. We report here our final results for the azimuthally averaged cross-section and the $\langle \cos(2\phi)\rangle$ anisotropy. The higher order harmonics are provided in appendix~\ref{app:cosnphi}. For the convenience of the reader we recall the Sudakov factor introduced in Eq.\,\eqref{eq:CSS}:
\begin{equation}
   \mathcal{S}(\Pt^2,\rbbpt^2)=\exp\left(-\int_{c_0^2/\rbbpt^2}^{\Pt^2}\frac{\der\mu^2}{\mu^2}\frac{\alpha_s(\mu^2)N_c}{\pi}\left[\frac{1}{2}\ln\left(\frac{\Pt^2}{\mu^2}\right)+\frac{C_F}{N_c}s_0-s_f\right]\right) \,,
   \label{eq:CSS_2}
\end{equation}
where the single log coefficients $s_0$ and $s_f$ read
\begin{equation}
    s_0 = \ln\left(\frac{2(1+\cosh(\Delta \eta_{12}))}{R^2}\right)+\mathcal{O}(R^2)\,,\quad s_f=\ln\left(\frac{\Pt^2x_{\rm Bj}}{z_1z_2 Q^2 c_0^2x_f}\right)\,,\label{eq:s02}
\end{equation}
and $x_f = Q_f^2 x_{\mathrm{Bj}}/ (Q^2 z_f)$.

Since the Sudakov logarithms are now included in the Sudakov exponentiation factor $\mathcal{S}(\Pt^2,\rbbpt^2)$, we remove them from the impact factor. We obtain for the azimuthally averaged cross-section, up to terms of order $\mathcal{O}(R^2)$,
\begin{align}
    \der \sigma^{(0),\lambda=\rm L}&=\alpha_{\rm em}\alpha_s e_f^2\deltatwo\Hcal_{\rm LO}^{0,\lambda=\rm L}(\Pt)\int\frac{\der^2\rbbpt}{(2\pi)^4}e^{-i\qt\cdot\rbbpt}\hat G^0_{Y_f}(\rbbpt)\mathcal{S}(\Pt^2,\rbbpt^2) \nonumber\\
    \times&\left\{1+\frac{\alpha_sC_F}{\pi}\left[\frac{3}{2}\ln(c_0^2)-3\ln(R)+\frac{1}{2}\ln^2\left(\frac{z_1^2}{z_2^2}\right)+\frac{11}{2}+3\ln(2)-\frac{\pi^2}{2}\right]\right.\nonumber\\
    &\left.+\frac{\alpha_sN_c}{2\pi}\left[\ln\left(\frac{z_f^ 2}{z_1z_2}\right)\ln(c_0^2)-\ln^2\left(\frac{Q_f^2c_0^2}{\Pt^2}\right)\right]\right\}\,\nonumber\\
    &+\alpha_{\rm em}\alpha_s e_f^2\deltatwo\Hcal_{\rm LO}^{0,\lambda=\rm L}(\Pt)\int\frac{\der^2\rbbpt}{(2\pi)^4}e^{-i\qt\cdot\rbbpt}\hat h^0_{Y_f}(\rbbpt)\mathcal{S}(\Pt^2,\rbbpt^2)\nonumber\\
    &\times\frac{\alpha_sN_c}{2\pi}\left\{1+\frac{2C_F}{N_c}\ln(R^2)-\frac{1}{N_c^2}\ln(z_1z_2)\right\}\nonumber\\
    &+\alpha_{\rm em}\alpha_se_f^2\deltatwo\left[\frac{1}{2}\Hcal_{\rm NLO,1}^{\lambda=\textrm {L},ii}(\Pt)\right]\int\frac{\der^2 \rbbpt}{(2\pi)^4}e^{-i\qt\cdot\rbbpt}\hat G^{0}_{Y_f}(\rbbpt)\mathcal{S}(\Pt^2,\rbbpt^2)\nonumber\\
    &\times\frac{\alpha_sN_c}{2\pi}\left[\frac{1}{2}\ln\left(\frac{z_1z_2}{z_f^2}\right)-\frac{3C_F}{2N_c}\right]\,\nonumber\\
    &+\left\{\alpha_{\rm em}\alpha_se_f^2\deltatwo\left[\frac{1}{2}\Hcal_{\rm NLO,2}^{\lambda=\textrm {L},ii}(\Pt)\right]\int\frac{\der^2 \rbbpt}{(2\pi)^4}e^{-i\qt\cdot\rbbpt}\hat G^{0}_{Y_f}(\rbbpt)\mathcal{S}(\Pt^2,\rbbpt^2)\right.\nonumber\\
    &\left.\times\left(\frac{-\alpha_s}{2\pi N_c}\right)+c.c.\right\}+\der \sigma^{(0),\lambda=\rm L}_{\rm other}\,,
    \label{eq:zeroth-moment-final}
\end{align}
where repeated indices are summed over. 
The first term, proportional to the LO hard factor and unpolarized WW gluon TMD comes from the LO cross-section (the $1+$ term without $\alpha_s$ suppression) and the part of $\der\sigma_{\rm V,no-sud,LO}$ cross-section which is proportional to the LO hard factor\footnote{This last piece is dependent on the jet definition, see appendix~\ref{app:jet-algo} for more details.}. It also displays a rapidity factorization scheme dependence through $z_f$ and $Q_f$. 

While a full two-loop computation will certainly help to clarify the optimal value of the $z_f$, our study of the Sudakov form factor suggests that a natural physical choice for the factorization scales $z_f$ and $Q_f$ is
\begin{align}
    z_f&=z_1z_2\,, \quad 
    Q_f=P_\perp/c_0 \,,
\end{align}
so that the coefficient $s_f$ of the single Sudakov log cancels (cf.\,Eqs.\,\eqref{eq:CSS_2} and \eqref{eq:s02}) and only the $C_F$ dependent single log $s_0$, familiar from collinear factorization, remains. These choices correspond to
\begin{equation}
    x_f = \frac{Q^2}{c_0^2\Pt^2}\frac{x_1x_2}{x_{\rm Bj}}\,,\label{eq:xf-choice_2}
\end{equation}
where $x_1=k_1^+/P^+\sim x_{\rm Bj}$ and $x_2=k_2^+/P^+\sim x_{\rm Bj}$.

The sensitivity of our result to higher orders can then be studied by varying $z_f$ or $x_f$ around this central value.

The second term, also proportional to the LO hard factor but depending on the linearly polarized WW gluon TMD comes from the effect of final state soft gluon radiation. The two terms proportional to $\Hcal_{\rm NLO,1}$ and $\Hcal_{\rm NLO,2}$ are pure $\alpha_s$ corrections, and involve additional NLO hard factors. Finally, the last term noted $\der \sigma^{(0),\lambda=\rm L}_{\rm other}$ is defined as the correlation limit of the azimuthally averaged terms in the full cross-section that depend on color correlators that do not naturally collapse to the WW gluon TMD, namely:
\begin{equation}
    \der \sigma^{(0),\lambda=\rm L}_{\rm other}\equiv\lim\limits_{P_\perp\gg q_\perp}\frac{1}{2\pi}\int_0^{2\pi}\der\phi  \  \left[\der\sigma^{\lambda=\rm L}_{\rm R, no-sud,other}+\der\sigma^{\lambda=\rm L}_{\rm V, no-sud,other}\right]\,,
\end{equation}
with $\der\sigma_{\rm R, no-sud,other}$ and $\der\sigma_{\rm V, no-sud,other}$ given respectively by Eqs.\,\eqref{eq:dijet-NLO-long-real-other-final} and \eqref{eq:V-other}. We have used the notation $\lim\limits_{P_\perp\gg q_\perp}$ as a shortcut for the correlation limit. This term has been discussed in section~\ref{sub:other-corr}.

The finite terms for the $\langle \cos(2\phi)\rangle$ anisotropy are obtained in a similar fashion:
\begin{align}
    \der \sigma^{(2),\lambda=\rm L}&=\alpha_{\rm em}\alpha_s e_f^2\deltatwo\Hcal_{\rm LO}^{0,\lambda=\rm L}(\Pt)\int\frac{\der^2\rbbpt}{(2\pi)^4}e^{-i\qt\cdot\rbbpt}\frac{\cos(2\theta)}{2}\hat h^0_{Y_f}(\rbbpt)\mathcal{S}(\Pt^2,\rbbpt^2) \nonumber\\
    \times&\left\{1+\frac{\alpha_sC_F}{\pi}\left[\frac{3}{2}\ln(c_0^2)-4\ln(R)+\frac{1}{2}\ln^2\left(\frac{z_1^2}{z_2^2}\right)+\frac{11}{2}+3\ln(2)-\frac{\pi^2}{2}\right]\right.\nonumber\\
    &\left.+\frac{\alpha_sN_c}{2\pi}\left[-\frac{5}{4}+\ln\left(\frac{z_f^ 2}{z_1z_2}\right)\ln(c_0^2)-\ln^2\left(\frac{Q_f^2c_0^2}{\Pt^2}\right)\right]+\frac{\alpha_s}{4\pi N_c}\ln(z_1 z_2)\right\}\,\nonumber\\
    &+\alpha_{\rm em}\alpha_s e_f^2\deltatwo\Hcal_{\rm LO}^{0,\lambda=\rm L}(\Pt)\int\frac{\der^2\rbbpt}{(2\pi)^4}e^{-i\qt\cdot\rbbpt}\frac{\cos(2\theta)}{2}\hat G^0_{Y_f}(\rbbpt)\mathcal{S}(\Pt^2,\rbbpt^2)\nonumber\\
    &\times\frac{\alpha_sN_c}{\pi}\left\{1+\frac{2C_F}{N_c}\ln(R^2)-\frac{1}{N_c^2}\ln(z_1z_2)\right\}\nonumber\\
    &+\alpha_{\rm em}\alpha_se_f^2\deltatwo\left[\frac{1}{2}\Hcal_{\rm NLO,1}^{\lambda=\textrm{L},ii}(\Pt)\right]\int\frac{\der^2 \rbbpt}{(2\pi)^4}e^{-i\qt\cdot\rbbpt}\frac{\cos(2\theta)}{2}\hat h^{0}_{Y_f}(\rbbpt)\mathcal{S}(\Pt^2,\rbbpt^2)\nonumber\\
    &\times\frac{\alpha_sN_c}{2\pi}\left[\frac{1}{2}\ln\left(\frac{z_1z_2}{z_f^2}\right)-\frac{3C_F}{2N_c}\right]\,\nonumber\\
    &+\left\{\alpha_{\rm em}\alpha_se_f^2\deltatwo\left[\frac{1}{2}\Hcal_{\rm NLO,2}^{\lambda=\textrm{L},ii}(\Pt)\right]\int\frac{\der^2 \rbbpt}{(2\pi)^4}e^{-i\qt\cdot\rbbpt}\frac{\cos(2\theta)}{2}\hat h^{0}_{Y_f}(\rbbpt)\mathcal{S}(\Pt^2,\rbbpt^2)\right.\nonumber\\
    &\left.\times\left(\frac{-\alpha_s}{2\pi N_c}\right)+c.c.\right\}+\der \sigma^{(2),\lambda=\rm L}_{\rm other}\,.
    \label{eq:second-moment-final}
\end{align}
where, as above, $\der \sigma^{(2),\lambda=\rm L}_{\rm other}$ is defined by
\begin{equation}
    \der \sigma^{(2),\lambda=\rm L}_{\rm other}\equiv\lim\limits_{P_\perp\gg q_\perp}\frac{1}{2\pi}\int_0^{2\pi}\der\phi\cos(2\phi)  \  \left[\der\sigma^{\lambda=\rm L}_{\rm R, no-sud,other}+\der\sigma^{\lambda=\rm L}_{\rm V, no-sud,other}\right]\,,
\end{equation}
and correspond to the finite $\mathcal{O}(\alpha_s)$ terms proportional to color structures which are not Weizs\"{a}cker-Williams like.

The first term in the curly brackets here is nearly identical to the azimuthally averaged cross-section $\der \sigma^{(0)}$; the only difference comes from the $\langle \cos(2\phi)\rangle$ anisotropy generated by soft gluon radiation that also affects the contribution proportional to the linearly polarized WW TMD $\hat h^0$. In particular, the cone size dependence goes like $\propto -4\ln(R) \hat h^0$ compared to $\propto -3\ln(R)\hat G^0$ in $\der \sigma^{(0),\lambda=\rm L}$. The second term is also interesting, as it corresponds to a $\langle\cos(2\phi)\rangle$ anisotropy proportional to the unpolarized TMD $\hat G^0$ coming from final state gluon radiation. 

Eqs.\,\eqref{eq:zeroth-moment-final} and \eqref{eq:second-moment-final} (as well as 
Eq.\,\eqref{eq:R2R2b2b-cn-final} for the higher harmonics) together with Eq.\,\,\eqref{eq:CSS_2} for the Sudakov form factor $\mathcal{S}(\Pt^2,\rbbpt^2)$ are the principal new results of this paper. As noted earlier, some pieces of these results have been presented previously. The contribution proportional to the unpolarized WW TMD $\hat G^0$ in the fourth and fifth lines of the expression for the  $\langle \cos(2\phi)\rangle $ anisotropy is also present and discussed in \cite{Hatta:2021jcd}.
By comparing our expression for the coefficient of this anisotropy
\begin{equation}
    \frac{\alpha_sN_c}{\pi}\left[1+\frac{2C_F}{N_c}\ln(R^2)-\frac{1}{N_c^2}\ln(z_1z_2)\right]\,,
    \label{eq:c2g-our}
\end{equation}
with the value
\begin{equation}
    \frac{\alpha_sN_c}{\pi}\left[1+\frac{2C_F}{N_c}\ln(R^2)+\frac{1}{N_c^2}\left(\frac{z_1}{z_2}\ln(z_1)+\frac{z_2}{z_1}\ln(z_2)\right)\right]\,,\label{eq:c2g-hattaetal}
\end{equation}
obtained in \cite{Hatta:2021jcd}, 
we find that the two expressions agree up to the $1/N_c^2$ suppressed term\footnote{Not only is the sign of this $1/N_c^2$ term  different but its behavior as well when the rapidity difference $\Delta \eta_{12}$ between the two jets becomes large. Specifically, if we substitute the relation between $\Delta \eta_{12}$ and $z_{1,2}$ given by Eq.\,\eqref{eq:dY12-id} into Eq.\,\eqref{eq:c2g-our}, one observes that it goes to $+\infty$ when $\Delta \eta_{12}\gg 1$ as the single log coefficient $s_0$ given by Eq.\,\eqref{eq:s0}.
In contrast, the $1/N_c^2$ term in Eq.\,\eqref{eq:c2g-hattaetal} converges to $2\alpha_sC_F\ln(eR^2)/\pi$. This difference between the two computations likely 
comes from the fact that \cite{Hatta:2020bgy,Hatta:2021jcd} assumes at the outset momentum space factorization between the initial state distributions and the soft gluon eikonal factor, and consequently, some kind of ``factorization" of the anisotropies. Our computations of Eqs.\,\eqref{eq:R2R2-soft-b2b} and \eqref{eq:R2R2'-soft-b2b} shows that the convolution is more complex. One might worry that the coefficient of the single Sudakov logarithm Eq.\,\eqref{eq:s0} and Eq.\,\eqref{eq:c2g-our} diverges as $\Delta \eta_{12}\to\infty$, or equivalently, as  $z_1\to 0$ or $z_2\to0$. One should however note that the factorization of the rapidity divergence in the NLO cross-section applies when $z_1, z_2$ are not small, typically much larger than $z_0$ and of the order of the factorization scale $z_f$. When $z_{1,2}\sim z_0$, we are not in the small $x$ regime anymore and our formalism breaks down.}. The rest of the contributions in Eq.~\eqref{eq:second-moment-final}, which are of comparable magnitude to Eq.~\eqref{eq:c2g-our}, are presented here for the first time. 

We close this section by reminding the reader that we have not included corrections of the form $Q_s/P_\perp$ which are not captured by the correlation limit. Thus our results are valid in the regions shown in Eqs.\,\eqref{eq:dilute-regime} and \eqref{eq:correlation-regime}. For a more detailed discussion see Sections\,\ref{subsub:corlimit} and \ref{sub:other-corr}.

\section{Conclusions}

In \cite{Caucal:2021ent}, we computed the NLO impact factor for the fully inclusive DIS dijet cross-section at small $x$ in the Color Glass Condensate Effective Field Theory. These results were updated and reorganized in this paper to ensure smooth extraction of the large Sudakov double and single logarithms that appear in the impact factor in back-to-back kinematics. While naively of $\mathcal{O}(\alpha_s)$, such logarithms can give $\mathcal{O}(1)$ contributions when the large dijet transverse momentum $P_\perp$ is much greater than the momentum imbalance $q_\perp$ between the jets. 
Though large, these contributions are known to be universal and can be exponentiated to 
all orders in perturbation theory; they contribute to an overall Sudakov form factor that suppresses the formation of inclusive dijets in the back-to-back limit. 
This suppression of the cross-section complements the suppression arising from the 
many-body scattering and screening effects that are especially large in the gluon saturation regime in large nuclei at small $x$. 

Employing the aforementioned reorganization of the NLO impact factor, we performed a systematic study of the back-to-back limit of the inclusive dijet cross-section. We extracted, for arbitrary $N_c$, all the Sudakov double and single logarithmic terms, as well as all the finite $\mathcal{O}(\alpha_s)$ contributions that contribute at this NLO order to the cross-section. These expressions depend nontrivially on unpolarized and linearly polarized Weizs\"{a}cker-Williams gluon TMDs. Such terms at small $x$ will therefore be essential for precision extraction of novel gluon TMDs at the Electron-Ion Collider.

Our principal results for the dijet differential cross-section in the back-to-back kinematics are shown in Section \ref{sec:final_results}. The Sudakov form factor at single logarithmic accuracy is shown in Eq.\,\eqref{eq:CSS_2}. The zeroth and second azimuthal harmonics of the dijet differential cross-section are 
presented in Eqs.\,\eqref{eq:zeroth-moment-final} and \eqref{eq:second-moment-final}, and higher harmonics are presented in Appendix \ref{app:cosnphi}. Our results are valid in the narrow jet cone approximation, and in the regions shown in Eqs.\,\eqref{eq:dilute-regime} and \eqref{eq:correlation-regime}. A numerical study of the quantitative impact of these results is underway and will be presented separately.

On a conceptual level, our work clarifies the nontrivial interplay between Sudakov resummation and rapidity evolution. It provides a factorization scheme that allows for the simultaneous resummation of the large logarithms of $P_\perp/q_\perp$ and the large rapidity (or ``small $x$") logarithms. This factorization scheme follows from rapidity evolution of the WW gluon TMD along the projectile (virtual photon) direction whose evolution kernel is shown to be the leading order B-JIMWLK kernel. However as we showed explicitly, this leading order kernel must be modified to satisfy a kinematic constraint that enforces lifetime ordering of successive gluon emissions. This modified kernel is precisely that which resums the large double transverse logs that are known to stabilize BFKL evolution beyond leading logarithmic accuracy. Our results are therefore  a nontrivial confirmation of the importance of lifetime ordering in small $x$ evolution even at leading logarithmic accuracy.

The resummation  of Sudakov logarithms in the NLO impact factor is performed by an exponentiation of the double and single logarithms following the Collins-Soper-Sterman  formalism. A novel feature is the dependence of the single log coefficients on the rapidity factorization variable $Y_f$. Whether such contributions also exponentiate at small $x$, is an open question that can be addressed by exploring the matching of 
the CSS and CGC formalism at larger $x$ and $Q^2$. A first step would be to compare our improved rapidity evolution of the WW gluon TMD with the evolution equation derived in \cite{Balitsky:2015qba}, which accounts for the exponentiation of the Sudakov double logarithms in the soft regime. 

Another observation in this direction regarding the Sudakov single logarithms is the absence of a term proportional to the leading order coefficient $\beta_0$ in the QCD $\beta$-function. Such a term is obtained in the collinear factorization framework. We argued that this term can be extracted in the dilute and collinear limit of the CGC EFT; it vanishes beyond this limit and likely 
will therefore not contribute in the typical small $x$ kinematics available at colliders.

As noted, an important result of this study is the calculation of the finite $\alpha_s$ corrections that are not suppressed by powers $q_\perp/P_\perp$ in back-to-back kinematics. Conceptually, these finite $\alpha_s$ corrections are two types. The first of these is from terms that involve convolutions with the unpolarized and linearly polarized WW gluon TMDs, thus do not break the TMD factorization that holds at leading order. Specifically, such  $\mathcal{O}(\alpha_s)$ corrections are from the virtual diagrams where the gluon does not scatter off the shockwave. They also appear in in-cone real gluon emission and in soft large angle radiation. In the latter case, we showed that they can bring sizeable azimuthal anisotropies which are then convoluted with the initial state anisotropy associated with the gluon TMD distributions.

The other type of  finite terms at NLO appear in diagrams where the gluon scatters off the shockwave. These terms break TMD factorization beyond leading order since new operators (distinct from the polarized and linearly polarized WW gluon TMDs) appear in the correlation limit. They are associated with $\mathcal{O}(Q_s/q_\perp)$ corrections that are important in Regge asymptotics. However if the saturation scale is smaller than the momentum imbalance of the dijets ($Q_s\ll q_\perp$), we show that these higher order operators collapse to the leading order gluon TMDs, restoring TMD factorization at NLO.

An important extension of our work is to perform it to two-loop accuracy,  the state-of-the-art for fully inclusive DIS in the CGC EFT. This would significantly expand the demonstration of rapidity factorization to next-to-leading-log accuracy shown there. In particular, it would provide a much deeper understanding of the role of kinematic constraints in small $x$ evolution and the interplay of this with 
Sudakov resummation. From a practical standpoint, such a computation would help clarify  the renormalization scale ambiguities in the running coupling and in the choice of the rapidity factorization scale. 

We can use the results derived in this manuscript for other related processes like dijet production in the photo-production limit $Q^2\to0$ as in \cite{Taels:2022tza} and  dihadron production in DIS \cite{Bergabo:2022tcu}. Such final states may be more easily accessible (albeit less precise) probes at the EIC of the many-body dynamics of interest. We also note that it is feasible to extend our results to the massive quark case \cite{Beuf:2021qqa,Beuf:2022ndu} and thereby study back-to-back heavy quark pair-production in DIS. It is also interesting to study the consequences\footnote{The interplay between the kinematic constraint and threshold resummation has been discussed in \cite{Kang:2019ysm,Shi:2021hwx,Liu:2022xsc}.} of  kinematically constrained small $x$ evolution in semi-inclusive hadron/jet production in DIS \cite{Marquet:2009ca,Iancu:2020jch,Hatta:2022lzj}. 

\section*{Acknowledgements}

We are grateful to Guillaume Beuf, Yoshitaka Hatta, Edmond Iancu, Zhongbo Kang, Xiaohui Liu, Cyrille Marquet, Al Mueller, Tomasz Stebel, Bowen Xiao and Feng Yuan for valuable discussions. This material is based on work supported by the U.S. Department of Energy, Office of Science, Office of Nuclear Physics, under Contracts No. DE-SC0012704 and for R.V. within the framework of the TMD Theory Topical Collaboration. F.S. is supported by the National Science Foundation under grant No. PHY-1945471, and partially supported by the UC Southern California Hub, with funding from the UC National Laboratories division of the University of California Office of the President.

\appendix   

\section{Notations and Conventions}
\label{app:convention}
We work in lightcone coordinates,
\begin{align}
    x^+ = \frac{1}{\sqrt{2}}\left(x^0 + x^3 \right), \quad x^- = \frac{1}{\sqrt{2}}\left(x^0 - x^3 \right)\,,
\end{align}
with the transverse momenta components defined as in Minkowski space. Four-vectors are defined as $a^\mu = (a^+,a^-,\at)$, where $\at$ denote the two-dimensional transverse components. The magnitude of the two-dimensional (Euclidean) vector $\at$ is denoted as $a_\perp = |\at|$. Following these conventions, the scalar product of two four-vectors is $a_\mu b^\mu = a^+b^- + a^- b^+ - \at \cdot \bt$. We use $\at^i$ to denote the $i$th component of the two dimensional Euclidean vector $\at$.

We use the same convention for the gamma matrices $\gamma^+$ and $\gamma^-$, with the anti-commutation relations satisfying
\begin{align}
    \left\{\gamma^\mu,\gamma^\nu \right\} = 2 g^{\mu\nu} \mathbbm{1}_{4}\,,
    \label{eq:ACR}
\end{align}
where the only non-zero entries in the metric are $g^{+-}=g^{-+}=1$ and $g^{ij}=-\delta^{ij}$.

The CGC effective vertices for the eikonal interaction of the quark or gluon (moving with large minus lightcone momentum component) with the background gauge field of the classical small $x$ gluons of the nucleus reads
\begin{align}
    \mathcal{T}^q_{\sigma\sigma',ij}(l,l') &=  (2\pi) \delta(l^--l'^-) \gamma^-_{\sigma\sigma'} \,\mathrm{sgn}(l^-) \int \der^2\vect{z} e^{-i(\lt-\lt')\cdot \vect{z}} V_{ij}^{\mathrm{sgn}(l^-)}(\vect{z}) \,, \\
    \mathcal{T}^g_{\mu\nu,ab}(l,l') &=  -(2\pi) \delta(l^--l'^-) (2l^-)  g_{\mu\nu} \,\mathrm{sgn}(l^-) \int \der^2\vect{z} e^{-i(\lt-\lt')\cdot \vect{z}} U_{ab}^{\mathrm{sgn}(l^-)}(\vect{z})\,,
\end{align}
respectively, where $l$ and $l'$ are the outgoing and incoming momenta of the quark/gluon. The
superscript $\mathrm{sgn}(l^-)$ denotes the color matrix or its inverse  $V
^{+1}(\zt) = V(\zt)$ and $V^{-1}
(\zt) = V^\dagger
(\zt)$. The lightlike Wilson lines in the fundamental and adjoint representations appearing in the effective CGC vertices are given by the SU(3) matrices
\begin{align}
    V_{ij}(\vect{z}) &= \Pcal \exp{ \left( ig \int_{-\infty}^\infty dz^- A^{+,c}_{\mathrm{cl}}  (z^-,\vect{z}) t^c_{ij}  \right)}\,, \\
    U_{ab}(\vect{z}) &= \Pcal \exp{ \left( ig \int_{-\infty}^\infty dz^- A^{+,c}_{\mathrm{cl}}  (z^-,\vect{z}) T^c_{ab}  \right)} \,,
\end{align}
where $t^c_{ij}$ and $T^c_{ab}$ are the generators of SU(3) in the fundamental and adjoint representations respectively. The classical field $A^+_\mathrm{cl}$ is in Lorenz gauge. Here $\Pcal$ stands for path ordering such that the operator at $z=-\infty$ is in the rightmost position, while that at $z=+\infty$ is in the leftmost position.

\section{Reorganization and update to the NLO impact factor in Ref.~\cite{Caucal:2021ent}
}
\label{app:xs-dec}

This appendix deals with the derivation of the three terms $\der\sigma_{\rm sud2}$, $\der\sigma_{\rm sud1}$ and $\der\sigma_{\rm no-sud}$ in the reorganization of the NLO impact factor shown in Eq.\,\eqref{eq:xsec-decomposition}. This reorganization makes it easier to extract the large Sudakov logarithms that appear in back-to-back kinematics. 
In sections \ref{subsub:sud2}, \ref{subsub:sud1}, \ref{subsub:no-sud}, we will provide explicit expressions, respectively, for each of the stated terms. The derivation of the final term in Eq.~\eqref{eq:xsec-decomposition}, corresponding to the JIMWLK evolved  leading order cross-section, was given in Section \ref{subsub:JIMWLK}.

We will work out in detail here the longitudinally polarized photon cross-section to illustrate how one extracts the $\rm sud2$ and $\rm sud1$ terms. Since these terms depend only trivially on the photon polarization, the computations go through analogously for transversely polarized photons. Expressions for the finite terms for the transversely polarized photon cross-section will be given in Appendix~\ref{app:transverse}. 

\subsection{Derivation of $\der\sigma_{\rm sud2}$}
\label{subsub:sud2}

The back-to-back limit of the inclusive dijet cross-section is strongly suppressed by  large logarithms corresponding to soft gluon radiation. The amplitude-level factorization of soft gluon emission comes from diagram $\rm R_2$ and $\rm R_2'$ only \cite{Roy:2019hwr}. Indeed in these diagrams, since the additional gluon emitted does not scatter off the shockwave, the quark or antiquark emitting the gluon go on-shell as the four-momentum $k_g^\mu$ of the gluon goes to zero. For this reason, the relevant diagrams to extract the ``$\rm sud2$" terms are the diagrams $\rm R_2\times \rm R_2$ and $\rm R_2\times R_2'$ at the cross-section level.

\paragraph*{Diagram $\rm R_2\times R_2$.} Let us first recall the full expression for diagram $\rm R_2\times \rm R_2$ before integrating over the phase-space of the gluon.  In \cite{Caucal:2021ent}, we obtained for longitudinally polarized photons,
\begin{align}
    &\der\sigma^{\gamma_{\rm L}^\star+A\to q \bar qg+X}_{\mathrm{R}2\times \mathrm{R}2}=\frac{\alpha_{\rm em}e_f^2N_c\deltathree}{(2\pi)^8}\int\der^8\Xt e^{-i\ktone\cdot\rxxtp-i\kttwo\cdot\ryytp}\Xi_{\rm LO}(\xt,\yt;\xt',\yt')\nonumber\\
    &\times\alpha_sC_F\left\{ 32z_1z_{2}^3(1-z_{2})^2Q^2\left(1+\frac{z_g}{z_1}+\frac{z_g^2}{2z_1^2}\right)K_0(\bar Q_{\mathrm{R}2}r_{xy})K_0(\bar Q_{\mathrm{R}2}r_{x'y'})\frac{e^{-i\kgt \cdot\rxxtp}}{(\kgt-\frac{z_g}{z_1}\ktone)^2}\right\}\,,
    \label{eq:dijet-NLO-long-R2R2-final}
\end{align}
with $\deltathree=\delta(1-z_1-z_2-z_g)$ and $\bar Q_{\rm R2}^2=z_2(1-z_2)Q^2$. When taking the $z_g\to0$ limit inside the curly bracket, one notices that the expression reduces to the leading order perturbative factor $\Rcal^{\rm L}_{\rm LO}$ except for the additional $\kgt$ dependent factor:
\begin{equation}
    \frac{4\alpha_s C_F}{(2\pi)^2} \times\frac{e^{-i\kgt\cdot\rxxtp}}{\left(\kgt-\frac{z_g}{z_1}\ktone\right)^2}\,.\label{eq:kgt-prefactor}
\end{equation}
This expression must be treated with care, given that we aim at extracting the dominant contributions of the NLO impact factor in the back-to-back limit. In this limit, $|\ktone|\sim |\Pt|$. On the other hand, the momentum imbalance is driven by the soft gluon so the relevant $\kgt$ phase-space corresponds to $|\kgt|\sim |\qt|$. In this phase-space, $\kgt$ and $z_g/z_1 \ktone\sim z_g \Pt$ can be of the same order even for small $z_g$ and therefore, it is convenient to keep the $z_g$ dependent term in the denominator.\footnote{However, if one is only interested in the leading $z_g\to 0$ divergence, it is sufficient to set $z_g=0$ in Eq.\,\eqref{eq:kgt-prefactor}.}

To address this point, one can decompose  Eq.\,\eqref{eq:dijet-NLO-long-R2R2-final} to  extract the leading divergence in the product  $z_g\ktone$ inside the denominator as $z_g\rightarrow 0$:
\begin{equation}
    \der\sigma^{\gamma_{\lambda}^\star+A\to q \bar qg+X}_{\mathrm{R}2\times \mathrm{R}2}=\der\sigma^{\gamma_{\lambda}^\star+A\to q \bar qg+X}_{\mathrm{R}2\times \mathrm{R}2,\rm soft-div}+\der\sigma^{\gamma_{\lambda}^\star+A\to q \bar qg+X}_{\mathrm{R}2\times \mathrm{R}2,\rm sud2}+\der\sigma^{\gamma_{\lambda}^\star+A\to q \bar qg+X}_{\mathrm{R}2\times \mathrm{R}2,\rm no-Sud}\,,
    \label{eq:R2R2-dec}
\end{equation}
with 
\begin{align}
    \der\sigma^{\gamma_{\rm \lambda}^\star+A\to q \bar qg+X}_{\mathrm{R}2\times \mathrm{R}2,\rm soft-div}&=\frac{\alpha_{\rm em}e_f^2N_c\deltatwo}{(2\pi)^8}\int\der^8\Xt e^{-i\ktone\cdot\rxxtp-i\kttwo\cdot\ryytp}\Rcal_{\mathrm{LO}}^{\lambda}(\rxyt,\rxytp)\nonumber\\
    \times &\Xi_{\rm LO}(\xt,\yt;\xt',\yt')\times 4\alpha_sC_F\frac{e^{-i(\kgt-z_g/z_1\ktone)\cdot\rxxtp}}{\left(\kgt-\frac{z_g}{z_1}\ktone\right)^2}\,,\\
    \der\sigma^{\gamma_{\rm \lambda}^\star+A\to q \bar qg+X}_{\mathrm{R}2\times \mathrm{R}2,\rm sud2}&=\frac{\alpha_{\rm em}e_f^2N_c\deltatwo}{(2\pi)^8}\int\der^8\Xt e^{-i\ktone\cdot\rxxtp-i\kttwo\cdot\ryytp}\Rcal_{\mathrm{LO}}^{\lambda}(\rxyt,\rxytp)\nonumber\\
    \times &\Xi_{\rm LO}(\xt,\yt;\xt',\yt')\times 4\alpha_sC_F\frac{e^{-i\kgt\cdot\rxxtp}\left(1-e^{iz_g/z_1\ktone\cdot\rxxtp}\right)}{\left(\kgt-\frac{z_g}{z_1}\ktone\right)^2}\,.
\end{align}
We recall that the perturbative factors $\Rcal_{\rm LO}^{\lambda=\rm L,T}$ are defined by Eqs.\,\eqref{eq:dijet-NLO-LLO} and \eqref{eq:dijet-NLO-TLO}. The ``no-Sud" piece in Eq.~\eqref{eq:R2R2-dec} is the difference between the full result and the above two terms. (This term has a more complex dependence on the polarization of the virtual photon - see Appendix~\ref{app:transverse} for the transverse photon case.)
The exact expression will be provided in section \ref{subsub:no-sud}. The divergent 
term above corresponds to the leading slow gluon divergence that gives rise to the JIMWLK kernel but does not account correctly for the $\kgt\sim z_g\Pt$ phase-space relevant in the back-to-back limit. (Indeed, one can simply shift the transverse momentum $\kgt\to \kgt-\frac{z_g}{z_1}\ktone$ when performing the $\kgt$ integral to remove the $\frac{z_g}{z_1}\ktone$ term.) The ``sud2" term  is given by the subtraction between the expression containing the exact phase factor in  Eq.~\eqref{eq:kgt-prefactor} and that of the ``div" contribution, and corresponds precisely to the $\kgt\sim z_g\Pt$ phase-space.

Equipped with this decomposition, we define the out-of cone leading divergence of the $\rm R_2\times R_2$ diagram as the gluon phase-space integration of $\der\sigma^{\gamma_{\lambda}^\star+A\to q \bar qg+X}_{\mathrm{R}2\times \mathrm{R}2,\rm soft-div}$ above with the quark, antiquark and gluon forming three separated jets:
\begin{align}
    \der\sigma_{\rm R_2\times R_2, out, soft-div}&\equiv\int_{z_0}^{z_1}\frac{\der z_g}{z_g}\int\der^2\kgt\der\sigma^{\gamma_{\lambda}^\star+A\to q \bar qg+X}_{\mathrm{R}2\times \mathrm{R}2,\rm div}\Theta\left(C_{qg\perp}^2-R^2\ktone^2\mathrm{min}\left(1,\frac{z_g^2}{z_1^2}\right)\right)\,,\label{eq:R2R2-out-div}
\end{align}
where $z_0=\Lambda^-/q^-$ is the longitudinal momentum fraction between the rapidity cut-off $\Lambda^-$ and $q^-$. The case where the quark and the gluon form a single jet (``in-cone" contribution) will be addressed in the next section since it contributes to the ``sud1" term in the NLO impact factor.

In principle, momentum conservation of the minus lightcone momentum imposes $z_g\le 1-z_1\simeq z_2$. It turns out that the choice of the upper limit of the $z_g$ integral in Eq.\,\eqref{eq:R2R2-out-div} only affects the finite terms included in the ``no-sud" component of the cross-section associated with $\rm R_2\times \rm R_2$. 
Hence we have the freedom to choose the upper boundary of the $z_g$ integral and it is convenient to choose $z_1$ in order to avoid the complications introduced by the $\rm min$ function in the out-of-cone jet condition discussed in section~\ref{sec:Jet-def}. It is then straightforward to perform the phase-space integral and one gets:
\begin{align}
    &\der\sigma_{\rm R_2\times R_2, out, soft-div}=\frac{\alpha_{\rm em}e_f^2N_c\deltatwo}{(2\pi)^6}\int\der^8\Xt e^{-i\ktone\cdot\rxxtp-i\kttwo\cdot\ryytp}\Rcal_{\mathrm{LO}}^{\lambda}(\rxyt,\rxytp)\nonumber\\
    \times &\Xi_{\rm LO}(\xt,\yt;\xt',\yt')\times \frac{\alpha_sC_F}{\pi}\left\{\ln^2\left(\frac{z_1}{z_0}\right)-\ln\left(\frac{z_1}{z_0}\right)\ln\left(\frac{\ktone^2\rxxtp^2R^2}{c_0^2}\right)+\mathcal{O}(R^2)\right\}\,,\label{eq:R2R2-outdiv}
\end{align}
with $c_0=2e^{-\gamma_E}$. This expression displays a soft double logarithmic divergence in $\ln^2(z_0)$. The cancellation of this soft divergence is addressed in the next sub-section: we will show that it cancels with the in-cone soft gluon divergence, namely, the integral of ``$\rm R_2\times \rm R_2$, soft-div" cross-section over the in-cone gluon phase-space.

The ``$\rm sud2$" dijet component of the diagram $\rm R_2\times \rm R_2$ can be defined in a straightforward way from our discussion of  $\rm R_2\times \rm R_2$ above as the integral over the out-of-cone gluon phase-space of the ``sud2" component of $\der\sigma^{\gamma_{\rm L}^\star+A\to q \bar qg+X}_{\mathrm{R}2\times \mathrm{R}2,\rm sud2}$:
\begin{align}
    \der\sigma_{\rm R_2\times R_2,\rm sud2}&\equiv\int_{0}^{z_1}\frac{\der z_g}{z_g}\int\der^2\kgt\der\sigma^{\gamma_{\lambda}^\star+A\to q \bar qg+X}_{\mathrm{R}2\times \mathrm{R}2,\rm sud2}\Theta\left(C_{qg\perp}^2-R^2\ktone^2\mathrm{min}\left(1,\frac{z_g^2}{z_1^2}\right)\right)\,,\label{eq:R2R2soft-def}
\end{align}
in the small-$R$ limit (up to powers of $R^2$ suppressed terms). We emphasize that since the leading rapidity divergence has already been subtracted (as seen from the decomposition in Eq.~\eqref{eq:R2R2-dec}), the integration over $z_g$ is convergent; therefore, no rapidity cut-off $z_0=\Lambda^-/q^-$ is needed. 

The $\kgt$ integration in Eq.\,\eqref{eq:R2R2soft-def} can be performed explicitly in the small $R$ limit using
\begin{equation}
    \int\frac{\der^2\Ccal_{qg\perp}}{(2\pi)^2}\frac{e^{-i\Ccal_{qg\perp}\cdot\Delta}}{\Ccal_{qg\perp}^2}\Theta\left(C_{qg\perp}^2-R^2\ktone^2\xi^2\right)=-\frac{1}{4\pi}\ln\left(\frac{\ktone^2\Delta^2R^2\xi^2}{c_0^2}\right)+\mathcal{O}(R^2)\,,
    \label{eq:integral1}
\end{equation}
so that the soft $\rm R_2\times \rm R_2$ cross-section reads, with $\xi=z_g/z_1$,
\begin{align}
    &\der\sigma_{\rm R_2\times R_2,\rm sud2}=\frac{\alpha_{\rm em}e_f^2N_c\deltatwo}{(2\pi)^6}\int\der^8\Xt e^{-i\ktone\cdot\rxxtp-i\kttwo\cdot\ryytp}\Rcal_{\mathrm{LO}}^{\lambda}(\rxyt,\rxytp)\nonumber\\
    \times &\Xi_{\rm LO}(\xt,\yt;\xt',\yt')\times \frac{\alpha_sC_F}{\pi}\int_0^1\frac{\der\xi}{\xi}\left[1-e^{-i\xi\ktone\cdot\rxxtp}\right]\ln\left(\frac{\ktone^2\rxxtp^2R^2\xi^2}{c_0^2}\right)+\mathcal{O}(R^2)\,,\label{eq:app-R2R2-soft}
\end{align}
up to powers of $R^2$ suppressed terms.
The expression for $\der\sigma_{\rm R_2'\times R_2',\rm sud2}$ can be obtained by quark-antiquark interchange, employing $\ktone\leftrightarrow\kttwo$, $\xt\leftrightarrow\yt$ and $\xt'\leftrightarrow\yt'$.

\paragraph*{Diagram $\rm R_2\times R_2'$.} The structure of the soft piece of the interference diagram $\rm R_2\times R_2'$ is slightly more complicated. We remind the reader that the full expression for this diagram in Ref.~\cite{Caucal:2021ent} is 
\begin{align}
\der\sigma^{\gamma_{\rm L}^\star+A\to q \bar qg+X}_{\mathrm{R}2\times \mathrm{R}2'}
    &=\frac{\alpha_{\rm em}e_f^2N_c\deltathree}{(2\pi)^8}\int\der^2\Xt e^{-i\ktone\cdot\rxxtp-i\kttwo\cdot\ryytp}\Xi_{\rm NLO,3}(\xt,\yt;\xt',\yt')\nonumber\\
    &\times(-\alpha_s)32z_1^3z_{2}^3(1-z_{2})(1-z_1)Q^2K_0(\bar{Q}_{\rm R2}r_{xy})K_0(\bar{Q}_{\rm R2'}r_{x'y'})\left[1+\frac{z_g}{2z_1}+\frac{z_g}{2z_{2}}\right]\nonumber\\
    &\times e^{-i\kgt\cdot \rxypt}\frac{(z_1\kgt-z_g\ktone)\cdot(z_{2}\kgt-z_g\kttwo)}{(z_1\kgt-z_g\ktone)^2(z_{2}\kgt-z_g\kttwo)^2}\,,\label{eq:R2R2'-full}
\end{align}
with $\bar Q_{\rm R2'}^2=z_1(1-z_1)Q^2$ and the color correlator $\Xi_{\rm NLO,3}$ defined by
\begin{align}
    \Xi_{\rm NLO,3}(\xt,\yt;\xt',\yt')=\frac{N_c}{2}\left\langle 1-D_{xy}-D_{y'x'}+D_{xy}D_{y'x'}\right\rangle-\frac{1}{2N_c}\Xi_{\rm LO}(\xt,\yt;\xt',\yt')\,.
\end{align}
As in the case of  $\rm R_2\times \rm R_2$, we first extract the $z_g\to 0$ behavior, keeping terms of the form $z_g\ktone$ or $z_g\kttwo$. Since this diagram has no collinear divergence, the in-cone phase-space is suppressed by at least one power of $R^2$.  One can therefore freely integrate over the full gluon phase-space up to terms of order $\mathcal{O}(R^2)$. The object we need to study is then 
\begin{align}
    (-4\alpha_s)\int\frac{\der z_g}{z_g}\int\frac{\der \kgt^2}{(2\pi)^2}e^{-i\kgt\cdot\rxypt}&\frac{(\kgt-
    \frac{z_g}{z_1}\ktone)\cdot(\kgt-\frac{z_g}{z_2}\kttwo)}{(\kgt-\frac{z_g}{z_1}\ktone)^2(\kgt-\frac{z_g}{z_2}\kttwo)^2}\nonumber\\
    &=(-\alpha_s)\int\frac{\der z_g}{z_g} \ e^{-i\frac{z_g}{z_1}\ktone\cdot\rxypt}\Jcal_R\left(\rxypt,\frac{z_g}{z_1z_2}\Pt\right)\,,
\end{align}
where, as in \cite{Caucal:2021ent}, we introduced the function $\Jcal_R$,
\begin{equation}
    \Jcal_R(\rt,\Kt)\equiv\int\frac{\der^2 \lt}{(2\pi)^2} e^{-i\lt\cdot\rt}\frac{4\lt\cdot(\lt+\Kt)}{\lt^2(\lt+\Kt)^2}\,.
\end{equation}
This function is singular as $\Kt\to 0$, which is evidently the case when $z_g\to 0$. This singularity was studied and extracted in \cite{Caucal:2021ent}, where we found that 
\begin{align}
    \Jcal_R(\rt,\Kt)&=4\int\frac{\der^2\lt}{(2\pi)^2}\frac{1}{\lt^2}\Theta\left(\frac{c_0^2}{\rt^2}\ge\lt^2\ge \Kt^2\right)-i+\mathcal{O}(|\Kt|)\nonumber \\
    &=-\frac{1}{\pi}\left[\ln\left(\frac{\Kt^2\rt^2}{c_0^2}\right)+i\pi+\mathcal{O}(|\Kt|)\right]\,,\label{eq:JR-singular}
\end{align}
The $i\pi$ term is not important here since it cancels when we include the complex conjugate diagram $\rm R_2'\times R_2$. 

Following along the lines of the decomposition in Eq.~\eqref{eq:R2R2-dec} of the diagram $\rm R_2\times R_2$, we will likewise isolate three terms in the expression of $\rm R_2\times R_2'$ based on the observed singular behavior:
\begin{equation}
    \der\sigma^{\gamma_{\rm \lambda}^\star+A\to q \bar qg+X}_{\mathrm{R}2\times \mathrm{R}2'}=\der\sigma^{\gamma_{\rm \lambda}^\star+A\to q \bar qg+X}_{\mathrm{R}2\times \mathrm{R}2',\rm soft-div}+\der\sigma^{\gamma_{\rm \lambda}^\star+A\to q \bar qg+X}_{\mathrm{R}2\times \mathrm{R}2',\rm sud2}+\der\sigma^{\gamma_{\rm \lambda}^\star+A\to q \bar qg+X}_{\mathrm{R}2\times \mathrm{R}2',\rm no-Sud}\label{eq:R2R2'-dec} \,,
\end{equation}
with the soft divergent and ``sud2" terms defined respectively as 
\begin{align}
    & \der\sigma^{\gamma_{\rm \lambda}^\star+A\to q \bar qg+X}_{\mathrm{R}2\times \mathrm{R}2',\rm soft-div}\equiv\frac{\alpha_{\rm em}e_f^2N_c\deltatwo}{(2\pi)^8}\int\der^8\Xt e^{-i\ktone\cdot\rxxtp-i\kttwo\cdot\ryytp}\Rcal_{\mathrm{LO}}^{\lambda}(\rxyt,\rxytp)\nonumber\\
    \times &\Xi_{\rm NLO,3}(\xt,\yt;\xt',\yt')\times (-4\alpha_s)\frac{1}{\lt^2}\Theta\left(\frac{c_0^2}{\rxypt^2}\ge \lt^2\ge\frac{z_g^2\Pt^2}{z_1^2z_2^2}\right)\,,\\
    &\der\sigma^{\gamma_{\rm \lambda}^\star+A\to q \bar qg+X}_{\mathrm{R}2\times \mathrm{R}2',\rm sud2}\equiv\frac{\alpha_{\rm em}e_f^2N_c\deltatwo}{(2\pi)^8}\int\der^8\Xt e^{-i\ktone\cdot\rxxtp-i\kttwo\cdot\ryytp}\Rcal_{\mathrm{LO}}^{\lambda}(\rxyt,\rxytp)\nonumber\\
    \times &\Xi_{\rm NLO,3}(\xt,\yt;\xt',\yt')\times (-4\alpha_s)\frac{\left(e^{-i\frac{z_g}{z_1}\ktone\cdot\rxypt}-1\right)}{\lt^2}\Theta\left(\frac{c_0^2}{\rxypt^2}\ge\lt^2\ge\frac{z_g^2\Pt^2}{z_1^2z_2^2}\right)\,, 
\end{align}
with $\lt=\kgt-z_g/z_1\ktone$.
The $\Theta$-functions have been chosen to ensure that  Eq.\,\eqref{eq:JR-singular} is recovered once the transverse momentum of the gluon is integrated over. As previously, the ``no-sud" contribution is defined as the full expression for $\rm R_2\times \rm R_2'$ minus the soft divergent and ``sud2" terms. As we observed in section~\ref{sec:TMD-NLO}, this term has neither divergences nor large Sudakov logarithms in the back-to-back limit.

After integrating $\der\sigma^{\gamma_{\lambda}^\star+A\to q \bar qg+X}_{\mathrm{R}2\times \mathrm{R}2',\rm soft-div}$ over the full gluon phase-space, 
\begin{align}
    \der\sigma_{\rm R_2\times R_2',soft-div}&\equiv \int_0^{z_1}\frac{\der z_g}{z_g}\int\der^2\kgt\der\sigma^{\gamma_{\lambda}^\star+A\to q \bar qg+X}_{\mathrm{R}2\times \mathrm{R}2',\rm soft-div}\\
    &=\frac{\alpha_{\rm em}e_f^2N_c\deltatwo}{(2\pi)^6}\int\der^8\Xt e^{-i\ktone\cdot\rxxtp-i\kttwo\cdot\ryytp}\Rcal_{\mathrm{LO}}^{\lambda}(\rxyt,\rxytp)\nonumber\\
    &\times\frac{\alpha_s}{\pi}\int_{z_0}^{z_1}\frac{\der z_g}{z_g}\left[2\ln\left(\frac{z_g}{z_1z_2}\right)+\ln\left(\frac{\Pt^2\rxypt^2}{c_0^2}\right)\right]\Xi_{\rm NLO,3}(\xt,\yt,\xt',\yt')\,,
    \label{eq:R2R2'div}
\end{align}
one gets double and single logarithmic divergences in $\ln(z_0)$. 
As we showed in \cite{Caucal:2021ent}, the double logarithms $\ln^2(z_0)$ will cancel with the virtual contribution $\rm V_3\times LO$, and the single logarithms combine to give the JIMWLK kernel associated with the color correlator $\Xi_{\rm NLO,3}$. This 
is shown in detail in section \ref{subsub:sud1}.

We will here focus on the ``sud2" term and define $\der\sigma_{\rm R_2\times R_2',sud2}$. We use the same definition as for $\rm R_2\times R_2$:
\begin{align}
    \der\sigma_{\rm R_2\times R_2',\rm sud2}&\equiv\int_{0}^{z_1}\frac{\der z_g}{z_g}\int\der^2\kgt \ \der\sigma^{\gamma_{\lambda}^\star+A\to q \bar qg+X}_{\mathrm{R}2\times \mathrm{R}2',\rm sud2}\Theta\left(C_{qg\perp}^2-R^2\ktone^2\mathrm{min}\left(1,\frac{z_g^2}{z_1^2}\right)\right)\,,\label{eq:R2R2'soft-def}
\end{align}
but this time, since there is no collinear divergence, one can remove the out-of-cone condition and integrate over the full phase-space of the gluon in the small-cone approximation. (To account for the exact $R$ dependence of the cross-section, the out-of-cone constraint must still be imposed.) The upper limit of the $z_g$ integral is again arbitrary, and one can take $z_1$ in $\rm R_2\times R_2'$ and $z_2$ in $\rm R_2'\times R_2$ in order to preserve the quark-antiquark symmetry. Any other choice only affects the finite $\mathcal{O}(\alpha_s)$ term. Finally, one gets
\begin{align}
    &\der\sigma_{\rm R_2\times R_2',\rm sud2}=\frac{\alpha_{\rm em}e_f^2N_c\deltatwo}{(2\pi)^6}\int\der^8\Xt e^{-i\ktone\cdot\rxxtp-i\kttwo\cdot\ryytp}\Rcal_{\mathrm{LO}}^{\lambda}(\rxyt,\rxytp)\nonumber\\
    \times &\Xi_{\rm NLO,3}(\xt,\yt;\xt',\yt')\times \frac{(-\alpha_s)}{\pi}\int_0^1\frac{\der\xi}{\xi}\left[1-e^{-i\xi\ktone\cdot\rxypt}\right]\ln\left(\frac{\Pt^2\rxypt^2\xi^2}{z_2^2c_0^2}\right)+\mathcal{O}(R^2)\,,\label{eq:app-R2R2'soft}
\end{align}
The back-to-back part of $\rm R_2'\times R_2$ can be obtained from quark-antiquark interchange in Eq.\,\eqref{eq:app-R2R2'soft}.  Eqs.\,\eqref{eq:app-R2R2-soft} and \eqref{eq:app-R2R2'soft} are the key results of this section.

\subsection{Derivation of $\der\sigma_{\rm sud1}$}
\label{subsub:sud1}

Before we compute the $\rm sud1$ term in Eq.\,\eqref{eq:xsec-decomposition} from which one can extract the single Sudakov logarithm + finite terms in the back-to-back limit, we will  first consider how the double logarithmic divergence in $\ln^2(z_0)$ found in Eq.\,\eqref{eq:R2R2-outdiv} and in Eq.\,\eqref{eq:R2R2'div} cancels out.

\paragraph{Single logarithm from the ``no-pole" term.} The sum of all the virtual amplitudes displayed Fig.\,\ref{fig:NLO-dijet-all-diagrams} is infrared divergent, and this divergence manifests as a collinear single $1/\varepsilon$ pole in two transverse dimensions. This divergence was isolated in \cite{Caucal:2021ent}, and one obtains 
\begin{align}
    &\der\sigma_{\rm IR\times LO}=\frac{\alpha_{\rm em}e_f^2N_c\deltatwo}{(2\pi)^6}\int\der^8\Xt e^{-i\ktone\cdot\rxxtp-i\kttwo\cdot\ryytp}\Rcal_{\mathrm{LO}}^{\lambda}(\rxyt,\rxytp)\nonumber\\
    \times &\Xi_{\rm LO}(\xt,\yt;\xt',\yt')\times\frac{\alpha_sC_F}{2\pi}\left\{\left(\ln\left(\frac{z_1}{z_0}\right)+\ln\left(\frac{z_2}{z_0}\right)-\frac{3}{2}\right)\left(\frac{2}{\varepsilon}+\ln(e^{\gamma_E}\pi\mu^2\rxyt^2)\right)\right.\nonumber\\
    &\left.+\frac{1}{2}\ln^2\left(\frac{z_2}{z_1}\right)-\frac{\pi^2}{6}+\frac{5}{2}-\frac{1}{2}\right\}\,,
    \label{eq:dijet-NLO-finite-SE1-V1-SE2uv}
\end{align}
with $\der\sigma_{\rm IR\times LO}$ defined as the sum of the virtual contributions $\rm SE_2\times LO$, $\rm SE_3\times LO$, $\rm V_2\times LO$ and the UV divergent component of $\rm SE_1\times LO$. 
The surviving pole in this expression cancels in the IRC safe cross-section once combined with the collinear divergence of the real cross-section. 

Only the diagrams $\rm R_2\times \rm R_2$ and $\rm R_2'\times R_2'$ are singular when the gluon is collinear to the quark or the antiquark. Once the C/A algorithm (for example) is applied to the $q\bar qg$ final state of these two diagrams, the in-cone contributions (where the quark or antiquark, and the gluon, lie inside the same jet) develop another collinear $1/\varepsilon$ pole which exactly cancels against the one in Eq.\,\eqref{eq:dijet-NLO-finite-SE1-V1-SE2uv}. The in-cone $\rm R_2\times \rm R_2$ cross-section reads
\begin{align}
    \der\sigma_{\rm R_2\times \rm R_2, in}&=\der\sigma_{\rm LO}\times \frac{\alpha_sC_F}{\pi}\left\{\left(\frac{3}{4}-\ln\left(\frac{z_1}{z_0}\right)\right)\frac{2}{\varepsilon}+\ln^2(z_1)-\ln^2(z_0)-\frac{\pi^2}{6}\right.\nonumber\\
    &\left.+\left(\ln\left(\frac{z_1}{z_0}\right)-\frac{3}{4}\right)\ln\left(\frac{R^2\ktone^2}{\tilde{\mu}^2z_1^2}\right)+\frac{1}{4}+\frac{3}{2}\left(1-\ln\left(\frac{z_1}{2}\right)\right)+\mathcal{O}(R^2)\right\}\,,
\label{eq:dijet-NLO-real-collinardiv}
\end{align}
and similarly for the $\rm R_2'\times R_2'$ in-cone term. The $\overline{\rm MS}$ momentum scale $\tilde\mu$ is defined by $\tilde\mu^2=4\pi e^{-\gamma_E}\mu^2$. Importantly, in contrast to \cite{Caucal:2021ent}, we have not subtracted the rapidity divergence yet.

Indeed in \cite{Caucal:2021ent}, we defined the ``$\rm IRC$ safe" contribution to the inclusive dijet cross-section as the sum respectively of the virtual and real terms, $\der\sigma_{\rm IR\times LO}$, $\der\sigma_{\rm R_2\times \rm R_2, in}$ and $\der\sigma_{\rm R_2'\times \rm R_2', in}$ \textit{after} subtracting  the logarithmic rapidity divergence. We found further that this $\rm IRC$ term has terms proportional to $\alpha_s\ln^2(k_f^-/q^-)$. Though these terms break the expected single logarithmic factorization of rapidity divergences, these double logarithmic terms however cancel once combined with the out-of-cone contributions from diagram $\rm R_2\times R_2$ and $\rm R_2'\times R_2'$, again \textit{after} subtraction of the rapidity divergence. 

As noted in \cite{Taels:2022tza}, one can also implement the jet algorithm first, and then combine the in-cone contribution with the rapidity divergent contribution of the 
out-cone contribution to see the cancellation of the $\ln^2(z_0)$ term appearing in Eq.\,\eqref{eq:dijet-NLO-real-collinardiv}. We will adopt here the same organizing principle as \cite{Taels:2022tza} to display our cross-section and include in the $\rm IRC$ safe term the logarithmic divergent phase-space of the out-of-cone $\rm R_2 \times R_2$ and $\rm R_2'\times R_2'$ diagrams. We will then have to subtract the leading logarithmic rapidity divergence which belongs to the last term in Eq.\,\eqref{eq:xsec-decomposition}.

We can now combine the $\rm IR$ divergent piece of the virtual cross-section $\der\sigma_{\rm IR\times LO}$ (Eq.\,\eqref{eq:dijet-NLO-finite-SE1-V1-SE2uv}), the in-cone contributions of diagrams $\rm R_2\times R_2$ and $\rm R_2'\times R_2'$ (Eq.\,\eqref{eq:dijet-NLO-real-collinardiv}), and the out-of-cone soft divergent piece of $\rm R_2\times \rm R_2$ and $\rm R_2'\times R_2'$ that was extracted and computed in the previous subsection (see Eq.\,\eqref{eq:R2R2-outdiv}). These combine to give
\begin{align}
    & \der\sigma_{\rm IR\times LO}+\der\sigma_{\rm LO\times IR}+\der\sigma_{\rm R_2\times \rm R_2, in}+\der\sigma_{\rm R_2\times R_2, out, soft-div.}+(\mathrm{R}_2\to\mathrm{R}_2')=\nonumber\\
    & \frac{\alpha_{\rm em}e_f^2N_c\deltatwo}{(2\pi)^6}\int\der^8\Xt e^{-i\ktone\cdot\rxxtp-i\kttwo\cdot\ryytp}\Rcal_{\mathrm{LO}}^{\lambda}(\rxyt,\rxytp)\Xi_{\rm LO}(\xt,\yt;\xt',\yt')\nonumber\\
    &\times\frac{\alpha_sC_F}{\pi}\left\{-\ln\left(\frac{z_1}{z_0}\right)\ln\left(\frac{\rxxtp^2}{|\rxyt||\rxytp|}\right)-\ln\left(\frac{z_2}{z_0}\right)\ln\left(\frac{\ryytp^2}{|\rxyt||\rxytp|}\right)-\frac{3}{4}\ln\left(\frac{\ktone^2\kttwo^2\rxyt^2\rxytp^2}{c_0^4}\right)\right.\nonumber\\
    &\left.-3\ln(R)+\frac{1}{2}\ln^2\left(\frac{z_1}{z_2}\right)+\frac{11}{2}+3\ln(2)-\frac{\pi^2}{2}+\mathcal{O}(R^2)\right\}\,.
    \label{eq:sum_IR_R2R2in_R2R2out}
\end{align}
This expression contains a rapidity divergence, which is isolated by introducing the factorization scale $k_f^-$ and $z_f=k_f^-/q^-$:
\begin{align}
    \der\sigma_{\rm LO,slow}=\frac{\alpha_{\rm em}e_f^2N_c\deltatwo}{(2\pi)^6}&\int\der^8\Xt e^{-i\ktone\cdot\rxxtp-i\kttwo\cdot\ryytp}\Rcal_{\mathrm{LO}}^{\lambda}(\rxyt,\rxytp)\nonumber\\
    &\times\frac{\alpha_sC_F}{\pi}\ln\left(\frac{k_f^-}{\Lambda^-}\right)\ln\left(\frac{\rxyt^2\rxytp^2}{\rxxtp^2\ryytp^2}\right)\Xi_{\rm LO}(\xt,\yt;\xt',\yt')\,.\label{eq:slowLL-LO}
\end{align}
This result is absorbed into the last term of Eq.\,\eqref{eq:xsec-decomposition} building the JIMWLK kernel associated with the color correlator $\Xi_{\rm LO}$ in the leading logarithmic rapidity evolution.

The expression resulting from the subtraction of the rapidity divergence in Eq.\,\eqref{eq:slowLL-LO} from the sum in Eq.\,\eqref{eq:sum_IR_R2R2in_R2R2out} 
is given by\footnote{We use the label ``no-pole" to remind the reader that this sum is performed to cancel the infrared $1/\varepsilon$ poles between real and virtual NLO corrections.}
\begin{align}
    &\der\sigma_{\rm no-pole}=\frac{\alpha_{\rm em}e_f^2N_c\deltatwo}{(2\pi)^6}\int\der^8\Xt e^{-i\ktone\cdot\rxxtp-i\kttwo\cdot\ryytp}\Rcal_{\mathrm{LO}}^{\lambda}(\rxyt,\rxytp)\Xi_{\rm LO}(\xt,\yt;\xt',\yt')\nonumber\\
    &\times\frac{\alpha_sC_F}{\pi}\left\{-\ln\left(\frac{z_1}{z_f}\right)\ln\left(\frac{\rxxtp^2}{|\rxyt||\rxytp|}\right)-\ln\left(\frac{z_2}{z_f}\right)\ln\left(\frac{\ryytp^2}{|\rxyt||\rxytp|}\right)-\frac{3}{4}\ln\left(\frac{\ktone^2\kttwo^2\rxyt^2\rxytp^2}{c_0^4}\right)\right.\nonumber\\
    &\left.-3\ln(R)+\frac{1}{2}\ln^2\left(\frac{z_1}{z_2}\right)+\frac{11}{2}+3\ln(2)-\frac{\pi^2}{2}+\mathcal{O}(R^2)\right\}\,.
    \label{eq:IRC-final}
\end{align}
This is the principal update of our calculation of the NLO inclusive dijet cross-section subsequent to the results published in \cite{Caucal:2021ent}. The constant terms in the last line of Eq.\,\eqref{eq:IRC-final} depend on the jet algorithm; further, $R^2$  and higher power terms are neglected in this computation.

We will see that $\der\sigma_{\rm no-pole}$  contains a single Sudakov logarithm in the back-to-back limit which arises from the first two terms inside the curly brackets. Indeed since parametrically $r_{xy}\sim r_{x'y'}\sim 1/P_\perp$ and $r_{xx'}\sim r_{yy'}\sim 1/q_\perp$,  we can write the contribution from these two terms as 
\begin{align}
    \der\sigma_{\rm no-pole, sud1}&=\frac{\alpha_{\rm em}e_f^2N_c\deltatwo}{(2\pi)^6}\int\der^8\Xt e^{-i\ktone\cdot\rxxtp-i\kttwo\cdot\ryytp}\Rcal_{\mathrm{LO}}^{\lambda}(\rxyt,\rxytp)\Xi_{\rm LO}(\xt,\yt;\xt',\yt')\nonumber\\
    &\times\frac{\alpha_sC_F}{\pi}\left\{-\ln\left(\frac{z_1}{z_f}\right)\ln\left(\frac{\rxxtp^2}{|\rxyt||\rxytp|}\right)-\ln\left(\frac{z_2}{z_f}\right)\ln\left(\frac{\ryytp^2}{|\rxyt||\rxytp|}\right)\right\}\,.
    \label{eq:IRC-Sud1}
\end{align}

\paragraph{Single Sudakov logarithm from the diagram $\rm V_3\times LO$.} We turn now to the regular contributions of the virtual cross-section  computed in \cite{Caucal:2021ent}. Amongst those, only  $\rm V_3\times \rm LO$ requires additional discussion; this is because its  singularity structure is tied to that of $\rm R_2\times R_2'$. The full expression for $\rm V_3\times \rm LO$  is
\begin{align}
    &\der\sigma_{\rm V3\times \rm LO}= \frac{\alpha_{\rm em}e_f^2N_c\deltatwo}{(2\pi)^6}\int\der^8\Xt e^{-i\ktone\cdot\rxxtp-i\kttwo\cdot\ryytp} 8z_1^3z_{2}^3Q^2K_0(\bar Qr_{x'y'})K_0(\bar Q_{\rm V3} r_{xy})\Xi_{\rm NLO,3}
    \nonumber \\
    & \times \frac{\alpha_s}{\pi} \int_0^{z_1}\frac{\der z_g}{z_g}\left\{  \left[\left(1-\frac{z_g}{z_1}\right)^2\left(1+\frac{z_g}{z_2}\right)(1+z_g) e^{i(\Pt+z_g\qt)\cdot \rxyt} K_0(-i\Delta_{\mathrm{V}3}r_{xy}) \right. \right. \nonumber\\
    &\left.-\left(1-\frac{z_g}{2z_1}+\frac{z_g}{2z_2}-\frac{z_g^2}{2z_1z_2}\right) e^{i\frac{z_g}{z_1}\ktone \cdot \rxyt} \Jcal_{\odot}\left(\rxyt,\left(1-\frac{z_g}{z_1}\right)\Pt,\Delta_{\mathrm{V}3}\right)\right]+(1\leftrightarrow 2)\Bigg\}  \,.
    \label{eq:Vif-b}
\end{align}
The $z_g\to0$ singularity structure of  Eq.\,\eqref{eq:Vif-b} is given by  \cite{Caucal:2021ent}
\begin{align}
    &\der\sigma_{\mathrm{V}3\times \rm LO, soft-div}+c.c.=\frac{\alpha_{\rm em}e_f^2N_c\deltatwo}{(2\pi)^6}\int\der^8\Xt e^{-i\ktone\cdot\rxxtp-i\kttwo\cdot\ryytp}\Rcal_{\mathrm{LO}}^{\lambda}(\rxyt,\rxytp)\nonumber\\
    &\times\frac{(-\alpha_s)}{\pi}\int_{z_0}^{z_1}\frac{\der z_g}{z_g}\left[2\ln\left(\frac{z_g}{z_1z_2}\right)+\ln\left(\frac{\Pt^2|\rxyt||\rxytp|}{c_0^2}\right)\right]\Xi_{\rm NLO,3}(\xt,\yt,\xt',\yt')+(1\leftrightarrow 2)\,.
    \label{eq:dijet-NLO-slow-xsection3}
\end{align}
This expression contains a double logarithmic $\ln^2(z_0)$ singularity which cancels against an identical logarithmic singularity of the phase-space integral of $\der\sigma^{\gamma_{\rm L}^\star+A\to q \bar qg+X}_{\mathrm{R}2\times \mathrm{R}2',\rm div}$.
Combining Eq.\,\eqref{eq:dijet-NLO-slow-xsection3} with Eq.\,\eqref{eq:R2R2'div} (and its $\rm R_2'\times R_2$ mirror contribution), we find
\begin{align}
    &(\der\sigma_{\mathrm{V}3\times \rm LO, soft-div}+c.c.)+\der\sigma_{\rm R_2\times R_2',soft-div}+\der\sigma_{\rm R_2'\times R_2,soft-div}\nonumber\\
    &=\frac{\alpha_{\rm em}e_f^2N_c\deltatwo}{(2\pi)^6}\int\der^8\Xt e^{-i\ktone\cdot\rxxtp-i\kttwo\cdot\ryytp}\Rcal_{\mathrm{LO}}^{\lambda}(\rxyt,\rxytp)\nonumber\\
    &\times\frac{\alpha_s}{\pi}\left\{\ln\left(\frac{z_1}{z_0}\right)\ln\left(\frac{\rxypt^2}{|\rxyt||\rxytp|}\right)+\ln\left(\frac{z_2}{z_0}\right)\ln\left(\frac{\ryxpt^2}{|\rxyt||\rxytp|}\right)\right\}\Xi_{\rm NLO,3}(\xt,\yt;\xt',\yt')\,.
\end{align}
Introducing the factorization scale $k_f^-$ to isolate the rapidity divergence, we get the contribution building the JIMWLK kernel of the color correlator $\Xi_{\rm NLO,3}$:
\begin{align}
    \der\sigma_{\rm NLO_3, slow}=\frac{\alpha_{\rm em}e_f^2N_c\deltatwo}{(2\pi)^6}&\int\der^8\Xt e^{-i\ktone\cdot\rxxtp-i\kttwo\cdot\ryytp}\Rcal_{\mathrm{LO}}^{\lambda}(\rxyt,\rxytp)\nonumber\\
    &\times\frac{\alpha_s}{\pi}\ln\left(\frac{k_f^-}{\Lambda^-}\right)\ln\left(\frac{\rxypt^2\ryxpt^2}{\rxyt^2\rxytp^2}\right)\Xi_{\rm NLO,3}(\xt,\yt;\xt',\yt')\,,\label{eq:slowLL-NLO3}
\end{align}
with a finite leftover term leading to a single Sudakov logarithm in the back-to-back limit given by
\begin{equation}
    \frac{\alpha_s}{\pi}\left\{\ln\left(\frac{z_1}{z_f}\right)\ln\left(\frac{\rxypt^2}{|\rxyt||\rxytp|}\right)+\ln\left(\frac{z_2}{z_f}\right)\ln\left(\frac{\ryxpt^2}{|\rxyt||\rxytp|}\right)\right\}\,,
\end{equation}
which also leads to a single Sudakov in the back-to-back limit.

The term ``$\rm sud1$" is then defined as the sum of the $\der\sigma_{\rm no-pole, sud1}$ (Eq.\,\eqref{eq:IRC-Sud1}) and this leftover piece:
\begin{align}
    &\der\sigma_{\rm sud1}=\frac{\alpha_{\rm em}e_f^2N_c\deltatwo}{(2\pi)^6}\int\der^8\Xt e^{-i\ktone\cdot\rxxtp-i\kttwo\cdot\ryytp}\Rcal_{\mathrm{LO}}^{\lambda}(\rxyt,\rxytp)\times \frac{\alpha_s}{\pi} \nonumber\\
    &\times\left\{C_F\Xi_{\rm LO}(\xt,\yt;\xt',\yt')\left[\ln\left(\frac{z_f}{z_1}\right)\ln\left(\frac{\rxxtp^2}{|\rxyt||\rxytp|}\right)+\ln\left(\frac{z_f}{z_2}\right)\ln\left(\frac{\ryytp^2}{|\rxyt||\rxytp|}\right)\right]\right.\nonumber\\
    &\hspace{0.35cm}+\left.\Xi_{\rm NLO,3}(\xt,\yt;\xt',\yt')\left[\ln\left(\frac{z_1}{z_f}\right)\ln\left(\frac{\rxypt^2}{|\rxyt||\rxytp|}\right)+\ln\left(\frac{z_2}{z_f}\right)\ln\left(\frac{\ryxpt^2}{|\rxyt||\rxytp|}\right)\right]\right\}\,.
    \label{eq:app-sigma_Sud1}
\end{align}
The expression Eq.\,\eqref{eq:app-sigma_Sud1} is our final result for the term in the NLO impact factor leading to at most a single Sudakov logarithm in the back-to-back limit.

\subsection{Derivation of $\der\sigma_{\rm no-sud}$}
\label{subsub:no-sud}

Finally, the ``no-Sud" term groups together all $\mathcal{O}(\alpha_S)$ in the inclusive NLO impact factor which do not give rise to Sudakov logarithms in the back-to-back limit. We again decompose the virtual and real ``no-Sud" terms depending on the color correlator that contributes: either $C_F\Xi_{\rm LO}$, $\Xi_{\rm NLO,3}$ or any other correlators,
\begin{align}
    \der\sigma_{\rm R/V, no-sud}\equiv \der\sigma_{\rm R/V,no-sud,LO}+\der\sigma_{\rm R/V,no-sud, NLO_3}+\der\sigma_{\rm R/V, no-sud,other}\,.
    \label{eq:app-sigma-nosud}
\end{align}
In the following two paragraphs, we will provide the results for the individual terms in this expression. 

\paragraph{Virtual finite terms.} The virtual contribution without Sudakov enhancement in the back-to-back limit gathers the leftover of $\der\sigma_{\rm no-pole}$, the finite part of $\der\sigma_{\rm V_3}$, as well as the UV finite component of $\rm SE_1$ and diagram $\rm V_1$.

The left over of $\der\sigma_{\rm no-pole}$ gives the term proportional to $C_F\Xi_{\rm LO}$, with dependence on the photon polarization entering through the leading order perturbative factor only:
\begin{align}
    &\der\sigma_{\rm V,no-sud,LO}=\frac{\alpha_{\rm em}e_f^2N_c\deltatwo}{(2\pi)^6}\int\der^8\Xt e^{-i\ktone\cdot\rxxtp-i\kttwo\cdot\ryytp}\Rcal_{\mathrm{LO}}^{\lambda}(\rxyt,\rxytp)\Xi_{\rm LO}(\xt,\yt;\xt',\yt')\nonumber\\
    &\times\frac{\alpha_sC_F}{\pi}\left\{-\frac{3}{4}\ln\left(\frac{\ktone^2\kttwo^2\rxyt^2\rxytp^2}{c_0^4}\right)-3\ln(R)+\frac{1}{2}\ln^2\left(\frac{z_1}{z_2}\right)+\frac{11}{2}+3\ln(2)-\frac{\pi^2}{2}+\mathcal{O}(R^2)\right\}\,.
    \label{eq:app-V-CF}
\end{align}
The finite part of diagram $\rm V_3\times LO$ (plus its complex conjugate) defines the cross-section $\der\sigma_{\rm V,no-sud,NLO_3}$
\begin{align}
    \der\sigma_{\rm V,no-sud,NLO_3} = \left(\der\sigma_{\rm V3\times \rm LO}  - \der\sigma_{\mathrm{V}3\times \rm LO, soft-div} \right) + c.c \,,
    \label{eq:V-NLO3full}
\end{align}
which for longitudinally polarized photons is given by Eq.\,\eqref{eq:V-NLO3}.

Finally, the term involving other color correlators (namely $\Xi_{\rm NLO,1}$ and $\Xi_{\rm NLO,2}$) coming from the UV finite part of $\rm SE_1\times LO$, and $\rm V_1\times LO$ (plus their complex conjugates) defines
\begin{align}
    \der\sigma_{\rm V, no-sud,other}  =& \left( \left.\der\sigma_{\rm SE1 \times \rm LO}\right|_{\rm UV-fin.} + \left.\der\sigma_{\rm SE1' \times \rm LO}\right|_{\rm UV-fin.} + \der\sigma_{\rm V1 \times \rm LO} + \der\sigma_{\rm V1' \times \rm LO} \right. \nonumber \\
    & -\left.\der\sigma_{\rm SE1 \times \rm LO, slow}\right|_{\rm UV-fin.} -\left.\der\sigma_{\rm SE1' \times \rm LO, slow}\right|_{\rm UV-fin.}\nonumber\\
    &\left.- \der\sigma_{\rm V1 \times \rm LO, slow} - \der\sigma_{\rm V1' \times \rm LO, slow} \right) + c.c. \,,
    \label{eq:V-otherfull}
\end{align}
which for longitudinally polarized photons is given by Eq.\,\eqref{eq:V-other}. The subscript ``slow" refers to the part of these contributions which is absorbed into the JIMWLK evolution and thus must be subtracted from the impact factor.

The corresponding expressions for Eqs.\,\eqref{eq:V-NLO3full} and \eqref{eq:V-otherfull} for a transversely polarized photon are given in Appendix~\ref{app:transverse}. 

\paragraph{Real finite terms.} 
The sum of the ``no-Sud" component of $\rm R_2\times \rm R_2$ (see Eq.\,\eqref{eq:R2R2-dec}) and that of $\rm R_2'\times \rm R_2'$ defines
\begin{align}
    \der\sigma_{\rm R,no-sud,LO} = \der\sigma^{\gamma_{\lambda}^\star+A\to q \bar qg+X}_{\mathrm{R}2\times \mathrm{R}2,\rm no-Sud} + \der\sigma^{\gamma_{\lambda}^\star+A\to q \bar qg+X}_{\mathrm{R}2'\times \mathrm{R}2',\rm no-Sud} \,,
\end{align}
which expression for longitudinally polarized photon is given Eq.\,\eqref{eq:R-CF}.

Similarly, the sum of the regular ``no-sud" component of $\rm R_2\times \rm R_2'$ (see Eq.~\eqref{eq:R2R2'-dec}) and its complex conjugate defines
\begin{align}
    \der\sigma_{\rm R,no-sud,NLO_{3}} = \der\sigma^{\gamma_{\lambda}^\star+A\to q \bar qg+X}_{\mathrm{R}2'\times \mathrm{R}2,\rm no-Sud} + \der\sigma^{\gamma_{\lambda}^\star+A\to q \bar qg+X}_{\mathrm{R}2\times \mathrm{R}2',\rm no-Sud} \,,
\end{align}
which for longitudinally polarized photons leads to Eq.\,\eqref{eq:R-NLO3}. Note that the "no-sud" $\rm R_2'\times \rm R_2'$ and $\rm R_2'\times \rm R_2$ contributions can easily be inferred from these two formulas from quark-antiquark (or $1\leftrightarrow2$ interchange).

Finally, the contribution from real diagrams in which the gluon crosses the shockwave (in the amplitude, complex conjugate amplitude or both) gives Eq.\,\eqref{eq:dijet-NLO-long-real-other-final} for longitudinally polarized photons. The corresponding expressions for transversely polarized photons are given in Appendix~\ref{app:transverse}.

\section{NLO cross-section for transversely polarized virtual photons}
\label{app:transverse}

\subsection{Virtual cross-section without Sudakov enhancement}

The ``no-sud" virtual cross-section for transversely polarized photons is decomposed into three terms,
\begin{equation}
    \der\sigma^{\lambda=\rm T}_{\rm V,no-sud}=\der\sigma^{\lambda=\rm T}_{\rm V,no-sud,LO}+\der\sigma^{\lambda=\rm T}_{\rm V,no-sud,NLO_3}+\der\sigma^{\lambda=\rm T}_{\rm V, no-sud,other}\,,\label{eq:regvir-dec-T}
\end{equation}
with $\der\sigma^{\lambda=\rm T}_{\rm V,no-sud,LO}$ given by Eq.\,\eqref{eq:V-CF}, and
\begin{align}
    &\der\sigma^{\lambda=\rm T}_{\rm V,no-sud,other}=\frac{\alpha_{\rm em}e_f^2N_c\deltatwo}{(2\pi)^6}\int\der^8\Xt e^{-i\ktone\cdot\rxxtp-i\kttwo\cdot\ryytp} \ 2z_1^2z_2^2\frac{QK_1(\bar Qr_{x'y'})}{r_{x'y'}}\nonumber\\
    &\times\frac{\alpha_s}{\pi}\Bigg\{ \int_0^{z_1}\frac{\der z_g}{z_g}\int\frac{\der^2\zt}{\pi}\left\{e^{-i\frac{z_g}{z_1}\ktone\cdot\rzxt}\frac{\bar Q K_1(QX_V)}{X_V}\Xi_{\rm NLO,1}\left[-\frac{z_g(z_g-z_1)^2z_2}{2 z_1^3}\frac{\rzxt\cdot\rxytp}{\rzxt^2}\right.\right.\nonumber\\
    &\left.+(z_1^2+z_2^2)\left(1-\frac{z_g}{z_1}+\frac{z_g^2}{2z_1^2}\right)\frac{\RtS\cdot\rxytp}{\rzxt^2}\right]-(z_1^2+z_2^2)e^{-\frac{\rzxt^2}{\rxyt^2e^{\gamma_E}}}\left(1-\frac{z_g}{z_1}+\frac{z_g^2}{2z_1^2}\right)\frac{\rxyt\cdot\rxytp}{\rzxt^2}\nonumber\\
    &\times QK_1(\bar Qr_{xy})C_F\Xi_{\rm LO}-e^{-i\frac{z_g}{z_1}\ktone\cdot\rzxt}\frac{\bar QK_1( Q X_V)}{X_V}\Xi_{\rm NLO,1}\left[\frac{z_g(z_1-z_g)}{2(z_g+z_2)}\frac{\rzxt\cdot\rxytp}{\rzxt^2}\right.\nonumber\\
    &+[z_1(z_1-z_g)+z_2(z_2+z_g)]\left(1-\frac{z_g}{z_1}\right)\left(1+\frac{z_g}{z_2}\right)\left(1-\frac{z_g}{2z_1}-\frac{z_g}{2(z_2+z_g)}\right)\nonumber\\
    &\left.\left.\times\frac{(\RtV\cdot\rxytp)(\rzxt\cdot\rzyt)}{\rzxt^2\rzyt^2}+\frac{z_g(z_1-z_g)(z_g+z_2-z_1)^2}{2z_1^2z_2}\frac{(\RtV\times\rxytp)(\rzxt\times\rzyt)}{\rzxt^2\rzyt^2}\right]\right\}
    \nonumber\\
    &+(1\leftrightarrow 2)\Bigg\}+c.c.-\frac{\alpha_{\rm em}e_f^2N_c}{(2\pi)^6}\deltatwo\Theta(z_g-z_f)\times \textrm{``slow"}\,,
\end{align}
where we introduce the two transverse vectors:
\begin{align}
    \RtS&=\rxyt+\frac{z_g}{z_1}\rzxt \,,\\
    \RtV&=\rxyt-\frac{z_g}{z_{2}+z_g}\rzyt \,,
\end{align}
Finally, the second term in Eq.\,\eqref{eq:regvir-dec-T} reads
\begin{align}
    &\der\sigma^{\lambda=\rm T}_{\rm V,no-sud,NLO_3}=\frac{\alpha_{\rm em}e_f^2N_c\deltatwo}{(2\pi)^6}\int\der^8\Xt e^{-i\ktone\cdot\rxxtp-i\kttwo\cdot\ryytp}\ (2z_1z_2)\frac{\bar QK_1(\bar Qr_{x'y'})}{r_{x'y'}}\nonumber\\
    &\times\frac{\alpha_s}{\pi}\Bigg\{\int_0^{z_1}\frac{\der z_g}{z_g}\frac{\bar Q_{\rm V3}K_1(\bar{Q}_{\mathrm{V3}} r_{xy})}{r_{xy}}\Xi_{\rm NLO,3}(\xt,\yt;\xt',\yt')\nonumber\\
    &\times \left[\left[z_1(z_1-z_g)+z_2(z_2+z_g)\right](1+z_g)\left(1-\frac{z_g}{z_1}\right)e^{i(\Pt+z_g\qt)\cdot\rxyt} (\rxyt\cdot\rxytp)K_0(-i\Delta_{\rm V3}r_{xy})\right.\nonumber\\
    &-\left[z_1(z_1-z_g)+z_2(z_2+z_g)\right]\left(1-\frac{z_g}{2z_1}+\frac{z_g}{2z_2}-\frac{z_g^2}{2z_1z_2}\right)e^{i\frac{z_g}{z_1}\ktone\cdot\rxyt}(\rxyt\cdot\rxytp)\nonumber\\
    &\times\Jcal_{\odot}\left(\rxyt,\left(1-\frac{z_g}{z_1}\right)\Pt,\Delta_{\rm V3}\right)\nonumber\\
    &\left.-i\frac{z_g(z_g+z_2-z_1)^2}{z_1z_2}e^{i\frac{z_g}{z_1}\ktone\cdot\rxyt}(\rxyt\times\rxytp)\Jcal_{\otimes}\left(\rxyt,\left(1-\frac{z_g}{z_1}\right)\Pt,\Delta_{\rm V3}\right)\right]\nonumber\\
    &+[z_1^2+z_2^2]\frac{\rxyt\cdot\rxytp }{r_{xy}}\bar QK_1(\bar Qr_{xy}) \ln\left(\frac{z_g P_\perp r_{xy}}{c_0z_1z_2}\right)+(1\leftrightarrow 2)\Bigg\}+c.c. \,.\label{eq:V-nosud-NLO3-transverse}
\end{align}
with the $\Jcal_{\otimes}$ function computed in \cite{Caucal:2021ent} defined as
\begin{equation}
        \Jcal_{\otimes}(\rt,\Kt,\Delta)=\int\frac{\der^2 \lt}{(2\pi)}\frac{(-i)\lt \times \Kt \ e^{i\lt \cdot \rt}}{\lt^2\left[(\lt-\Kt)^2-\Delta^2- i \epsilon\right]}\,.
    \label{eq:Jtimes-def}
\end{equation}
The cross-section $\der\sigma_{\rm V, no-sud,NLO_3}$ contributes to the hard factor $\Hcal_{\rm NLO,2}^{\lambda=\textrm{T},ij}$, which can be obtained by taking the correlation limit of Eq.\,\eqref{eq:V-nosud-NLO3-transverse}:
\begin{align}
    &\Hcal_{\rm NLO,2}^{\lambda=\textrm{T},ij}(\Pt)=\frac{1}{2}\int\frac{\der^2\ut}{(2\pi)}\int\frac{\der^2\ut'}{(2\pi)}e^{-i\Pt\ruupt} \ \ut^i\ut'^j\Rcal_{\rm LO}^{\lambda=\rm T}(\ut,\ut')\nonumber\\
    &\times\int_0^{z_1}\frac{\der z_g}{z_g}\left\{\frac{\bar Q_{\rm V3}K_1(\bar{Q}_{\mathrm{V3}} u_\perp)}{\bar QK_1(\bar Q u_\perp)}\right.\nonumber\\
    &\times \left[\frac{\left[z_1(z_1-z_g)+z_2(z_2+z_g)\right]}{[z_1^2+z_2^2]}(1+z_g)\left(1-\frac{z_g}{z_1}\right)e^{i\Pt\cdot\ut} K_0(-i\Delta_{\rm V3}u_\perp)\right.\nonumber\\
    &-\frac{\left[z_1(z_1-z_g)+z_2(z_2+z_g)\right]}{[z_1^2+z_2^2]}\left(1-\frac{z_g}{2z_1}+\frac{z_g}{2z_2}-\frac{z_g^2}{2z_1z_2}\right)e^{i\frac{z_g}{z_1}\Pt\cdot\ut}\Jcal_{\odot}\left(\ut,\left(1-\frac{z_g}{z_1}\right)\Pt,\Delta_{\rm V3}\right)\nonumber\\
    &\left.-i\frac{z_g(z_g+z_2-z_1)^2}{z_1z_2(z_1^2+z_2^2)}e^{i\frac{z_g}{z_1}\Pt\cdot\ut}\frac{(\ut\times\ut')}{\ut\cdot\ut'}\Jcal_{\otimes}\left(\ut,\left(1-\frac{z_g}{z_1}\right)\Pt,\Delta_{\rm V3}\right)\right]\nonumber\\
    &+\ln\left(\frac{z_gP_\perp u_\perp}{c_0z_1z_2}\right)\Bigg\}+(1\leftrightarrow 2)\,.\label{eq:HNLO2-transverse}
\end{align}

\subsection{Real cross-section without Sudakov enhancement}

The real cross-section without Sudakov enhanced logarithms is decomposed in a similar fashion:
\begin{equation}
        \der\sigma^{\lambda=\rm T}_{\rm R,no-sud}=\der\sigma^{\lambda=\rm T}_{\rm R,no-sud,LO}+\der\sigma^{\lambda=\rm T}_{\rm R,no-sud,NLO_3}+\der\sigma^{\lambda=\rm T}_{\rm R, no-sud,other}\,,\label{eq:regreal-dec-T}
\end{equation}
with
\begin{align}
    &\der\sigma^{\gamma_{\rm T}^*+A\to q\bar qg+X}_{\rm R,no-sud,LO}=\frac{\alpha_{\rm em}e_f^2N_c}{(2\pi)^8}\int\der^8\Xt e^{-i\ktone\cdot\rxxtp-i\kttwo\cdot\ryytp}(4\alpha_s)C_F\Xi_{\rm LO}\frac{e^{-i\kgt \cdot (\xt-\xt')}}{(\kgt-\frac{z_g}{z_1}\ktone)^2}\nonumber\\
    &\times \left\{2z_1z_{2}\bar Q_{\rm R2}^2\left[z_{2}^2+(1-z_{2})^2\right]\left(1+\frac{z_g}{z_1}+\frac{z_g^2}{2z_1^2}\right)\frac{\rxyt\cdot\rxytp}{r_{xy}r_{x'y'}}K_1(\bar Q_{\mathrm{R}2}r_{xy})K_1(\bar Q_{\mathrm{R}2}r_{x'y'})\deltathree\right.\nonumber\\
    &\left.-\Rcal^{\rm T}_{\rm LO}(\rxyt,\rxytp)\Theta(z_1-z_g)\deltatwo\right\}+(1\leftrightarrow2)\,,
    \label{eq:dijet-NLO-trans-R2R2-final}
\end{align}
and
\begin{align}
    &\der\sigma^{\gamma_{\rm T}^*+A\to q\bar qg+X}_{\rm R,no-sud,NLO_3}=\frac{\alpha_{\rm em}e_f^2N_c}{(2\pi)^8}\int\der^8\Xt e^{-i\ktone\cdot\rxxtp-i\kttwo\cdot\ryytp}(-4\alpha_s)\Xi_{\rm NLO,3}\frac{e^{-i\frac{z_g}{z_1}\ktone\cdot\rxypt}}{\lt^2}\nonumber\\
    &\times\Bigg\{2z_1z_{2}\bar Q_{\rm R2}K_1(\bar Q_{\mathrm{R}2}r_{xy})\bar Q_{\rm R2'}K_1(\bar Q_{\mathrm{R}2'}r_{x'y'})\deltathree\nonumber\\
    &\times \left[\left(z_1+z_{2}-2z_1z_{2}\right)\left(1+\frac{z_g}{2z_1}+\frac{z_g}{2z_{2}}\right)e^{-i\lt\cdot\rxypt}\frac{\lt\cdot(\lt+\Kt)}{(\lt+\Kt)^2}\frac{\rxyt\cdot\rxytp}{r_{xy}r_{x'y'}}\right.\nonumber\\
    &\left.\left.-\frac{z_g}{8z_1z_{2}}(z_1-z_{2})^2e^{-i\lt\cdot\rxypt}\frac{\lt\times\Kt}{(\lt+\Kt)^2}\frac{\rxyt\times\rxytp}{r_{xy}r_{x'y'}}\right]\right.\nonumber\\
    &\left.-\Rcal_{\rm LO}^{\rm T}(\rxyt,\rxytp)\Theta\left(\frac{c_0^2}{\rxypt^2}\ge\lt^2\ge\Kt^2\right)\Theta(z_1-z_g)\deltatwo\right\}+(1\leftrightarrow2)\,.
    \label{eq:dijet-NLO-trans-R2R2'-final}
\end{align}
Finally, the term $\der\sigma^{\lambda=\rm T}_{\rm R,no-sud,other}$ is given by the sum of Eq.\,(B8)-(B9)-(B10)-(B11) in \cite{Caucal:2021ent}.

\section{Dependence of NLO impact factor on jet definitions}
\label{app:jet-algo}

In this appendix, we provide the value of the NLO impact factor for jet algorithms in the generalized $k_t$ family and in the small $R$ limit (meaning, up to power of $R^2$ suppressed corrections).

The in-cone condition is given by Eq.\,\eqref{eq:CA-alg}. In the small $R$ limit, this constraint only modifies the in-cone phase-space integral of diagram $\rm R_2\times R_2$ and $\rm R_2'\times R_2
'$. We define $\Delta\sigma^{k_t\rm-alg.}$ as the difference between the cross-section $\der\sigma_{\rm V,no-sud,LO}$ computed using generalized $k_t$ jet algorithms and the cross-section computed using the jet definition of Ivanov \& Papa \cite{Ivanov:2012ms}. We have, using Eqs.\,\eqref{eq:jetcone-smallR-2}-\eqref{eq:CA-alg} in Eq.\,\eqref{eq:dijet-NLO-long-R2R2-final},
\begin{align}
    \Delta\sigma^{k_t\rm-alg.}&=\der\sigma_{\rm LO}\times\frac{\alpha_s C_F}{\pi}\int_0^{z_1}\der z_g\left[\frac{1}{z_g}-\frac{1}{z_1}+\frac{z_g}{2z_1^2}\right]\ln\left(\frac{z_g^2(z_1-z_g)^2}{\textrm{min}(z_g^2,(z_1-z_g)^2)z_1^2}\right) + (1 \leftrightarrow 2) \nonumber \\
    &=\der\sigma_{\rm LO}\times\frac{\alpha_s C_F}{\pi}\left[3-\frac{\pi^2}{3}-3\ln(2)\right] \,. \label{eq:B-1}
\end{align}
This result is valid up to power of $R^2$ suppressed terms.
Eq.\,\eqref{eq:B-1} corresponds to the difference between the in-cone gluon phase-space integral of the $\rm R_2\times R_2$ and  $\rm R_2'\times R_2'$ contributions for the two jet definitions, which are manifest in the argument of the logarithm. This result agrees  with the calculations of NLO jet functions in SCET performed in \cite{Ellis:2010rwa,Kang:2016mcy}. 

If one uses a jet algorithm from the generalized $k_t$ family to define the dijet cross-section, one should add the term $\Delta\sigma^{k_t\rm-alg.}$ to the impact factor presented in this paper computed using the Ivanov \& Papa jet definition. Note that this NLO contribution completely factorizes from the LO cross-section and is a finite $\alpha_s$ correction that does not introduce additional $\langle \cos(n\phi)\rangle$ anisotropies relative to that given by the LO cross-section.

\section{$\langle \cos(n\phi)\rangle$ anisotropies for $n\ge4$}
\label{app:cosnphi}

In this Appendix, we compute the finite terms in the $\langle \cos(n\phi)\rangle$ anisotropies for even $n\ge4$. Only the soft terms labeled ``sud2" in the NLO cross-section contribute to these anisotropies. For a longitudinally polarized photon, one gets, using the Jacobi-Anger identity (Eq.~\eqref{eq:Jacobi-Anger}),
\begin{align}
    \der \sigma^{(n=2p),\lambda=\rm L}&= \alpha_{\rm em}\alpha_s e_f^2\deltatwo\Hcal_{\rm LO}^{0,\lambda=\rm L}(\Pt)\int\frac{\der^2\rbbpt}{(2\pi)^4}e^{-i\qt\cdot\rbbpt}\frac{\alpha_s}{\pi}\cos(n\theta)\left\{\hat G^0_{Y_f}(\rbbpt)\right.\nonumber\\
     &\times\int_0^1\frac{\der\xi}{\xi}(-1)^{p+1}\BesselJ_{n}(\xi|\Pt||\rbbpt|) \left[N_c\ln\left(\frac{\Pt^2\rbbpt^2\xi^2}{c_0^2}\right)+2C_F\ln(R^2)-\frac{1}{N_c}\ln(z_1z_2)\right]\nonumber\\
     &+\frac{1}{2}\hat h^0_{Y_f}(\rbbpt)\int_0^1\frac{\der\xi}{\xi}(-1)^p\left[\BesselJ_{n-2}(\xi|\Pt||\rbbpt|)+\BesselJ_{n+2}(\xi|\Pt||\rbbpt|)\right]\nonumber\\
     &\left.\times\left[N_c\ln\left(\frac{\Pt^2\rbbpt^2\xi^2}{c_0^2}\right)+2C_F\ln(R^2)-\frac{1}{N_c}\ln(z_1z_2)\right]\right\}\label{eq:R2R2b2b-cn}\,.
\end{align}
In the limit $P_\perp/q_\perp\to\infty$, the leading term in this expression reads
\begin{align}
     \der \sigma^{(n=2p),\lambda=\rm L}&= \alpha_{\rm em}\alpha_s e_f^2\deltatwo\Hcal_{\rm LO}^{0,\lambda=\rm L}(\Pt)\int\frac{\der^2\rbbpt}{(2\pi)^4}e^{-i\qt\cdot\rbbpt}\cos(n\theta) \nonumber\\
     &\times \frac{\alpha_s}{\pi}\left\{ \hat G^0_{Y_f}(\rbbpt)\frac{(-1)^{p+1}}{n}\left[2N_c\left(\mathfrak{H}(p)-\frac{1}{n}\right)+\left(2C_F\ln(R^2)-\frac{1}{N_c}\ln(z_1z_2)\right)\right]\right.\nonumber\\
     &+\hat h^0_{Y_f}(\rbbpt)\frac{(-1)^p}{n^2-4}\left[N_c\left((n+2)\mathfrak{H}(p-1)+(n-2)\mathfrak{H}(p+1)-\frac{2(n^2+4)}{n^2-4}\right)\right.\nonumber\\
     &\left.\left.+n\left(2C_F\ln(R^2)-\frac{1}{N_c}\ln(z_1z_2)\right)\right]\right\}\label{eq:R2R2b2b-cn-final}
\end{align}
where we used the identities (given below) in Appendix \ref{app:math-id}, and $\mathfrak{H}(p)$ is the $p^{\rm th}$ harmonic number defined by
\begin{equation}
    \mathfrak{H}(p)=\sum_{k=1}^p \frac{1}{k} \,.\label{eq:def-harmonicp}
\end{equation}
In the limit $n\to\infty$, using $\mathfrak{H}(p)\sim \ln(p)$, one finds
\begin{align}
     \der \sigma^{(n),\lambda=\rm L}\underset{n\to\infty}{\sim} &\alpha_{\rm em}\alpha_s e_f^2\deltatwo\Hcal_{\rm LO}^{0,\lambda=\rm L}(\Pt)\int\frac{\der^2\rbbpt}{(2\pi)^4}e^{-i\qt\cdot\rbbpt}\cos(n\theta)\left\{-\hat G^0_{Y_f}(\rbbpt)+\hat h^0_{Y_f}(\rbbpt)\right\}\nonumber\\
     &\times\frac{\alpha_sN_c}{\pi}\frac{2(-1)^{n/2}\ln(n)}{n}\label{eq:R2R2b2b-cn-infty}\,.
\end{align}
Unlike the zeroth and second harmonic of the differential cross-section, we note that the higher harmonics $\langle \cos(n\phi)\rangle$ with $n \geq 4$ do not possess double or single Sudakov logarithms. Furthermore, due to soft gluon radiation these harmonics are suppressed at large $n$, and the Fourier series should converge rapidly.

\section{Useful integrals}
\label{app:math-id}

\subsection{Integrals with Bessel functions}

For all $a>0$,
\begin{align}
    \int_0^a\frac{\der\xi}{\xi}\left(1-\BesselJ_0(x\xi)\right)&=\frac{1}{2}\ln\left(\frac{a^2x^2}{c_0^2}\right)+\mathcal{O}\left(x^{-3/2}\right)\,,\\
    \int_0^a\frac{\der\xi}{\xi}\left(1-\BesselJ_0(x\xi)\right)\ln\left(\frac{x^2\xi^2}{c_0^2}\right)&=\frac{1}{4}\ln^2\left(\frac{a^2x^2}{c_0^2}\right)+\mathcal{O}\left(\ln(x)x^{-3/2}\right)\,.
\end{align}
Recall $c_0=2e^{-\gamma_E}$, with $\gamma_E$ the Euler-Mascheroni constant.
For $n\ge 1$ and all $a>0$,
\begin{align}
    \lim\limits_{x\to \infty} \ \int_0^a\frac{\der\xi}{\xi}\BesselJ_{n}(x\xi)&=\frac{1}{n}\,,\\
   \lim\limits_{x\to\infty} \  \int_0^a\frac{\der\xi}{\xi}\BesselJ_{n}(x\xi)\ln\left(\frac{x^2\xi^2}{c_0^2}\right)&=\frac{2}{n^2}\left[n\mathfrak{H}\left(\frac{n}{2}\right)-1\right]\,.
\end{align}
where $\mathfrak{H}(p)$ is the $p^{\rm th}$ harmonic number defined in Eq.\,\eqref{eq:def-harmonicp}, which can also be defined for all real positive numbers in terms of the digamma function $\psi(z)$ using the identity
\begin{equation}
    \mathfrak{H}(p)=\gamma_E+\psi(p+1)\,.
\end{equation}

\subsection{Integral of the JIMWLK kernel with kinematic constraint}

We will compute here the integral 
\begin{equation}
    \mathcal{I}(X=|\rbbpt|Q)=\int_0^1\frac{\der z}{z}\int \der^2 \zt\frac{\rbbpt^2}{\rzbt^2\rzbpt^2}\Theta\left(\textrm{min}\left(\rzbt^2,\rzbpt^2\right)-\frac{1}{zQ^2}\right)\,.
\end{equation}
A priori, this integral depends on $\rbbpt$ and $Q$ but we shall see that it is actually a function of $|\rbbpt|Q$ only. This is expected from dimensional analysis since the integral is dimensionless.
After the change of variable $u=|\rzbt|/|\rbbpt|$, using the identity
\begin{align}
    \textrm{min}(\rzbt^2,\rzbpt^2)&=\frac{\rzbt^2+\rzbpt^2-|\rzbt^2-\rzbpt^2|}{2}\nonumber\\
    &=\frac{\rbbpt^2}{2}\left(1+2 u^2-2u\cos(\theta)-|1-2u\cos(\theta)|\right),
\end{align}
in polar coordinates, and the symmetry of the integral, one can write $\mathcal{I}(X)$ more simply as,
\begin{align}
    \mathcal{I}(X)&=2\int_{1/X}^\infty\frac{\der u}{u}\int_0^{2\pi}\der\theta \ \frac{\Theta(1-2u\cos(\theta))}{u^2-2u\cos(\theta)+1}\ln\left(u^2X^2\right)\,.
\end{align}
When $u\le1/2$, one has $1-2u\cos(\theta)\ge 0$. Then, the $u$ integral between $1/X$ and $1/2$ contributes as (assuming $X\ge 2$)
\begin{align}
    4\pi\int_{1/X}^{1/2}\frac{\der u}{u(1-u^2)}\ln(u^2X^2)=\pi&\left[\ln^2(X^2)-2\ln(3)\ln(X^2)+4\ln\left(\frac{3}{2}\right)\ln(2)\right.\nonumber\\
    &\left.-2\mathrm{Li}_2\left(\frac{1}{4}\right)+2\mathrm{Li}_2\left(\frac{1}{X^2}\right)\right]\,,
\end{align}
where $\mathrm{Li}_2$ is the dilogarithmic function. 
For $u\ge 1/2$, the integral is more complicated as one needs to cut the $\theta$ integration at $\theta=\arccos(1/(2u))$ and $\theta=2\pi-\arccos(1/(2u))$, and use
\begin{align}
    \int_{\arccos(1/(2u))}^{2\pi-\arccos(1/(2u))}  \frac{\der\theta}{u^2-2u\cos(\theta)+1} = \frac{4}{u^2-1}\arctan\left( \frac{u-1}{u+1} \sqrt{\frac{2u+1}{2u-1}} \right) \,.
\end{align}
After this rather tedious calculation, we end up with the  expression,
\begin{align}
    \mathcal{I}(X)&=\pi\left[\ln^2(X^2)-2\ln(3)\ln(X^2)+4\ln\left(\frac{3}{2}\right)\ln(2)-2\mathrm{Li}_2\left(\frac{1}{4}\right)+2\mathrm{Li}_2\left(\frac{1}{X^2}\right)\right]\nonumber\\
    &+8\int_{1/2}^\infty\frac{\der u}{u(u^2-1)}\ln(u^2X^2)\arctan\left(\frac{u-1}{1+u}\sqrt{\frac{2u+1}{2u-1}}\right)\,.
\end{align}
This formula can be simplified further using the identities
\begin{align}
    \int_{1/2}^\infty\frac{\der u}{u(u^2-1)}\arctan\left(\frac{u-1}{1+u}\sqrt{\frac{2u+1}{2u-1}}\right)&=\frac{\ln(3)\pi}{4}\,,\\
    \int_{1/2}^\infty\frac{\der u}{u(u^2-1)}\ln(u^2)\arctan\left(\frac{u-1}{1+u}\sqrt{\frac{2u+1}{2u-1}}\right)&=-\frac{\pi}{2}\left[\ln\left(\frac{3}{2}\right)\ln(2)-\frac{1}{2}\mathrm{Li}_2\left(\frac{1}{4}\right)\right]\,.
\end{align}
Our final expression for the kinematically constrained JIMWLK kernel $\mathcal{I}(X)$ is therefore (for $X\ge 2$),
\begin{align}
    \mathcal{I}(X)&=\pi\left[\ln^2(X^2)+2\mathrm{Li}_2\left(\frac{1}{X^2}\right)\right] \nonumber\\
    &=\pi\ln^2(X^2)+\mathcal{O}\left(\frac{1}{X^2}\right)\,,
\end{align}
where we have taken the limit $X\to\infty$ in the second line.
\newpage
\bibliographystyle{utcaps}
\bibliography{sudakov-ref}

\end{document}